\documentclass[preprint,amsmath,amssymb,aps,a4]{revtex4}

\usepackage{graphicx}
\usepackage{bm}
\usepackage{txfonts}
\usepackage{hyperref}
\newcommand{\be}{\begin{eqnarray}}
	\newcommand{\ee}{\end{eqnarray}}

\newcommand{\bfP}{{\bf P}_{\perp}}

\newcommand{\bfp}{{\bf p}_{\perp}}

\newcommand{\Dp}{{\bf \Delta}_{\perp}}

\usepackage{xcolor}
\usepackage{rotating}

\begin{document}
	\title{Unraveling sub-leading twist GTMDs of proton using LFQDM}
	\author{Shubham Sharma}
	\email{s.sharma.hep@gmail.com}
	\affiliation{Department of Physics, Dr. B. R. Ambedkar National Institute of Technology, Jalandhar 144008, India}
	\author{Sameer Jain}
	\email{sameerjainofficial@gmail.com}
	\affiliation{Department of Physics, Dr. B. R. Ambedkar National Institute of Technology, Jalandhar 144008, India}
	\author{Harleen Dahiya}
	\email{dahiyah@nitj.ac.in}
	\affiliation{Department of Physics, Dr. B. R. Ambedkar National Institute of Technology, Jalandhar 144008, India}
	
	\date{\today}
	%
\begin{abstract}
\noindent
This study investigates the sub-leading twist generalized transverse momentum dependent distributions (GTMDs) of a proton within the light-front quark-diquark model (LFQDM) framework. We solve the parametrization equations for the Dirac matrix structure to yield the explicit GTMD expressions for scalar as well as vector diquark configurations for active $u$ and $d$ quarks. This analysis addresses the multi-dimensional nature of GTMDs by exploring their dependencies on one or two variables while keeping others fixed. Additionally, we extract transverse momentum dependent form factors (TMFFs) from GTMDs by integrating over the longitudinal momentum fraction $x$. The study uses $3$-dimensional plots to illustrate the variation of TMFFs with the quark's transverse momentum $p_{\perp}$ and the transverse momentum transfer to the proton $\Delta_{\perp}$.
\vspace{0.1cm}
\end{abstract}
%
\maketitle
%
\section{Introduction\label{secintro}}
\noindent
At the constituent level, matter can be described as being composed of partons, which include quarks and gluons. This is where quantum chromodynamics (QCD) emerges as an essential tool, allowing us to characterize the dynamics of elementary particles using quark and gluon parameters while preserving their quantum nature \cite{Harindranath:1996hq, Brodsky:1997de}.
The structure of the proton is studied by analyzing the cross-section data from various scattering processes \cite{Chay:2013zya} including deep inelastic scattering (DIS) \cite{Collins:1981uk, Ji:2004wu, Collins:2004nx}, Drell-Yan (DY) \cite{Falciano:1986wk, Conway:1989fs, Zhu:2006gx}, semi-inclusive deep inelastic scattering (SIDIS) \cite{Arneodo:1986cf, Airapetian:1999tv, Avakian:2003pk, Airapetian:2004tw, Alexakhin:2005iw, Gregor:2005qv, Ageev:2006da, Airapetian:2005jc, Kotzinian:2007uv, Diefenthaler:2005gx, Airapetian:2008sk, Osipenko:2008rv, Giordano:2009hi, Gohn:2009, Airapetian:2009jy}, and deeply virtual Compton scattering (DVCS). The cross-sections of these scattering processes can be parameterized in terms of distribution functions (DFs). For example, DIS is associated with parton distribution functions (PDFs), which provide information about partons with longitudinal momentum fraction $x$ inside the proton \cite{Gluck:1994uf, Collins:1981uw, 9803445}. Transverse momentum dependent parton distributions (TMDs) encode the multi-dimensional structure and angular momentum information of the proton, corresponding to SIDIS and DY processes \cite{Collins:1981uk, Ji:2004wu, Collins:2004nx, Cahn:1978se, Konig:1982uk, Chiappetta:1986yg, Collins:1984kg, Sivers:1989cc, Efremov:1992pe, Collins:1992kk, Collins:1993kq, Kotzinian:1994dv, Mulders:1995dh, Boer:1997nt, Boer:1997mf, Boer:1999mm, Bacchetta:1999kz, Brodsky:2002cx, Collins:2002kn, Belitsky:2002sm, Burkardt:2002ks, Pobylitsa:2003ty, Goeke:2005hb, Bacchetta:2006tn, Cherednikov:2007tw, Brodsky:2006hj, Avakian:2007xa, Miller:2007ae, Arnold:2008kf, Brodsky:2010vs, lattice-TMD}. Generalized parton distributions (GPDs), which depend on the longitudinal momentum fraction $x$ and the momentum transfer to the hadron $\bf{\Delta}$, are related to the DVCS process \cite{Mueller:1998fv, Goeke:2001tz, Diehl03, Ji04, Belitsky05, Boffi07, Ji96, Brodsky06, Radyushkin97, Burkardt00, Diehl02, DC05, Hagler03, Kanazawa14, Rajan16, Ji97, Radyushkin:1996nd}.
%
%
%
%
\par
The generalized transverse momentum dependent distributions (GTMDs), also identified as ``mother distributions", are resultant functions when generalized parton correlation functions (GPCFs) are integrated over the quark's light-front momentum component \cite{Meissner:2008ay, Meissner:2009ww, Echevarria:2016mrc, Lorce13}. GPCFs are fully un-integrated, off-diagonal quark-quark correlators that depend on the quark's four-momentum and the momentum transfer to the hadron, hence  containing  complete theoretical information about the parton within the parent hadron. GTMDs are functions of the quark's longitudinal momentum fraction $x$, transverse momentum $\bfp$, and the four-momentum transfer to the hadron $\bf{\Delta}$. GTMDs can be obtained by the Fourier transform of Wigner distributions, which are the closest objects to classical phase space distributions in quantum mechanics \cite{Ji:2003ak,Belitsky:2003nz,Lorce11,Wigner32}. Being mother distributions, various DFs can be derived from GTMDs; for example, the application of the TMD limit and GPD limit yields TMDs and GPDs, respectively. The relationship between GPCFs, GTMDs, and various DFs is illustrated in Fig. (\ref{figtree}). The connection between GTMDs and spin-orbit correlations is detailed in Refs. \cite{Lorce:2011ni, Hatta:2011ku}. In Ref. \cite{Echevarria:2022ztg,Bhattacharya:2017bvs}, sufficient linkage between pion-nucleon double Drell-Yan (DDY) process, $\pi N \rightarrow (l_1^- + l_1^+) + (l_2^- + l_2^+) + N'$ and GTMDs calculations has been claimed. It is important to mention that considering the principle of uncertainty, the aforementioned quantum distributions are quasi-probabilistic in nature.
\begin{figure*}
	\centering
	\begin{minipage}[c]{0.98\textwidth}
		\includegraphics[width=0.96\textwidth]{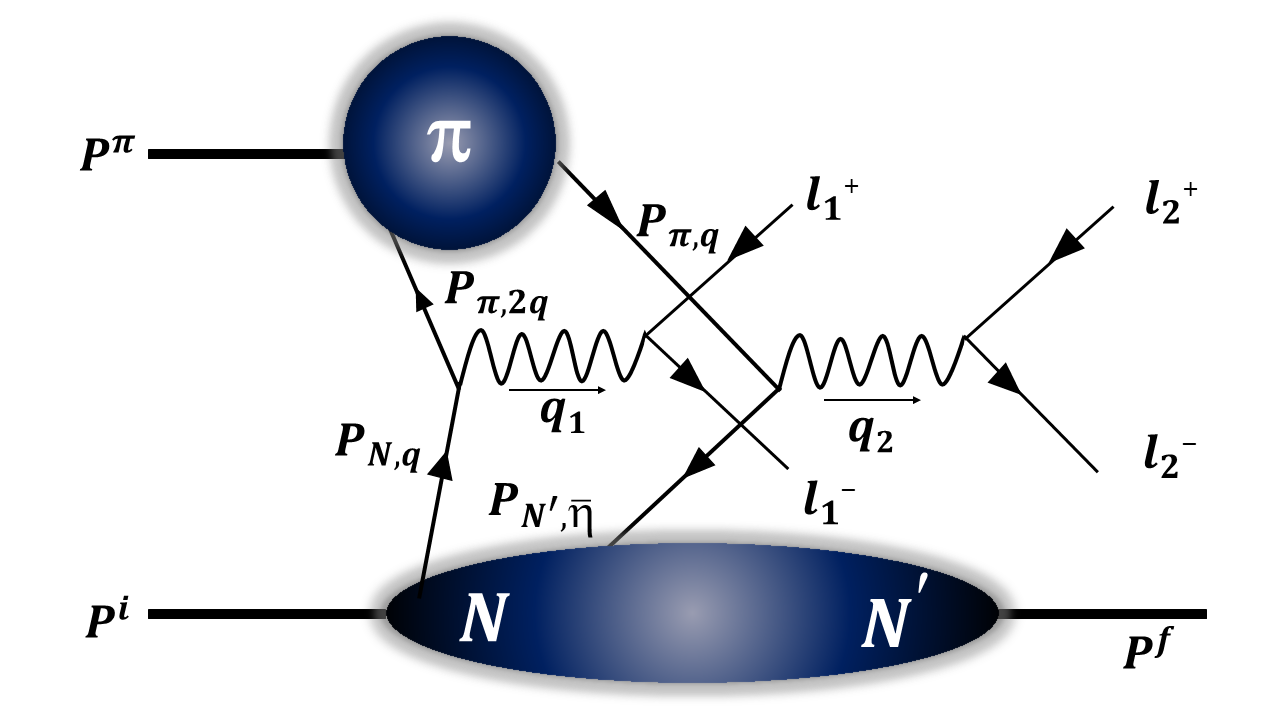}
	\end{minipage}
	\caption{\label{DDY} Visualization of the GTMDs-linked DDY process involving a pion and proton, $\pi + P^i \rightarrow (l_1^- + l_1^+) + (l_2^- + l_2^+) + P^f$.
	}
\end{figure*}
	\begin{figure*}
	\centering
	\begin{minipage}[c]{0.98\textwidth}
		\includegraphics[width=17cm]{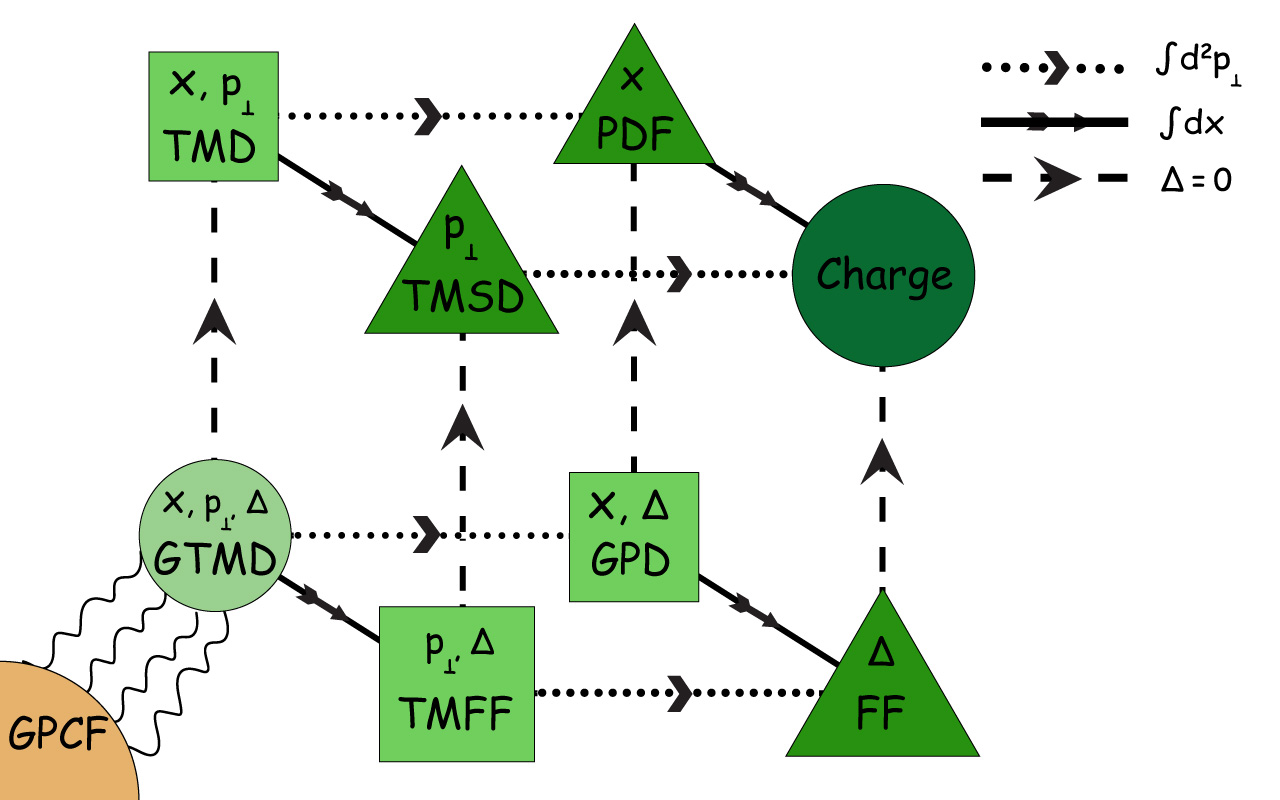}
		\hspace{0.05cm}\\
	\end{minipage}
	\caption{\label{figtree}
		The hierarchy of generalized parton correlation functions (GPCFs) is depicted, where various arrows on the GTMDs represent distinct limits. The integration over the quark's transverse momentum $\bfp$ is represented by a dotted line. The integration over the longitudinal momentum fraction $x$ is indicated by a solid line. The case with no momentum transfer is represented by a dashed line.
	}
\end{figure*}
%
%
\par
As a result of the non-perturbative aspects of QCD Lagrangian, analyzing the proton structure and the dynamics of its constituents using QCD is an exhaustive project. Consequently, many QCD-inspired models have been developed. For instance, the Valon model conceptualizes hadrons as bound states of valence and sea quarks \cite{Yazdi:2014zaa}. The light-front constituent quark model (LFCQM) characterizes hadrons in relation to constituent quarks using the light-front approach \cite{Pasquini:2008ax, Pasquini:2010af, Lorce:2011dv, Boffi:2009sh, Pasquini:2011tk, Lorce:2014hxa, Kofler:2017uzq, Pasquini:2018oyz, Rodini:2019ktv, Ma:2018ysi}. Quark-diquark models view hadrons as quark-diquark bound states and are particularly useful for baryon spectroscopy \cite{Jakob:1997wg, Gamberg:2007wm, Cloet:2007em, Bacchetta:2008af, She:2009jq, Lu:2012gu, Maji:2015vsa, Maji:2016yqo, Maji:2017bcz}. The bag models envision hadrons as bound states within a constant energy density ``bag," providing insights into hadron structure and properties \cite{Jaffe:1991ra, Yuan:2003wk, Courtoy:2008vi, Avakian:2008dz, Courtoy:2008dn, Avakian:2010br}.
%
%
\par
Light-front quark-diquark model (LFQDM) is one such model that assumes a proton to be composed primarily of an active quark along with a spectator diquark \cite{Maji:2016yqo,Chakrabarti:2019wjx}. The dynamics of this two-body system is described using light-front Anti-de Sitter (AdS) space/QCD which itself is very favorable while calculating different proton phenomena, for example, its agreement with Drell-Yan-West relation \cite{DY70, West70} as well as it follows the quark-counting rule \cite{Maji:2016yqo}. LFQDM has emerged as a successful model to unravel the intricate structure of the proton, demonstrating its application in analyzing experimental data about SIDIS spin asymmetry from well-known experiments such as HERMES and COMPASS \cite{Gurjar:2022rcl} and comparing the flavor combination of PDF $e(x)$ with recent CLAS data \cite{sstwist3}. In LFQDM, various aspects of proton structure have been extensively explored. This includes the analysis of transversity and helicity PDFs, demonstrating agreement with experimental findings as well as confirming the Soffer bound and empirical findings of axial and tensor charges. GPDs for quarks within the proton have been investigated in both the position and momentum spaces, showing diffraction patterns and fulfilling diquark model inequalities \cite{Mondal:2015uha}. Furthermore, gravitational form factors and mechanical properties of the proton, like mechanical radius, shear forces, and pressure distributions, have been studied \cite{Chakrabarti:2020kdc}. The transverse structure of the proton has been thoroughly examined in the  LFQDM which includes extensive studies of leading twist, sub-leading twist, twist-$4$ T-even TMDs \cite{Maji:2017bcz,sstwist3,sstwist4}. Investigations into the average square momentum and average square transverse momentum for higher twist TMDs have been undertaken in the LFQDM, demonstrating notable agreement with results from bag models and LFCQM. In the LFQDM, leading twist GTMDs and twist-$4$ GTMDs with zero skewness have been studied in Refs. \cite{majigtmd} and \cite{Sharma:2023tre}, respectively. 
%
%
%
\par
DF can be categorized by their twist, which indicates their appearance in the Operator Product Expansion (OPE) and their scaling behavior with respect to the hard scale $Q$ of a physical process \cite{Bhattacharya:2023jsc}. While leading twist contributions have been extensively studied, sub-leading twist contributions remain relatively understudied despite their relevance at current energy scales in experiments \cite{Zhu:2009uq}. Sub-leading twist and higher twist DFs for hadrons have been addressed in various works \cite{Avakian:2010br, Jakob:1997wg, Lorce:2014hxa, Pasquini:2018oyz, Kundu:2001pk, Mukherjee:2010iw, PhysRevLett.67.552, SIGNAL1997415, PhysRevD.95.074017, ELLIS19821, ELLIS198329, QIU1991105, QIU1991137, PhysRevD.83.054010, liu21, sstwist4}. GTMDs and the parameterization of quark GTMDs for spin-$0$ and spin-$\frac{1}{2}$ hadrons up to twist-$4$ have been detailed in Ref. \cite{Meissner:2008ay, Meissner:2009ww, Echevarria:2016mrc, Lorce13}. Leading twist quark GTMDs have been computed using models such as the LFCQM, light-cone quark model \cite{Ma:2018ysi}, chiral quark soliton model \cite{Lorce:2011dv}, and spectator model \cite{Meissner:2008ay, Meissner:2009ww}.
%
%
\par  
As modern experiments push the limits of energy, the significance of higher twist distributions in DFs related to cross-sections of various scattering processes increases, demanding greater precision. Several studies have focused on higher twist DFs, including chiral even GPDs at twist-$4$ \cite{Sharma:2023ibp}; T-even TMDs at sub-leading twist \cite{sstwist3} and twist-$4$ \cite{sstwist4} within the framework of LFQDM. Recently, twist-$4$ GTMDs have also been computed within LFQDM \cite{Sharma:2023tre}. Given that GTMDs provide different DFs under different limits, understanding sub-leading twist GTMDs is crucial for advancing our understanding of proton structure. Presently, no model has systematically examined GTMDs at sub-leading twist. This study aims to investigate sub-leading twist GTMDs for the proton using LFQDM. Within LFQDM, we derive sub-leading twist GTMDs for the proton, decoding the fully un-integrated quark-quark GTMD correlator for the sub-leading twist Dirac matrix structure. Explicit equations for sub-leading twist GTMDs are obtained by comparing them with parameterization equations. We derive these equations for both $u$ and $d$ quark possibilities from scalar and vector diquark components. Considering GTMDs as multi-dimensional functions, we explore their dependencies by varying one or two variables while holding others constant or integrated.
\par
Our work is organized as follows. Sec. \ref{secmodel} discusses the key components of the LFQDM, including its input parameters and other constants. Sec. \ref{seccor} contains details of the sub-leading twist quark-quark GTMD correlator and matching parameterization equations. Sec. \ref{secresults} presents the explicit results of the sub-leading twist GTMDs. Sec. \ref{secdiscussion} offers an interpretation of GTMDs using $2$-dimensional and $3$-dimensional plots. Sec. \ref{seccon} presents the conclusions.
%
%
	\section{Light-Front Quark-Diquark Model \label{secmodel}}
\noindent 
LFQDM is based on a straightforward concept that considers the interactions of incident particles with an active quark while the remaining two quarks act as spectators. The light-front wave functions (LFWFs) of this model not only account for nonperturbative effects through the effective degrees of freedom of the spectator diquark but also incorporate the SU$(4)$ structure of the nucleon. This structure includes the isoscalar-scalar diquark singlet $|u~ S^0\rangle$, isoscalar-vector diquark $|u~ A^0\rangle$, and isovector-vector diquark $|d~ A^1\rangle$ states \cite{Jakob:1997wg, Bacchetta:2008af}
	\begin{equation}
		|P; \pm\rangle = C_S|u~ S^0\rangle^\pm + C_V|u~ A^0\rangle^\pm + C_{VV}|d~ A^1\rangle^\pm. \label{PS_state}
	\end{equation}
In the above expression, $S$ and $A=V,VV$ represent the scalar and vector diquarks respectively, with their superscripts $(0,1)$ indicating their corresponding isospins. Calculations in Ref. \cite{Maji:2016yqo} suggest the values of the coefficients $C_{i}$ $\left(i = S, V, VV\right)$ to be
	\begin{equation}
		\begin{aligned}
			C_{S}^{2} &=1.3872, \\
			C_{V}^{2} &=0.6128, \\
			C_{V V}^{2} &=1.
		\end{aligned}
		\label{Eq3d1}
	\end{equation}
Standard light-front notations are used where $x$ denotes the longitudinal momentum fraction of the active quark, defined as $x = p^+/P^+$. We consider $p$ and $P_X$ to be the four-momentum of the active quark and diquark, respectively, which in light-front notations read as
\begin{eqnarray}
		p~ &&\equiv \bigg(xP^+, p^-,\bfp \bigg)\,,\label{qu} \\
		P_X~ &&\equiv \bigg((1-x)P^+,P^-_X,-\bfp\bigg). \label{diq}
\end{eqnarray}
In the case of a scalar diquark, the nucleon Fock-state for the angular momentum component $J^z = \pm 1/2$ can be formulated \cite{majiref25}
%
\begin{eqnarray}
		|u~ S\rangle^\pm &=&\sum_{\lambda^q}  \int \frac{dx~ d^2\bfp}{2(2\pi)^3\sqrt{x(1-x)}}  \psi^{\pm(\nu)}_{\lambda^q}\left(x,\bfp\right)\bigg|\lambda^{q},\lambda^{S}; xP^+,\bfp\bigg\rangle .\label{fockSD}
\end{eqnarray}
Similarly, for the case of a vector diquark, the state is
	\begin{eqnarray}
		|\nu~ A \rangle^\pm &=&\sum_{\lambda^q} \sum_{\lambda^D} \int \frac{dx~ d^2\bfp}{2(2\pi)^3\sqrt{x(1-x)}} \Bigg[ \psi^{\pm(\nu)}_{\lambda^q \lambda^D }\left(x,\bfp\right)\bigg|\lambda^{q},\lambda^{D}; xP^+,\bfp\bigg\rangle ,\label{fockVD}
	\end{eqnarray}
where $|\lambda^q,~\lambda^{Sp}; xP^+, \bfp\rangle$ is a two-particle state. Here, $\lambda^q = \pm \frac{1}{2}$ is the helicity of the active quark, while $\lambda^{Sp}$ represents the helicity of the diquark: $\lambda^{Sp}= \lambda^{S} = 0$ for scalar diquark, and $\lambda^{Sp}=\lambda^{D} = 0, \pm 1$ for vector diquark. Corresponding to the Fock-states considered in the above expressions of the nucleon state, the LFWFs are given in Table \ref{tab_LFWF} for reference, where $J^z = \pm 1/2$ and the function $\varphi_i^{(\nu)}(x,\bfp)$ introduced in Table \ref{tab_LFWF} is expressed as
	\begin{eqnarray}
		\varphi_i^{(\nu)}(x,\bfp)=\frac{4\pi}{\kappa}\sqrt{\frac{\log(1/x)}{1-x}}x^{a_i^\nu}(1-x)^{b_i^\nu}\exp\Bigg[-\delta^\nu\frac{\bfp^2}{2\kappa^2}\frac{\log(1/x)}{(1-x)^2}\bigg].
		\label{LFWF_phi}
	\end{eqnarray}
	\begin{table}[h]
		\centering 
		\begin{tabular}{ |p{1.5cm}|p{1.3cm}|p{1.1cm}|p{1.8cm} p{3.9cm}|p{1.8cm} p{3.9cm}|  }
			\hline
			~~$\rm{Diquark}$&~~~~~$\lambda^q$~&~~~$\lambda^{Sp}$~~&\multicolumn{2}{c|}{LFWFs for $J^z=+1/2$} & \multicolumn{2}{c|}{LFWFs for $J^z=-1/2$}\\
			\hline
			~~$\rm{Scalar}$&~~$+1/2$~~&~~$~~0$~~&~~$\psi^{+(\nu)}_{+}(x,\bfp)$~&~~$=~N_S~ \varphi^{(\nu)}_{1}$~~&~~$\psi^{-(\nu)}_{+}(x,\bfp)$~&~~$=~N_S \bigg(\frac{p^1-ip^2}{xM}\bigg)~ \varphi^{(\nu)}_{2}$~~  \\
			&~~$-1/2$~~&~~$~~0$~~&~~$\psi^{+(\nu)}_{-}(x,\bfp)$~&~~$=~-N_S\bigg(\frac{p^1+ip^2}{xM} \bigg)~ \varphi^{(\nu)}_{2}$~~&~~$\psi^{-(\nu)}_{-}(x,\bfp)$~&~~$=~N_S~ \varphi^{(\nu)}_{1}$~~   \\
			\hline
			~~$\rm{Vector}$&~~$+1/2$~~&~~$+1$~~&~~$\psi^{+(\nu)}_{+~+}(x,\bfp)$~&~~$=~~N^{(\nu)}_1 \sqrt{\frac{2}{3}} \bigg(\frac{p^1-ip^2}{xM}\bigg)~  \varphi^{(\nu)}_{2}$~~&~~$\psi^{-(\nu)}_{+~+}(x,\bfp)$~&~~$=~~0$~~  \\
			~~~~&~~$-1/2$~~&~~$+1$~~&~~$\psi^{+(\nu)}_{-~+}(x,\bfp)$~&~~$=~~N^{(\nu)}_1 \sqrt{\frac{2}{3}}~ \varphi^{(\nu)}_{1}$~~&~~$\psi^{-(\nu)}_{-~+}(x,\bfp)$~&~~$=~~0$~~   \\
			&~~$+1/2$~~&~~$~~0$~~&~~$\psi^{+(\nu)}_{+~0}(x,\bfp)$~&~~$=~~-N^{(\nu)}_0 \sqrt{\frac{1}{3}}~  \varphi^{(\nu)}_{1}$~~&~~$\psi^{-(\nu)}_{+~0}(x,\bfp)$~&~~$=~~N^{(\nu)}_0 \sqrt{\frac{1}{3}} \bigg( \frac{p^1-ip^2}{xM} \bigg)~  \varphi^{(\nu)}_{2}$~~   \\
			&~~$-1/2$~~&~~$~~0$~~&~~$\psi^{+(\nu)}_{-~0}(x,\bfp)$~&~~$=~~N^{(\nu)}_0 \sqrt{\frac{1}{3}} \bigg(\frac{p^1+ip^2}{xM} \bigg)~ \varphi^{(\nu)}_{2}$~~&~~$\psi^{-(\nu)}_{-~0}(x,\bfp)$~&~~$=~~N^{(\nu)}_0\sqrt{\frac{1}{3}}~  \varphi^{(\nu)}_{1}$~~   \\
			&~~$+1/2$~~&~~$-1$~~&~~$\psi^{+(\nu)}_{+~-}(x,\bfp)$~&~~$=~~0$~~&~~$\psi^{-(\nu)}_{+~-}(x,\bfp)$~&~~$=~~- N^{(\nu)}_1 \sqrt{\frac{2}{3}}~  \varphi^{(\nu)}_{1}$~~   \\
			&~~$-1/2$~~&~~$-1$~~&~~$\psi^{+(\nu)}_{-~-}(x,\bfp)$~&~~$=~~0$~~&~~$\psi^{-(\nu)}_{-~-}(x,\bfp)$~&~~$=~~N^{(\nu)}_1 \sqrt{\frac{2}{3}} \bigg(\frac{p^1+ip^2}{xM}\bigg)~  \varphi^{(\nu)}_{2}$~~   \\
			\hline
		\end{tabular}
		\caption{The LFWFs for the two distinct diquark scenarios in which $J^z=\pm1/2$, for the spectrum of active quark $\lambda^q$ and spectator diquark $\lambda^{Sp}$ helicities. $N_S$, $N^{(\nu)}_0$, and $N^{(\nu)}_1$ are the normalization constants \cite{Maji:2016yqo}.}
		\label{tab_LFWF}
	\end{table}
The LFWFs considered in our calculations were first introduced in Ref. \cite{Maji:2016yqo} to account for the inclusion of $C_V|u~ A^0\rangle^\pm$ and $C_{VV}|d~ A^1\rangle^\pm$ vector diquark states. These wave functions were developed based on predictions from soft-wall AdS/QCD, under the parameter conditions $a_i^{\nu} = b_i^{\nu} = 0$ and $\delta^\nu = 1.0$, they match exactly with those predictions. The energy scale of this model is taken to be $\mu_0 = 0.313\ {\mathrm GeV}$ \cite{Maji:2016yqo}. The values for the proton mass ($M = 0.938\ \mathrm{GeV}$), constituent quark mass ($m = 0.4\ \mathrm{GeV}$), and AdS/QCD scale parameter ($\kappa = 0.4\ \mathrm{GeV}$) are adopted from Refs. \cite{Chakrabarti:2013dda, Chakrabarti:2013gra}. Table \ref{tab_par} provides the model parameter values for both $u$ and $d$ flavors along with the normalization constants $N_{i}^{2}$.
	\begin{table}[h]
		\centering 
		\begin{tabular}{ |c|c|c|c|c|c|c|c| }
			\hline
			~~$\nu$~~&~~$a_1^{\nu}$~~&~~$b_1^{\nu}$~~&~~$a_2^{\nu}$~~&~~$b_2^{\nu}$~~&~~$N_{S}$~~&~~$N_0^{\nu}$~~&~~$N_1^{\nu}$~~   \\
			\hline
			~~$u$~~&~~$0.280\pm 0.001$~~&~~$0.1716 \pm 0.0051$~~&~~$0.84 \pm 0.02$~~&~~$0.2284 \pm 0.0035$~~&~~$2.0191$~~&~~$3.2050$~~&~~$0.9895$~~  \\
			~~$d$~~&~~$0.5850 \pm 0.0003$~~&~~$0.7000 \pm 0.0002$~~&~~$0.9434^{+0.0017}_{-0.0013}$~~&~~$0.64^{+0.0082}_{-0.0022}$~~&~~$0$~~&~~$5.9423$~~&~~$1.1616$~~    \\
			\hline
		\end{tabular}
		\caption{Normalization constants $N_{i}^{2}$ and model parameter values for both $u$ and $d$ quarks \cite{Maji:2016yqo}.}
		\label{tab_par} 
	\end{table}
%
Any physical observable $O$ can be expressed for $u$ and $d$ quarks in the LFQDM as a sum of contributions from the isoscalar-scalar ($O^{u(S)}$), isoscalar-vector ($O^{u(V)}$), and isovector-vector ($O^{d(VV)}$) diquark parts, respectively as
		\begin{eqnarray} 
		O^{u}&=&  ~O^{u(S)} + ~O^{u(V)}\,,\label{gtmdu} \\
		O^{d} &=&  ~O^{d(VV)}\,\label{gtmdd}.
	\end{eqnarray}
%
%
	\section{Sub-leading twist GTMDs in LFQDM}\label{seccor}
\noindent
This section contains a comprehensive analysis of the sub-leading twist GTMDs in the LFQDM. Ref. \cite{Meissner:2009ww} defines the fully un-integrated quark-quark correlator for a spin-$\frac{1}{2}$ hadron at constant light-front time $z^+ = 0$ as 
\begin{eqnarray} 
	W^{\nu [\Gamma]}_{[\Lambda^{N_i}\Lambda^{N_f}]}(x, p_{\perp},\Delta_{\perp},\theta)=\frac{1}{2}\int \frac{dz^-}{(2\pi)} \frac{d^2z_T}{(2\pi)^2} e^{ip.z} 
	\langle P^{f}; \Lambda^{N_f} |\bar{\psi} (-z/2)\Gamma \mathcal{W}_{[-z/2,z/2]} \psi (z/2) |P^{i};\Lambda^{N_i}\rangle \bigg|_{z^+=0}\,. \hspace{0.75cm}
	\label{corr}
\end{eqnarray}
\noindent
We take $|P^{i};\Lambda^{N_i}\rangle$ and $|P^{f}; \Lambda^{N_f}\rangle$ to be the initial and final states of the proton with helicities $\Lambda^{N_i}$ and $\Lambda^{N_f}$, respectively, while $\psi$ denotes the quark field operator. The pictorial representation of the linked DDY process $\pi + P^i \rightarrow (l_1^- + l_1^+) + (l_2^- + l_2^+) + P^f$, where di-lepton pair is observed in the final state along with the proton, has been given in Fig. \ref{DDY}. Following Ref. \cite{sstwist4}, for skewness $\xi = -\Delta^+/2P^+ = 0$, we write the GTMD correlator $W^{\nu [\Gamma]}_{[\Lambda^{N_i}\Lambda^{N_f}]}(x,\xi, p_{\perp},\Delta_{\perp},\bfp \cdot \Dp)$ as $W^{\nu [\Gamma]}_{[\Lambda^{N_i}\Lambda^{N_f}]}(x, p_{\perp}, \Delta_{\perp}, \theta)$ or simply $W^{\nu [\Gamma]}_{[\Lambda^{N_i}\Lambda^{N_f}]}$, where $\Gamma$ represents sub-leading twist Dirac matrices. The Wilson line $\mathcal{W}_{[-z/2,z/2]}$ ensures $SU(3)$ color gauge invariance in the resulting bilocal quark operator, which we set as $1$ for simplicity. In the scenario of a symmetric frame, we define the average momentum of the proton as $P = \frac{1}{2} (P^{f} + P^{i})$, while the momentum transfer is indicated by $\Delta = (P^{f} - P^{i})$. Along with the four-momentum of initial and final states, the kinematics of the proton can be represented as
		\begin{eqnarray}
		P &\equiv& \bigg(P^+,\frac{M^2+\Dp^2/4}{P^+},\textbf{0}_\perp\bigg)\,,\\
		\Delta &\equiv& \bigg(0, 0,\Dp \bigg)\,,\\
		P^{i} &\equiv& \bigg(P^+,\frac{M^2+\Dp^2/4}{P^+},-\Dp/2\bigg)\,,\label{Pp}\\
		P^{f} &\equiv& \bigg(P^+,\frac{M^2+\Dp^2/4}{P^+},\Dp/2\bigg)\,. \label{Ppp}
	\end{eqnarray}
The momentum of the active quark ($p$) (diquark ($P_X$)) in the initial and final states can be obtained using Eq.~(\ref{qu}) (Eq.~(\ref{diq})). Here, $t = \Delta^2 = -\Dp^2$ denotes the square of the total transverse momentum transfer. By substituting the nucleon Fock-state expressions given in Eq.~(\ref{fockSD}) and Eq.~(\ref{fockVD}) into the quark-quark correlator in Eq.~(\ref{corr}) and using Eq.~(\ref{PS_state}), the GTMD correlator can be expressed for the scalar and vector diquark components in terms of the overlaps of the LFWFs provided in Table \ref{tab_LFWF}. This generalizes the solution of the GTMD correlator for scalar and vector diquarks as
	\begin{eqnarray} 
		W^{ [\Gamma](S)}_{[\Lambda^{N_i}\Lambda^{N_f}]}&=&\frac{C_{S}^{2}}{16\pi^3} \sum_{\lambda^{q_i}} \sum_{\lambda^{q_f}} \psi^{\Lambda^{N_f}\dagger}_{\lambda^{q_f}}\left(x,\bfp+(1-x)\frac{\Dp}{2}\right)\psi^{\Lambda^{N_i}}_{\lambda^{q_i}}\left(x,\bfp-(1-x)\frac{\Dp}{2}\right) \nonumber\\  &&\frac{u^{\dagger}_{\lambda^{q_f}}\left(x P^{+},\bfp+\frac{\Dp}{2}\right)\gamma^{0} \Gamma u_{\lambda^{q_i}}\left(x P^{+},\bfp-\frac{\Dp}{2}\right)}{2 x P^{+}}\,, \label{cors} \\
		W^{ [\Gamma](A)}_{[\Lambda^{N_i}\Lambda^{N_f}]}&=&\frac{C_{A}^{2}}{16\pi^3} \sum_{\lambda^{q_i}} \sum_{\lambda^{q_f}} \sum_{\lambda^{D}} \psi^{\Lambda^{N_f}\dagger}_{\lambda^{q_f} \lambda^D}\left(x,\bfp+(1-x)\frac{\Dp}{2}\right)\psi^{\Lambda^{N_i}}_{\lambda^{q_i}\lambda^D}\left(x,\bfp+(1-x)\frac{\Dp}{2}\right) \nonumber\\  &&\frac{u^{\dagger}_{\lambda^{q_f}}\left(x P^{+},\bfp+\frac{\Dp}{2}\right)\gamma^{0} \Gamma u_{\lambda^{q_i}}\left(x P^{+},\bfp+\frac{\Dp}{2}\right)}{2 x P^{+}}\,. \label{corv} 
	\end{eqnarray} 
In this case, $C_A = C_V, C_{VV}$ for the $u$ and $d$ quarks in sequential order. The spinor product associated with the Dirac matrices at sub-leading twist $(\Gamma= 1, \gamma_5, \gamma^j, \gamma^j \gamma_5, i\sigma^{ij} \gamma_5$, and $ i\sigma^{+-} \gamma_5)$ is shown by the symbol $u^{\dagger}_{\lambda^{q_f}}\left(x P^{+},\bfp+\frac{\Dp}{2}\right)\gamma^{0} \Gamma u_{\lambda^{q_i}}\left(x P^{+},\bfp-\frac{\Dp}{2}\right)$. Explicit expressions for Dirac spinors can be found in Refs. \cite{Harindranath:1996hq,Brodsky:1997de}. The quark helicity in both the beginning and the end state is represented by $\lambda^{q_i}$ and $\lambda^{q_f}$, respectively. An extra summing over diquark helicity $\lambda^{D}$ is required for the vector diquark.
%
%
%
%
%
For various Dirac matrix structure $(\Gamma)$ at sub-leading twist, there exist a total of 32 quark GTMDs $E_{2,1},~E_{2,2},~E_{2,3},~E_{2,4},~E_{2,5},~E_{2,6},~E_{2,7},~E_{2,8},$ $~F_{2,1},~F_{2,2},~F_{2,3},~F_{2,4},~F_{2,5},~F_{2,6},~F_{2,7},~F_{2,8},~G_{2,1},~G_{2,2},~G_{2,3},~G_{2,4},~G_{2,5},~G_{2,6},~G_{2,7},~G_{2,8},~H_{2,1},~H_{2,2},$ $~H_{2,3},~H_{2,4},~H_{2,5},~H_{2,6},~H_{2,7},$ and $H_{2,8}$. These GTMDs can be projected as \cite{Meissner:2009ww}
%
	\begin{eqnarray}
		W_{[\Lambda^{N_i}\Lambda^{N_f}]}^{[1]}
		&=& \frac{1}{2P^+} \, \bar{u}(P^{f}, \Lambda^{N_F}) \, \bigg[
		{\color{blue}E_{2,1}}
		+ \frac{i\sigma^{i+} p_{\perp}^i}{P^+} \, 	{\color{blue}E_{2,2}}
		+ \frac{i\sigma^{i+} \Delta_{\perp}^i}{P^+} \, 	{\color{blue}E_{2,3}} \nonumber\\*
		& &  + \frac{i\sigma^{ij} p_{\perp}^i \Delta_{\perp}^j}{M^2} \, 	{\color{blue}E_{2,4}}
		\bigg] \, u(P^{i}, \Lambda^{N_i})
		\,, \label{par1}  \\
		W_{[\Lambda^{N_i}\Lambda^{N_f}]}^{[\gamma_5]}
		&=& \frac{1}{2P^+} \, \bar{u}(P^{f}, \Lambda^{N_F}) \, \bigg[
		- \frac{i\varepsilon_{\perp}^{ij} p_{\perp}^i \Delta_{\perp}^j}{M^2} \,	{\color{blue} E_{2,5}}
		+ \frac{i\sigma^{i+}\gamma_5 p_{\perp}^i}{P^+} \, 	{\color{blue}E_{2,6}}
		+ \frac{i\sigma^{i+}\gamma_5 \Delta_{\perp}^i}{P^+} \, 	{\color{blue}E_{2,7}} \nonumber\\*
		& &  + i\sigma^{+-}\gamma_5 \,	{\color{blue} E_{2,8}}
		\bigg] \, u(P^{i}, \Lambda^{N_i})
		\,, \label{par2}  	\\ 
		W_{[\Lambda^{N_i}\Lambda^{N_f}]}^{[\gamma^{j}]}
		&=& \frac{1}{2P^+} \, \bar{u}(P^{f}, \Lambda^{N_F}) \, \bigg[
		\frac{ p_{\perp}^j}{M} \,	{\color{blue} F_{2,1}}
		+ \frac{\Delta_{\perp}^j}{M} \,	{\color{blue} F_{2,2}}
		+ \frac{M \, i\sigma^{j+}}{k^+} \,	{\color{blue} F_{2,3}} \nonumber\\*
		& &  + \frac{p_{\perp}^j \, i\sigma^{~\rho +} p_{\perp}^{~\rho}}{M \, P^+} \,	{\color{blue}F_{2,4}}
		+ \frac{{\Delta_\perp}^j \, i\sigma^{~\rho +} p_T^{~\rho}}{M \, P^+} \,	{\color{blue} F_{2,5}}
		+ \frac{\Delta_{\perp}^j \, i\sigma^{~\rho +} \Delta_{\perp}^\rho}{M \, P^+} \,	{\color{blue} F_{2,6}} \nonumber\\*
		& &  + \frac{p_{\perp}^i \, i\sigma^{ij}}{M} \, 	{\color{blue}F_{2,7}}
		+ \frac{\Delta_{\perp}^i \, i\sigma^{ij}}{M} \, 	{\color{blue}F_{2,8}}
		\bigg] \, u(P^{i}, \Lambda^{N_i})
		\,, \label{par3}  	\\ 
		W_{[\Lambda^{N_i}\Lambda^{N_f}]}^{[\gamma^{j}\gamma_5]}
		&=& \frac{1}{2P^+} \, \bar{u}(P^{f}, \Lambda^{N_F}) \, \bigg[
		- \frac{i\varepsilon_{\perp}^{ij} p_{\perp}^i}{M} \, 	{\color{blue}G_{2,1}}
		- \frac{i\varepsilon_{\perp}^{ij} \Delta_{\perp}^i}{M} \, 	{\color{blue}G_{2,2}}
		+ \frac{M \, i\sigma^{j+}\gamma_5}{P^+} \, 	{\color{blue}G_{2,3}}\nonumber
	\\*
		& &  + \frac{p_{\perp}^j \, i\sigma^{~\rho +}\gamma_5 p_{\perp}^{~\rho}}{M \, P^+} \,	{\color{blue} G_{2,4}}
		+ \frac{{\Delta_\perp}^j \, i\sigma^{~\rho +}\gamma_5 p_{\perp}^{~\rho}}{M \, P^+} \,	{\color{blue} G_{2,5}}
		+ \frac{{\Delta_\perp}^j \, i\sigma^{~\rho +}\gamma_5 \Delta_{\perp}^{~\rho}}{M \, P^+} \, 	{\color{blue}G_{2,6}} \nonumber\\*
		& &  + \frac{p_{\perp}^j \, i\sigma^{+-}\gamma_5}{M} \, 	{\color{blue}G_{2,7}}
		+ \frac{\Delta_{\perp}^j \, i\sigma^{+-}\gamma_5}{M} \, 	{\color{blue}G_{2,8}}
		\bigg] \, u(P^{i}, \Lambda^{N_i})\,, \label{par4} \\ W_{[\Lambda^{N_i}\Lambda^{N_f}]}^{[i\sigma^{ij}\gamma_5]}
		&=& -\frac{i\varepsilon_{\perp}^{ij}}{2P^+} \, \bar{u}(P^{f}, \Lambda^{N_F}) \, \bigg[
			{\color{blue}H_{2,1}}
		+ \frac{i\sigma^{~\rho +} p_{\perp}^{~\rho}}{P^+} \, 	{\color{blue}H_{2,2}}
		+ \frac{i\sigma^{~\rho +} \Delta_{\perp}^\rho}{P^+} \, 	{\color{blue}H_{2,3}} \nonumber\\*
		& &  + \frac{i\sigma^{~\rho r} p_{\perp}^{~\rho} \Delta_{\perp}^r}{M^2} \, 	{\color{blue}H_{2,4}}
		\bigg] \, u(P^{i}, \Lambda^{N_i})
		\,, \label{par5} \\ 
		W_{[\Lambda^{N_i}\Lambda^{N_f}]}^{[i\sigma^{+-}\gamma_5]}
		&=& \frac{1}{2P^+} \, \bar{u}(P^{f}, \Lambda^{N_F}) \, \bigg[
		- \frac{i\varepsilon_{\perp}^{ij} p_{\perp}^i \Delta_{\perp}^j}{M^2} \, 	{\color{blue}H_{2,5}}
		+ \frac{i\sigma^{i+}\gamma_5 p_{\perp}^i}{P^+} \, 	{\color{blue}H_{2,6}}
		+ \frac{i\sigma^{i+}\gamma_5 \Delta_{\perp}^i}{P^+} \, 	{\color{blue}H_{2,7}} \nonumber\\*
		& &  + i\sigma^{+-}\gamma_5 \,	{\color{blue} H_{2,8}}
		\bigg] \, u(P^{i}, \Lambda^{N_i}). \label{par6}
	\end{eqnarray}
\noindent
In the above parameterization equations, the {\color{blue}blue} color highlights the sub-leading twist GTMDs. While computing the aforementioned expressions, we have adopted the same conventions as those in Ref. \cite{sstwist4}. Here, the indices $i$, $j$, $\rho$, and $r$ denote transverse directions.
%
%
%
	\section{Results}\label{secresults}
	\noindent
We use the scalar and vector diquark Fock-state expressions specified in Eqs. \eqref{fockSD} and \eqref{fockVD}, incorporating proper polarization into the quark-quark correlator Eq. \eqref{corr} via proton state Eq. \eqref{PS_state} to derive the equations for the GTMDs corresponding to both types of diquarks. To specify a particular GTMD, we employ Eqs. \eqref{par1} to \eqref{par6} to select the appropriate correlation (for $\Gamma = 1, \gamma_5, \gamma^j, \gamma^j \gamma_5, i\sigma^{ij} \gamma_5$, and $i\sigma^{+-} \gamma_5$) and suitably combine the proton's polarization. To maintain completeness and clarity, let's consider the GTMDs $E_{2,1}^{\nu}$, $H_{2,8}^{\nu}$, and $F_{2,1}^{\nu}$, corresponding to the Dirac matrix structures $1$, $i\sigma^{+-} \gamma_5$, and $\gamma^j$ respectively. These expressions can be derived using Eqs. \eqref{par1}, \eqref{par3}, and \eqref{par6} as
	\begin{eqnarray}
		E_{2,1}^{\nu}(x, p_{\perp},\Delta_{\perp},\theta) &=& \frac{P^+}{2 M^2}\Bigg( ~W^{\nu[1]}_{[++]}(x, p_{\perp},\Delta_{\perp},\theta)+W^{\nu[1]}_{[--]}(x, p_{\perp},\Delta_{\perp},\theta)\Bigg)\, ,\\
		H_{2,8}^{\nu}(x, p_{\perp},\Delta_{\perp},\theta) &=& \frac{P^+}{4 M}\Bigg( ~W^{\nu[i\sigma^{+-} \gamma_5]}_{[++]}(x, p_{\perp},\Delta_{\perp},\theta)-W^{\nu[i\sigma^{+-}\gamma_5]}_{[--]}(x, p_{\perp},\Delta_{\perp},\theta)\Bigg)\, ,\\
		F_{2,1}^{\nu}(x, p_{\perp},\Delta_{\perp},\theta) &=& \frac{P^+}{2(p_x\Delta_y-p_y\Delta_x)} 
		\Bigg(
		\Delta_y
		\Big( ~W^{\nu[\gamma^1]}_{[++]}(x, p_{\perp},\Delta_{\perp},\theta)+W^{\nu[\gamma^1]}_{[--]}(x, p_{\perp},\Delta_{\perp},\theta)\Big) \nonumber \\ 
		&&- \Delta_x
		\Big( ~W^{\nu[\gamma^2]}_{[++]}(x, p_{\perp},\Delta_{\perp},\theta)+W^{\nu[\gamma^2]}_{[--]}(x, p_{\perp},\Delta_{\perp},\theta)\Big)
		\Bigg)
		\,.
	\end{eqnarray}

The remaining GTMDs are able to be defined similarly concerning the GTMD correlator (Eq. \eqref{corr}) and, as such, in terms of the overlap of LFWFs using Eqs. \eqref{cors} and \eqref{corv}. To simplify the expressions, we define
	\begin{eqnarray}
		T_{ij}^{(\nu)}\left(x,p_{\perp},\Delta_{\perp}\right)&=&\varphi_i^{(\nu) \dagger}\left(x,\bfp+(1-x)\frac{\Dp}{2}\right) \varphi_j^{(\nu)}\left(x,\bfp-(1-x)\frac{\Dp}{2}\right)
		\label{Tij1},
	\end{eqnarray}
here $i,j=1,2$. Above equation along with LFWF, given in Eq. \eqref{LFWF_phi}, directly lead to the relations
	\begin{eqnarray}
		T_{ij}^{(\nu)}(x,p_{\perp},\Delta_{\perp})&=&T_{ji}^{(\nu)}(x,p_{\perp},\Delta_{\perp})\label{Tij2},\\
		\varphi_i^{(\nu)\dagger}\left(x,\bfp+(1-x)\frac{\Dp}{2}\right)&=&\varphi_i^{(\nu)}\left(x,\bfp+(1-x)\frac{\Dp}{2}\right)\label{Tij3}.
	\end{eqnarray}
For both scalar and vector diquarks, the expressions of GTMDs for the Dirac matrix structure $\Gamma=1$ are given as
	\begin{eqnarray} 
		%
		%
		xE_{2,1}^{\nu(S)} &=&  \frac{C_{S}^{2} N_s^2}{32 \pi^3}\frac{1}{M}\Bigg[  2m\Bigg(T_{11}^{\nu} +  \bigg(\bfp^2-\left(1-x\right)^2\frac{\Dp^2}{4} \bigg) \frac{T_{22}^{\nu}}{x^2 M^2}\Bigg)+  \left(1-x\right)\Dp^2 \frac{T_{12}^{\nu}}{xM}\Bigg], \label{e21s}\\
		%
		%
		xE_{2,1}^{\nu(A)} &=&  \frac{C_{A}^{2}}{32 \pi^3}  \bigg(\frac{1}{3} |N_0^\nu|^2+\frac{2}{3}|N_1^\nu|^2 \bigg)\frac{1}{M}\Bigg[  2m\Bigg(T_{11}^{\nu} +  \bigg(\bfp^2-\left(1-x\right)^2\frac{\Dp^2}{4} \bigg) \frac{T_{22}^{\nu}}{x^2 M^2}\Bigg)+\nonumber\\
		&&   \left(1-x\right)\Dp^2 \frac{T_{12}^{\nu}}{xM}\Bigg], \label{e21a}	\\
		%
		%
		xE_{2,2}^{\nu(S)} &=&  -\frac{C_{S}^{2} N_s^2}{16 \pi^3}  \left( \bfp \cdot \Dp\right) \frac{T_{22}^{\nu}}{x^2 M^2}, \label{e22s}\\
		%
		%
		xE_{2,2}^{\nu(A)} &=&  \frac{C_{A}^{2}}{16 \pi^3}  \bigg(\frac{1}{3} |N_0^\nu|^2 \bigg)\left( \bfp \cdot \Dp\right)  \frac{T_{22}^{\nu}}{x^2 M^2},
	 \label{e22a}\\
		%
		%
		%
		xE_{2,3}^{\nu(S)} &=&  \frac{C_{S}^{2} N_s^2}{32 \pi^3}\Bigg[ -T_{11}^{\nu}+\Bigg( 2 m\left(1-x\right) \frac{T_{12}^{\nu}}{xM}+\bigg(\bfp^2+(1-x)^2\frac{\Dp^2}{4} \bigg) \frac{T_{22}^{\nu}}{x^2 M^2}\Bigg)+\nonumber
						\end{eqnarray}
		\begin{eqnarray}
		&& 
		\frac{1}{2M}\Bigg( 2m\Bigg(T_{11}^{\nu} +  \bigg(\bfp^2-(1-x)^2\frac{\Dp^2}{4} \bigg) \frac{T_{22}^{\nu}}{x^2 M^2}\Bigg)+  (1-x)\Dp^2 \frac{T_{12}^{\nu}}{xM}\Bigg)\Bigg], \label{e23s}		\\  
		%
		%
		xE_{2,3}^{\nu(A)} &=&  \frac{C_{A}^{2}}{32 \pi^3}  \Bigg[\left( \frac{-1}{3} |N_0^\nu|^2\right) \Bigg[ \Bigg( -T_{11}^{\nu}+2 m\left(1-x\right) \frac{T_{12}^{\nu}}{xM}+\bigg(\bfp^2+(1-x)^2\frac{\Dp^2}{4} \bigg) \frac{T_{22}^{\nu}}{x^2 M^2}\Bigg)+\nonumber\\
		&& \bigg(\frac{1}{3} |N_0^\nu|^2+\frac{2}{3}|N_1^\nu|^2 \bigg)
		\frac{1}{2M}\Bigg(  2m\Bigg(T_{11}^{\nu} +  \bigg(\bfp^2-(1-x)^2\frac{\Dp^2}{4} \bigg) \frac{T_{22}^{\nu}}{x^2 M^2}\Bigg)+
		 (1-x)\Dp^2 \frac{T_{12}^{\nu}}{xM}\Bigg)\Bigg], \label{e23a}\\
		%
		%
		%
		xE_{2,4}^{\nu(S)} &=&  \frac{C_{S}^{2} N_s^2}{16 \pi^3}M \Bigg[\frac{T_{12}^{\nu}}{xM}-m (1-x)\frac{T_{22}^{\nu}}{x^2M^2}  \Bigg], \label{e24s}\\
		%
		%
		%
		xE_{2,4}^{\nu(A)} &=&  \frac{C_{A}^{2}}{16 \pi^3}  \bigg(\frac{1}{3} |N_0^\nu|^2-\frac{2}{3}|N_1^\nu|^2 \bigg)M \Bigg[\frac{T_{12}^{\nu}}{xM}-m (1-x)\frac{T_{22}^{\nu}}{x^2M^2}  \Bigg]. \label{e24a}
	\end{eqnarray}
Similarly, the expressions of GTMDs corresponding to the scalar and vector diquark for the Dirac matrix structure $\Gamma=\gamma_5$ can be expressed as 
	\begin{eqnarray} 
		%
		%
		xE_{2,5}^{\nu(S)} &=& 0, \label{e25s}\\
		%
		%
		xE_{2,5}^{\nu(A)} &=& 0, \label{e25a}\\
		%
		%
		xE_{2,6}^{\nu(S)} &=&  \frac{C_{S}^{2} N_s^2}{16 \pi^3} \left(  \bfp \cdot \Dp\right)  \frac{T_{22}^{\nu}}{x^2 M^2}, \label{e26s}\\
		%
		%
		xE_{2,6}^{\nu(A)} &=&  \frac{C_{A}^{2}}{16 \pi^3}  \bigg(-\frac{1}{3} |N_0^\nu|^2 \bigg) \left(  \bfp \cdot \Dp\right)  \frac{T_{22}^{\nu}}{x^2 M^2}, \label{e26a}\\
		%
		%
		%
		xE_{2,7}^{\nu(S)} &=&-\frac{C_{S}^{2} N_s^2}{32 \pi^3}\Bigg[   T_{11}^{\nu} +  \bigg(\bfp^2+(1-x)^2\frac{\Dp^2}{4} \bigg) \frac{T_{22}^{\nu}}{x^2 M^2} \Bigg], \label{e27s}\\  
		%
		xE_{2,7}^{\nu(A)} &=&  \frac{C_{A}^{2}}{32 \pi^3}\bigg(\frac{1}{3} |N_0^\nu|^2 \bigg) \Bigg[   T_{11}^{\nu} +  \bigg(\bfp^2+(1-x)^2\frac{\Dp^2}{4} \bigg) \frac{T_{22}^{\nu}}{x^2 M^2} \Bigg], \label{e27a}\\
		%
		%
		%
		xE_{2,8}^{\nu(S)} &=&  \frac{C_{S}^{2} N_s^2}{32 \pi^3} \frac{\left(  \bfp \cdot \Dp\right) }{M} \frac{T_{12}^{\nu}}{xM}, \label{e28s}\\
		%
		%
		%
		xE_{2,8}^{\nu(A)} &=&  \frac{C_{A}^{2}}{32 \pi^3}\bigg(\frac{1}{3} |N_0^\nu|^2-\frac{2}{3}|N_1^\nu|^2 \bigg) \frac{\left(  \bfp \cdot \Dp\right) }{M} \frac{T_{12}^{\nu}}{xM}. \label{e28a}
	\end{eqnarray}
The GTMD expressions corresponding to the Dirac matrix structure $\Gamma=\gamma^j$, for each of the scalar and vector diquark, can be expressed as 
	\begin{eqnarray} 
		%
		%
		xF_{2,1}^{\nu(S)} &=& \frac{C_{S}^{2} N_s^2}{16 \pi^3}\Bigg[  T_{11}^{\nu} +  \bigg(\bfp^2-(1-x)^2\frac{\Dp^2}{4}+(1-x)\frac{\Dp^2}{2} \bigg) \frac{T_{22}^{\nu}}{x^2 M^2}\Bigg], \label{f21s}\\
		%
		%
		xF_{2,1}^{\nu(A)} &=& \frac{C_{A}^{2}}{16 \pi^3}  \bigg(\frac{1}{3} |N_0^\nu|^2+\frac{2}{3}|N_1^\nu|^2 \bigg)\Bigg[  T_{11}^{\nu} +  \bigg(\bfp^2-(1-x)^2\frac{\Dp^2}{4}+(1-x)\frac{\Dp^2}{2} \bigg) \frac{T_{22}^{\nu}}{x^2 M^2}\Bigg], \label{f21a}		\\
		%
		%
		xF_{2,2}^{\nu(S)} &=&  -\frac{C_{S}^{2} N_s^2}{32 \pi^3} \left( 1-x\right) \left(  \bfp \cdot \Dp\right)  \frac{T_{22}^{\nu}}{x^2 M^2}, \label{f22s}\\
		%
		%
		xF_{2,2}^{\nu(A)} &=&  -\frac{C_{A}^{2}}{32 \pi^3} \bigg(\frac{1}{3} |N_0^\nu|^2+\frac{2}{3}|N_1^\nu|^2 \bigg)  \left( 1-x\right) \left(  \bfp \cdot \Dp\right)  \frac{T_{22}^{\nu}}{x^2 M^2}, \label{f22a}
		%
					\end{eqnarray}
		\begin{eqnarray}
		%
		%
		xF_{2,3}^{\nu(S)} &=&\frac{C_{S}^{2} N_s^2}{16 \pi^3}\Bigg[ \frac{ \left(  \bfp \cdot \Dp\right)}{M} \frac{T_{12}^{\nu}}{xM} + \frac{\bigg(\bfp^2 \Dp^2-\left(  \bfp \cdot \Dp\right)^2 \bigg)}{ M^2\left( \bfp \cdot \Dp\right)} \frac{1}{2} \bigg(2 T_{12}^{\nu}- T_{11}^{\nu}-\bigg(\bfp^2-(1-x)^2\frac{\Dp^2}{4} \nonumber\\
		&&+(1-x)\frac{\Dp^2}{2} \bigg) \frac{T_{22}^{\nu}}{x^2 M^2}\bigg)  \Bigg], \label{f23s}\\  
		%
		xF_{2,3}^{\nu(A)} &=&  \frac{C_{A}^{2}}{16 \pi^3}\Bigg[  \bigg(\frac{-1}{3} |N_0^\nu|^2 \bigg)  \frac{\left(  \bfp \cdot \Dp\right)}{M} \frac{T_{12}^{\nu}}{xM} + \frac{\bigg(\bfp^2 \Dp^2-\left(  \bfp \cdot \Dp\right)^2 \bigg)}{M^2\left(  \bfp \cdot \Dp\right)}\Bigg( \left(\frac{-1}{3} |N_0^\nu|^2 \right)T_{12}^{\nu}\nonumber\\
		&&-\frac{1}{2}\bigg(\frac{1}{3} |N_0^\nu|^2+\frac{2}{3}|N_1^\nu|^2 \bigg)\bigg(  T_{11}^{\nu}+\bigg(\bfp^2-(1-x)^2\frac{\Dp^2}{4}+(1-x)\frac{\Dp^2}{2} \bigg) \frac{T_{22}^{\nu}}{x^2 M^2} \Bigg)  \Bigg], \label{f23a}\\
		%
		%
		%
		xF_{2,4}^{\nu(S)} &=&  \frac{C_{S}^{2} N_s^2}{32 \pi^3} \frac{\Dp^2}{\left(  \bfp \cdot \Dp\right) } \Bigg[-2 T_{12}^{\nu}+ T_{11}^{\nu} +\bigg(\bfp^2-(1-x)^2\frac{\Dp^2}{4}+(1-x)\frac{\Dp^2}{2} \bigg) \frac{T_{22}^{\nu}}{x^2 M^2}\Bigg], \label{f24s}\\
		%
		%
		%
		xF_{2,4}^{\nu(A)} &=&  \frac{C_{A}^{2}}{32 \pi^3}\frac{\Dp^2}{\left(  \bfp \cdot \Dp\right) }\Bigg[ \bigg(\frac{1}{3} |N_0^\nu|^2 \bigg)2T_{12}^{\nu}+\bigg(\frac{1}{3} |N_0^\nu|^2+\frac{2}{3}|N_1^\nu|^2 \bigg)
		\bigg(  T_{11}^{\nu}+\bigg(\bfp^2-(1-x)^2\frac{\Dp^2}{4} \nonumber\\
	&&
	+(1-x)\frac{\Dp^2}{2} \bigg) \frac{T_{22}^{\nu}}{x^2 M^2}\bigg)\Bigg], \label{f24a}	
		\\
		%
		%
		xF_{2,5}^{\nu(S)} &=&  \frac{C_{S}^{2} N_s^2}{32 \pi^3}  \Bigg[ 2 T_{12}^{\nu}- T_{11}^{\nu} -\bigg(\bfp^2-(1-x)^2\frac{\Dp^2}{4}+(1-x)\frac{\Dp^2}{2} \bigg) \frac{T_{22}^{\nu}}{x^2 M^2}\Bigg], \label{f25s}\\
		%
		%
		xF_{2,5}^{\nu(A)} &=& \frac{C_{A}^{2}}{32 \pi^3}\Bigg[ \bigg(\frac{-1}{3} |N_0^\nu|^2 \bigg)2T_{12}^{\nu}-\bigg(\frac{1}{3} |N_0^\nu|^2+\frac{2}{3}|N_1^\nu|^2 \bigg)\bigg(  T_{11}^{\nu}+\bigg(\bfp^2-(1-x)^2\frac{\Dp^2}{4}+(1-x)\frac{\Dp^2}{2} \bigg)		\nonumber\\
	&&
		 \frac{T_{22}^{\nu}}{x^2 M^2}\bigg)\Bigg], \label{f25a}\\
		%
		%
		F_{2,6}^{\nu(S)} &=& \frac{C_{S}^{2} N_s^2}{16 \pi^3}\Bigg(M(1-x)\frac{ \left(  \bfp \cdot \Dp\right)}{\Dp^2}\frac{T_{12}}{xM}-\frac{M^2}{\Dp^2}\Bigg(\frac{\left(  \bfp \cdot \Dp\right)}{M} \frac{T_{12}^{\nu}}{xM} + \frac{\bigg(\bfp^2 \Dp^2-\left(  \bfp \cdot \Dp\right)^2 \bigg)}{2M^2\left(  \bfp \cdot \Dp\right)}\nonumber\\
		&&\left( 2T_{12}^{\nu}- T_{11}^{\nu}-\bigg(\bfp^2-(1-x)^2\frac{\Dp^2}{4}+(1-x)\frac{\Dp^2}{2} \bigg) \frac{T_{22}^{\nu}}{x^2 M^2}\right)  \Bigg)+\frac{ \left(  \bfp \cdot \Dp\right)}{2\Dp^2}\Bigg[ T_{11}^{\nu} \nonumber\\
		&&+  \bigg(\bfp^2-(1-x)^2\frac{\Dp^2}{4}+(1-x)\frac{\Dp^2}{2} \bigg) \frac{T_{22}^{\nu}}{x^2 M^2}\Bigg] - \left( 1-x\right) \left(  \bfp \cdot \Dp\right)  \frac{T_{22}^{\nu}}{4x^2 M^2}\Bigg] , \label{f26s}\\
		%
		%
		F_{2,6}^{\nu(A)} &=& \frac{C_{A}^{2}}{16 \pi^3}\Bigg[ \bigg(\frac{-1}{3} |N_0^\nu|^2 \bigg)M(1-x)\frac{ \left(  \bfp \cdot \Dp\right)}{\Dp^2}\frac{T_{12}}{xM}-\frac{M^2}{\Dp^2}\Bigg[  \bigg(\frac{-1}{3} |N_0^\nu|^2 \bigg)  \frac{\left(  \bfp \cdot \Dp\right)}{M} \frac{T_{12}^{\nu}}{xM} \nonumber\\
		&&+ \frac{\bigg(\bfp^2 \Dp^2-\left(  \bfp \cdot \Dp\right)^2 \bigg)}{2M^2\left(  \bfp \cdot \Dp\right)}\bigg( \bigg(\frac{-1}{3} |N_0^\nu|^2 \bigg)2T_{12}^{\nu}-\bigg(\frac{1}{3} |N_0^\nu|^2+\frac{2}{3}|N_1^\nu|^2 \bigg)\bigg(  T_{11}^{\nu}+\bigg(\bfp^2-(1-x)^2\frac{\Dp^2}{4}\nonumber\\
		&&+(1-x)\frac{\Dp^2}{2} \bigg) \frac{T_{22}^{\nu}}{x^2 M^2}\bigg)\bigg)\Bigg]+\frac{ \left(  \bfp \cdot \Dp\right)}{2\Dp^2} \bigg(\frac{1}{3} |N_0^\nu|^2+\frac{2}{3}|N_1^\nu|^2 \bigg)\Bigg[  T_{11}^{\nu}+\bigg(\bfp^2-(1-x)^2\frac{\Dp^2}{4}+(1-x)\frac{\Dp^2}{2} \bigg) \nonumber\\
		&& \frac{T_{22}^{\nu}}{x^2 M^2} \Bigg]-\bigg(\frac{1}{3} |N_0^\nu|^2+\frac{2}{3}|N_1^\nu|^2 \bigg)  \left( 1-x\right) \left(  \bfp \cdot \Dp\right)  \frac{T_{22}^{\nu}}{4x^3 M^2}\Bigg], \label{f26a}\\ 
		%
		%
		%
		xF_{2,7}^{\nu(S)} &=& -\frac{C_{S}^{2} N_s^2}{16 \pi^3} \left( 1-x\right) \left(  \bfp \cdot \Dp\right)  \frac{T_{22}^{\nu}}{x^2 M^2}, \label{f27s}	
		%
								\end{eqnarray}
		\begin{eqnarray}
		xF_{2,7}^{\nu(A)} &=&   -\frac{C_{A}^{2}}{16 \pi^3}\bigg(\frac{1}{3} |N_0^\nu|^2-\frac{2}{3}|N_1^\nu|^2 \bigg) \left( 1-x\right) \left(  \bfp \cdot \Dp\right)  \frac{T_{22}^{\nu}}{x^2 M^2}, \label{f27a}		\\
		%
		%
		%
		xF_{2,8}^{\nu(S)} &=&  \frac{C_{S}^{2} N_s^2}{32 \pi^3}  \Bigg[ T_{11}^{\nu}+\bigg(2\bfp^2 (1-x)-\bigg(\bfp^2-(1-x)^2 \frac{\Dp^2}{4}\bigg)  \bigg) \frac{T_{22}^{\nu}}{x^2 M^2}\Bigg], \label{f28s}\\
		%
		%
		%
		xF_{2,8}^{\nu(A)} &=&  \frac{C_{A}^{2}}{32 \pi^3}\bigg(\frac{1}{3} |N_0^\nu|^2-\frac{2}{3}|N_1^\nu|^2 \bigg)  \Bigg[ T_{11}^{\nu}+\bigg(2\bfp^2 (1-x)-\bigg(\bfp^2 -(1-x)^2 \frac{\Dp^2}{4}\bigg) \bigg) \frac{T_{22}^{\nu}}{x^2 M^2}\Bigg]. \label{f28a}
	\end{eqnarray}
The GTMD expressions for the Dirac matrix structure $ \gamma^j \gamma_5$, corresponding to each of the scalar and vector diquark, can be expressed as 
	\begin{eqnarray} 
		%
		%
		xG_{2,1}^{\nu(S)} &=&-\frac{C_{S}^{2} N_s^2}{16\pi^3}\left( 1-x\right) \left(\bfp \cdot \Dp\right)  \frac{T_{22}^{\nu}}{x^2 M^2}, \label{g21s}\\
		%
		%
		xG_{2,1}^{\nu(A)} &=& -\frac{C_{A}^{2}}{16\pi^3}\bigg(\frac{1}{3} |N_0^\nu|^2+\frac{2}{3}|N_1^\nu|^2 \bigg) \left( 1-x\right) \left(\bfp \cdot \Dp\right)  \frac{T_{22}^{\nu}}{x^2 M^2}, \label{g21a}\\
		%
		%
		xG_{2,2}^{\nu(S)} &=&   \frac{C_{S}^{2} N_s^2}{32 \pi^3}  \Bigg[- T_{11}^{\nu}+\bigg(2\bfp^2(1-x)-\bigg( \bfP^2-(1-x)\frac{\Dp^2}{4}\bigg) \bigg) \frac{T_{22}^{\nu}}{x^2 M^2}+2m\left( 1-x\right) \frac{T_{12}^{\nu}}{xM}\Bigg], \label{g22s}\\
		%
		xG_{2,2}^{\nu(A)} &=& \frac{C_{A}^{2}}{32 \pi^3}\bigg(\frac{1}{3} |N_0^\nu|^2+\frac{2}{3}|N_1^\nu|^2 \bigg)   \Bigg[-T_{11}^{\nu}+\bigg(2\bfp^2(1-x)-\bigg( \bfP^2-(1-x)\frac{\Dp^2}{4}\bigg) \bigg) \frac{T_{22}^{\nu}}{x^2 M^2}+\nonumber\\
		&&2m\left( 1-x\right) \frac{T_{12}^{\nu}}{xM}\Bigg], \label{g22a}\\
		%
		%
		%
		xG_{2,3}^{\nu(S)} &=& \frac{C_{S}^{2} N_s^2}{32 \pi^3} \Bigg[\frac{1}{M} \Bigg(2 m T_{11}^{\nu}+\left( 1-x\right) \Dp^2  \frac{T_{12}^{\nu}}{ x M}  + 2 m \bigg(\bfp^2-(1-x)^2\frac{\Dp^2}{4} \bigg)\frac{T_{22}^{\nu}}{x^2 M^2}\Bigg)-\nonumber\\
		&&\left(\bfp \cdot \Dp\right)^2\left( 1-x\right)  \frac{1}{M^2} \frac{T_{22}^{\nu}}{x^2 M^2}-\frac{\Dp^2}{2 M^2}\Bigg(T_{11}^{\nu}-\bigg(2\bfp^2(1-x)-\bigg( \bfP^2-(1-x)\frac{\Dp^2}{4}\bigg) \bigg)\nonumber\\
		&&\frac{T_{22}^{\nu}}{x^2 M^2}-2m \left( 1-x\right) \frac{T_{12}^{\nu}}{xM} \Bigg) \Bigg], \label{g23s}\\  
		%
		%
		xG_{2,3}^{\nu(A)} &=&  \frac{C_{A}^{2}}{32 \pi^3}	\Bigg[\bigg(-\frac{1}{3} |N_0^\nu|^2\bigg)\frac{1}{M} \Bigg(2 m T_{11}^{\nu}+\left( 1-x\right) \Dp^2  \frac{T_{12}^{\nu}}{ x M}  + 2 m \bigg(\bfp^2-(1-x)^2\frac{\Dp^2}{4} \bigg)\frac{T_{22}^{\nu}}{x^2 M^2}\Bigg)-\nonumber\\
		&&+\bigg(\frac{1}{3} |N_0^\nu|^2+\frac{2}{3}|N_1^\nu|^2 \bigg) \left(\bfp \cdot \Dp\right)^2\left( 1-x\right)  \frac{1}{M^2} \frac{T_{22}^{\nu}}{x^2 M^2}-\frac{\Dp^2}{2 M^2}\Bigg(T_{11}^{\nu}-\bigg(2\bfp^2(1-x)-\nonumber\\
		&&\bigg( \bfP^2-(1-x)\frac{\Dp^2}{4}\bigg) \bigg)\frac{T_{22}^{\nu}}{x^2 M^2}-2m \left( 1-x\right) \frac{T_{12}^{\nu}}{xM} \Bigg)\Bigg) \Bigg], \label{g23a}\\
		%
		%
		%
		xG_{2,4}^{\nu(S)} &=&  \frac{C_{S}^{2} N_s^2}{8\pi^3}  M\Bigg[ \frac{T_{12}^{\nu}}{xM}- m\frac{T_{22}^{\nu}}{x^2M^2 }\Bigg], \label{g24s}\\
		%
		%
		%
		xG_{2,4}^{\nu(A)} &=&   \frac{C_{A}^{2}}{8\pi^3}M\Bigg[\bigg(-\frac{1}{3} |N_0^\nu|^2\bigg) \bigg(\frac{T_{12}^{\nu}}{xM}-m\frac{T_{22}^{\nu}}{x^2 M^2}\bigg)\Bigg], \label{g24a}\\
		%
		%
		xG_{2,5}^{\nu(S)} &=&-\frac{C_{S}^{2} N_s^2}{32\pi^3}\left( 1-x\right) \left(\bfp \cdot \Dp\right)  \frac{T_{22}^{\nu}}{x^2 M^2}, \label{g25s}		\\
		%
		%
		xG_{2,5}^{\nu(A)} &=& -\frac{C_{A}^{2}}{32\pi^3}\bigg(\frac{1}{3} |N_0^\nu|^2+\frac{2}{3}|N_1^\nu|^2 \bigg) \left( 1-x\right) \left(\bfp \cdot \Dp\right)  \frac{T_{22}^{\nu}}{x^2 M^2}, \label{g25a}
		%
										\end{eqnarray}
		\begin{eqnarray}
		%
		xG_{2,6}^{\nu(S)} &=&  \frac{C_{S}^{2} N_s^2}{16 \pi^3}M\Bigg[ \Bigg( \frac{2\left(  \bfp \cdot \Dp\right)^2}{\Dp^4}\frac{T_{12}^{\nu}}{xM}+ \frac{m}{\Dp^2 }T_{11}^{\nu}-\bigg(\frac{2\left(  \bfp \cdot \Dp\right)^2}{\Dp^4}-\frac{\bfp^2}{\Dp^2}-\frac{ \left( 1-x\right)^2}{4}\bigg)m\frac{T_{22}^{\nu}}{x^2 M^2}\Bigg)\nonumber		\\
		&&-\frac{1}{2\Dp^2} \Bigg(\Bigg(\left( 1-x\right) \Dp^2  \frac{T_{12}^{\nu}}{ x M} +2 m T_{11}^{\nu} + 2 m \bigg(\bfp^2-(1-x)^2\frac{\Dp^2}{4} \bigg)\frac{T_{22}^{\nu}}{x^2 M^2}\Bigg)-\left(\bfp \cdot \Dp\right)^2\left( 1-x\right) \nonumber		\\
		&& \frac{1}{M} \frac{T_{22}^{\nu}}{x^2 M^2}-\frac{\Dp^2}{2 M}\Bigg(T_{11}^{\nu}-\bigg(2\bfp^2(1-x)-\bigg( \bfP^2-(1-x)\frac{\Dp^2}{4}\bigg) \bigg)\frac{T_{22}^{\nu}}{x^2 M^2}-2m \left( 1-x\right) \frac{T_{12}^{\nu}}{xM} \Bigg) \Bigg)\nonumber\\
		&&-\frac{2 \left( \bfp \cdot \Dp\right)^2}{\Dp^4}\Bigg( \frac{T_{12}^{\nu}}{xM}- m\frac{T_{22}^{\nu}}{x^2M^2 }\Bigg)
		+\frac{\left( \bfp \cdot \Dp\right)}{2\Dp^2}\left( 1-x\right) \left(\bfp \cdot \Dp\right) \frac{1}{M} \frac{T_{22}^{\nu}}{x^2 M^2}\Bigg], \label{g26s}\\
		%
		%
		xG_{2,6}^{\nu(A)} &=&  \frac{C_{A}^{2}}{16 \pi^3}M\Bigg[\bigg(-\frac{1}{3} |N_0^\nu|^2\bigg) \Bigg( \frac{2\left(  \bfp \cdot \Dp\right)^2}{\Dp^4}\frac{T_{12}^{\nu}}{xM}+ \frac{m}{\Dp^2 }T_{11}^{\nu}-\bigg(\frac{2\left(  \bfp \cdot \Dp\right)^2}{\Dp^4}-\frac{\bfp^2}{\Dp^2}-\frac{ \left( 1-x\right)^2}{4}\bigg)m\frac{T_{22}^{\nu}}{x^2 M^2}\Bigg)\nonumber\\
		&&-\frac{1}{2\Dp^2} \Bigg(\bigg(-\frac{1}{3} |N_0^\nu|^2\bigg) \Bigg(\left( 1-x\right) \Dp^2  \frac{T_{12}^{\nu}}{ x M} +2 m T_{11}^{\nu} + 2 m \bigg(\bfp^2-(1-x)^2\frac{\Dp^2}{4} \bigg)\frac{T_{22}^{\nu}}{x^2 M^2}\Bigg)\nonumber\\
		&&-\bigg(\frac{1}{3} |N_0^\nu|^2+\frac{2}{3}|N_1^\nu|^2 \bigg) \left(\bfp \cdot \Dp\right)^2\left( 1-x\right)  \frac{1}{M} \frac{T_{22}^{\nu}}{x^2 M^2}-\frac{\Dp^2}{2 M}\Bigg(T_{11}^{\nu}-\bigg(2\bfp^2(1-x)-\bigg( \bfP^2\nonumber\\
		&&-(1-x)\frac{\Dp^2}{4}\bigg) \bigg)\frac{T_{22}^{\nu}}{x^2 M^2}-2m \left( 1-x\right) \frac{T_{12}^{\nu}}{xM} \Bigg)\Bigg) \Bigg)
		-\frac{2 \left( \bfp \cdot \Dp\right)^2}{\Dp^4}\Bigg(\bigg(-\frac{1}{3} |N_0^\nu|^2\bigg)\Bigg( \frac{T_{12}^{\nu}}{xM}- m\frac{T_{22}^{\nu}}{x^2M^2 }\Bigg)\nonumber\\
		&&+\frac{\left( \bfp \cdot \Dp\right)}{2\Dp^2})\bigg(\frac{1}{3} |N_0^\nu|^2+\frac{2}{3}|N_1^\nu|^2 \bigg) \left( 1-x\right) \left(\bfp \cdot \Dp\right)  \frac{T_{22}^{\nu}}{x^2 M^2}\Bigg], \label{g26a}\\
		%
		%
		%
		%
		xG_{2,7}^{\nu(S)} &=&  \frac{C_{S}^{2} N_s^2}{32 \pi^3}  \Bigg[ T_{11}^{\nu}-2m\frac{T_{12}^{\nu}}{xM }-\bigg(\bfp^2-(1-x)^2\frac{\Dp^2}{4}+(1-x)\frac{\Dp^2}{2} \bigg) \frac{T_{22}^{\nu}}{x^2 M^2}\Bigg], \label{g27s}\\  
		%
		xG_{2,7}^{\nu(A)} &=&    \frac{C_{A}^{2}}{32 \pi^3}\bigg(\frac{1}{3} |N_0^\nu|^2-\frac{2}{3}|N_1^\nu|^2 \bigg)\Bigg[ T_{11}^{\nu}-2m\frac{T_{12}^{\nu}}{xM }-\bigg(\bfp^2-(1-x)^2\frac{\Dp^2}{4}+(1-x)\frac{\Dp^2}{2} \bigg) \frac{T_{22}^{\nu}}{x^2 M^2}\Bigg], \label{g27a}\\
		%
		%
		%
		xG_{2,8}^{\nu(S)} &=&  \frac{C_{S}^{2} N_s^2}{64 \pi^3}\left( 1-x\right) \left(\bfp \cdot \Dp\right)  \frac{T_{22}^{\nu}}{x^2 M^2}, \label{g28s}\\
		%
		%
		%
		xG_{2,8}^{\nu(A)} &=&  \frac{C_{A}^{2}}{64\pi^3}\bigg(\frac{1}{3} |N_0^\nu|^2-\frac{2}{3}|N_1^\nu|^2 \bigg) \left( 1-x\right) \left(\bfp \cdot \Dp\right)  \frac{T_{22}^{\nu}}{x^2 M^2}. \label{g28a}
	\end{eqnarray}
Likewise, the GTMD expressions corresponding to the Dirac matrix structure $i\sigma^{ij} \gamma_5$, for each of the scalar and vector diquark, can be expressed as
	\begin{eqnarray} 
		%
		%
		xH_{2,1}^{\nu(S)} &=&    -\frac{C_{S}^{2} N_s^2}{16 \pi^3} \frac{\left(  \bfp \cdot \Dp\right)\left( 1-x\right) }{M} \frac{T_{12}^{\nu}}{xM}, \label{h21s}	\\
		%
		%
		xH_{2,1}^{\nu(A)} &=&-\frac{C_{A}^{2} }{16 \pi^3} \bigg(\frac{1}{3} |N_0^\nu|^2+\frac{2}{3}|N_1^\nu|^2 \bigg)\frac{\left(  \bfp \cdot \Dp\right)\left( 1-x\right) }{M} \frac{T_{12}^{\nu}}{xM}, \label{h21a}		\\
		%
		%
		%
		xH_{2,2}^{\nu(S)} &=&\frac{C_{S}^{2} N_s^2}{16 \pi^3}\Bigg[   T_{11}^{\nu} +  \bigg(\bfp^2+(1-x)^2\frac{\Dp^2}{4} \bigg) \frac{T_{22}^{\nu}}{x^2 M^2} \Bigg], \label{h22s}	\\  
		%
		xH_{2,2}^{\nu(A)} &=&  \frac{C_{A}^{2} }{16 \pi^3}\bigg(-\frac{1}{3} |N_0^\nu|^2\bigg)\Bigg[T_{11}^{\nu} +  \bigg(\bfp^2+(1-x)^2\frac{\Dp^2}{4} \bigg) \frac{T_{22}^{\nu}}{x^2 M^2} \Bigg], \label{h22a}\\
		%
		%
		%
		xH_{2,3}^{\nu(S)} &=&  -\frac{C_{S}^{2} N_s^2}{32 \pi^3}\Bigg[ \left( 1-x\right)^2 \left(  \bfp \cdot \Dp\right)  \frac{T_{22}^{\nu}}{x^2 M^2}+\frac{\left(  \bfp \cdot \Dp\right)\left( 1-x\right) }{M} \frac{T_{12}^{\nu}}{xM}\Bigg], \label{h23s}
		%
											\end{eqnarray}
		\begin{eqnarray}
		%
		xH_{2,3}^{\nu(A)} &=& - \frac{C_{A}^{2}}{32 \pi^3}\Bigg[\bigg(\frac{-1}{3} |N_0^\nu|^2\bigg) \left( 1-x\right)^2 \left(  \bfp \cdot \Dp\right)  \frac{T_{22}^{\nu}}{x^2 M^2}+\bigg(\frac{1}{3} |N_0^\nu|^2+\frac{2}{3}|N_1^\nu|^2 \bigg)\nonumber	\\
		&&\frac{\left(  \bfp \cdot \Dp\right)\left( 1-x\right) }{M} \frac{T_{12}^{\nu}}{xM}\Bigg], \label{h23a}	\\
		%
		%
		%
		xH_{2,4}^{\nu(S)} &=& 0, \label{h24s}\\
		%
		%
		%
		xH_{2,4}^{\nu(A)} &=& 0. \label{h24a}
	\end{eqnarray}
At last, the GTMD expressions corresponding to the Dirac matrix structure $i\sigma^{+-} \gamma_5$, for each of the scalar and vector diquark, may be expressed as
	\begin{eqnarray} 
		%
		%
		xH_{2,5}^{\nu(S)} &=&  \frac{C_{S}^{2} N_s^2}{8 \pi^3}M\left( 1-x\right)   \Bigg[ \frac{T_{12}^{\nu}}{xM}-m \frac{ T_{22}^{\nu}}{x^2M^2 }\Bigg], \label{h25s}\\
		%
		%
		xH_{2,5}^{\nu(A)} &=&  \frac{C_{A}^{2} }{8 \pi^3}\bigg(\frac{1}{3} |N_0^\nu|^2+\frac{2}{3}|N_1^\nu|^2 \bigg)M\left( 1-x\right)   \Bigg[ \frac{T_{12}^{\nu}}{xM }- m\frac{ T_{22}^{\nu}}{x^2M^2}\Bigg], \label{h25a}\\
		%
		%
		xH_{2,6}^{\nu(S)} &=&  \frac{C_{S}^{2} N_s^2}{8 \pi^3}\Bigg[ - T_{11}^{\nu}+2m\frac{T_{12}^{\nu}}{xM }+\bigg(\bfp^2+\left( 1-x\right)^2\frac{\Dp^2}{4} \bigg) \frac{T_{22}^{\nu}}{x^2 M^2}+\frac{\left( 1-x\right)\Dp^2}{2M}\Bigg( \frac{T_{12}^{\nu}}{xM}-m \frac{ T_{22}^{\nu}}{x^2M^2 }\Bigg)\Bigg], \label{h26s}\\
		%
		xH_{2,6}^{\nu(A)} &=&  \frac{C_{A}^{2} }{8 \pi^3} \Bigg[\bigg(\frac{-1}{3} |N_0^\nu|^2 \bigg)\bigg(- T_{11}^{\nu}+ 2m\frac{T_{12}^{\nu}}{xM }+\bigg(\bfp^2+(1-x)^2\frac{\Dp^2}{4} \bigg)\bigg) \frac{T_{22}^{\nu}}{x^2 M^2}+\nonumber\\
		&&\bigg(\frac{1}{3} |N_0^\nu|^2+\frac{2}{3}|N_1^\nu|^2 \bigg)\frac{\left( 1-x\right)\Dp^2}{2M} \bigg(\frac{T_{12}^{\nu}}{xM }-m\frac{ T_{22}^{\nu}}{x^2M^2}\Bigg)\Bigg], \label{h26a}\\
		%
		%
		%
		xH_{2,7}^{\nu(S)} &=& -\frac{C_{S}^{2} N_s^2}{16 \pi^3}\Bigg[ \left( 1-x\right)^2 \left( \bfp \cdot \Dp\right)  \frac{T_{22}^{\nu}}{x^2 M^2}+\frac{ \left( \bfp \cdot \Dp\right)\left( 1-x\right)}{M}   \Bigg( \frac{T_{12}^{\nu}}{xM}-m \frac{ T_{22}^{\nu}}{x^2M^2 } \Bigg)\Bigg], \label{h27s}\\  
		%
		xH_{2,7}^{\nu(A)} &=&  -\frac{C_{A}^{2}}{16 \pi^3}\Bigg[\bigg(\frac{-1}{3} |N_0^\nu|^2 \bigg)\left( 1-x\right)^2 \left( \bfp \cdot \Dp\right)  \frac{T_{22}^{\nu}}{x^2 M^2}+\frac{ \left( \bfp \cdot \Dp\right)\left( 1-x\right)}{M}\nonumber\\
		&&  \bigg(\frac{1}{3} |N_0^\nu|^2+\frac{2}{3}|N_1^\nu|^2 \bigg) \Bigg( \frac{T_{12}^{\nu}}{xM}- m \frac{ T_{22}^{\nu}}{x^2M^2 }\Bigg)\Bigg], \label{h27a}\\
		%
		%
		%
		xH_{2,8}^{\nu(S)} &=&  \frac{C_{S}^{2} N_s^2}{16 \pi^3} \frac{1}{M} \Bigg[m T_{11}^{\nu}+2\bfp^2\frac{ T_{12}^{\nu}}{xM}-\bigg(\bfp^2-(1-x)^2\frac{\Dp^2}{4} \bigg) m\frac{T_{22}^{\nu}}{x^2 M^2}\Bigg], \label{h28s}\\
		%
		%
		%
		xH_{2,8}^{\nu(A)} &=&  \frac{C_{A}^{2}}{16 \pi^3}\bigg(\frac{1}{3} |N_0^\nu|^2-\frac{2}{3}|N_1^\nu|^2 \bigg) \frac{1}{M} \Bigg[m T_{11}^{\nu}+2\bfp^2\frac{ T_{12}^{\nu}}{xM}-\bigg(\bfp^2-(1-x)^2\frac{\Dp^2}{4} \bigg) m \frac{T_{22}^{\nu}}{x^2 M^2}\Bigg].\label{h28a}
	\end{eqnarray}
	
	
	
	\section{Discussion}\label{secdiscussion}
	\noindent
Considering theoretical, experimental, and phenomenological relevance, we have focused on the GTMDs corresponding to the Dirac matrix structures $\Gamma = 1, \gamma_5$, along with GTMDs related to various TMDs via Eqs. \eqref{tmd5} to \eqref{tmd16}, in further discussions. We also discuss the corresponding TMFFs. At skewness $(\xi=0)$, the GTMDs are a function of four variables, denoted as $X^{\nu}(x, p_{\perp}, \Delta_{\perp}, \theta)$. Because it is impractical to illustrate the variations with all four variables concurrently, this study will examine their variations with one or two variables individually, while the others are kept constant or integrated. Although the angle $\theta$ formed between the $\bfp$ and $\Dp$ planes encompasses $0$ to $\pi$, which results in $\bfp \cdot \Dp$ spanning from $p_{\perp} \Delta_{\perp}$ to $-p_{\perp} \Delta_{\perp}$, we have fixed $\theta$ at $0$ for our analysis.
		\subsection{GTMDs}\label{ssgtmds}

%
	\subsubsection{Variation with $x$ and ${ p_\perp}$}\label{ssxp}
\begin{figure*}
		\centering
		\begin{minipage}[c]{0.98\textwidth}
		       (a)\includegraphics[width=7.3cm]{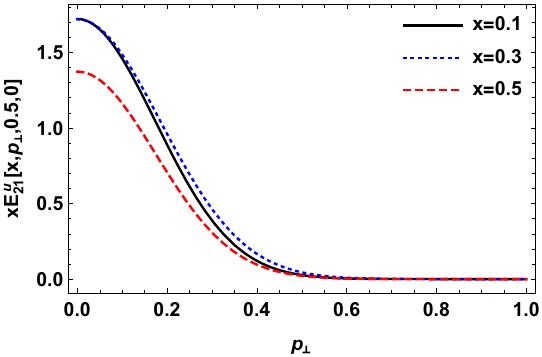}
				\hspace{0.05cm}
		       (b)\includegraphics[width=7.3cm]{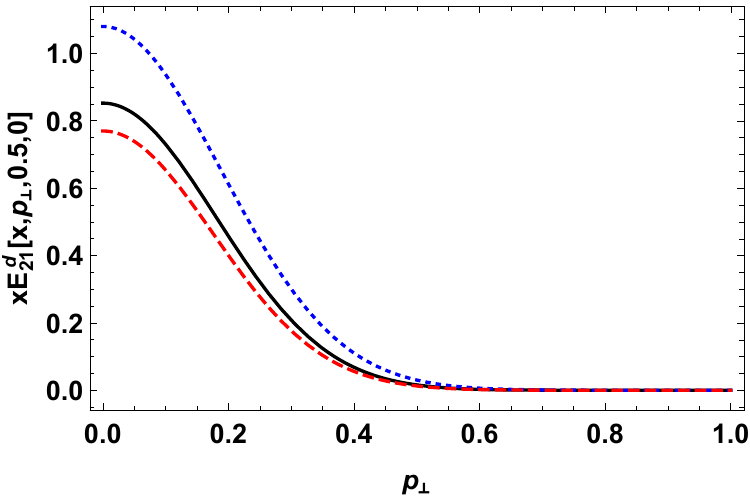}
				\hspace{0.05cm}
				(c)\includegraphics[width=7.3cm]{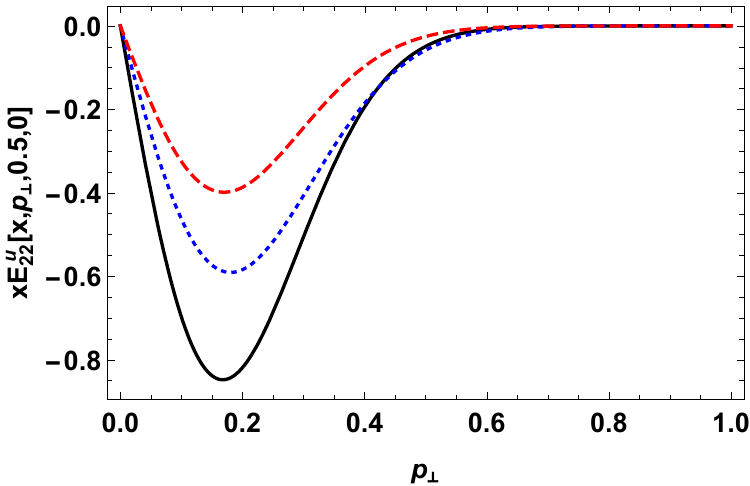}
				\hspace{0.05cm}
				(d)\includegraphics[width=7.3cm]{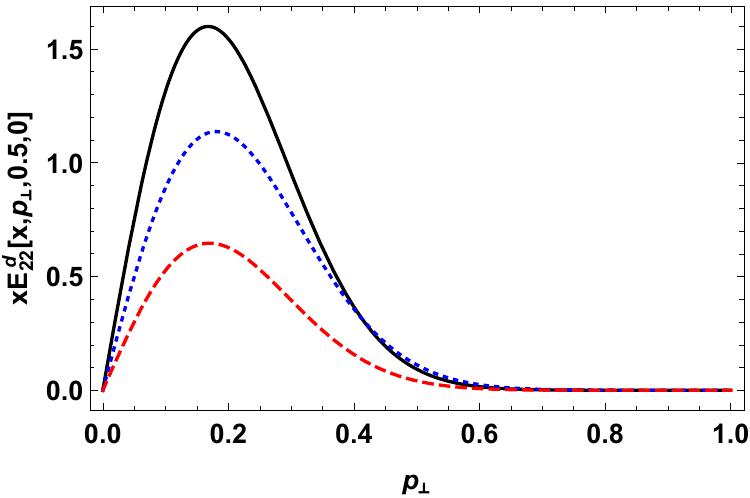}
				\hspace{0.05cm}
				(e)\includegraphics[width=7.3cm]{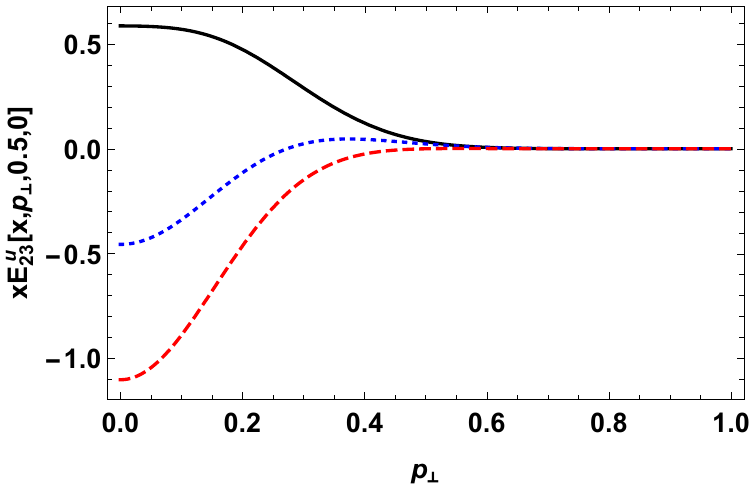}
				\hspace{0.05cm}
				(f)\includegraphics[width=7.3cm]{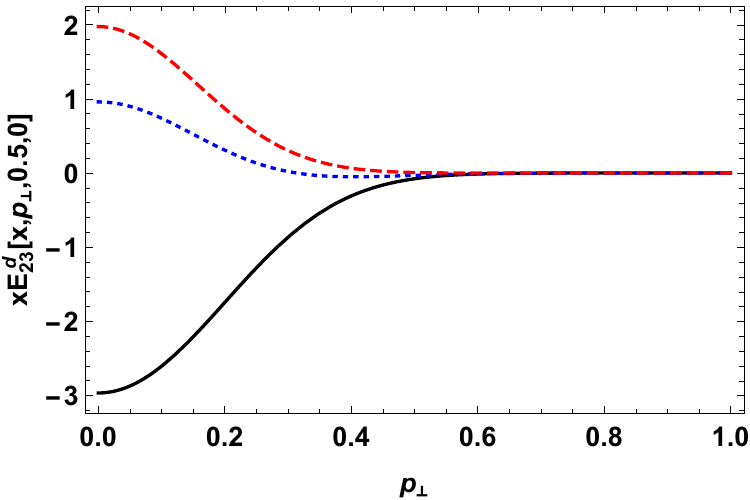}
				\hspace{0.05cm}
				(g)\includegraphics[width=7.3cm]{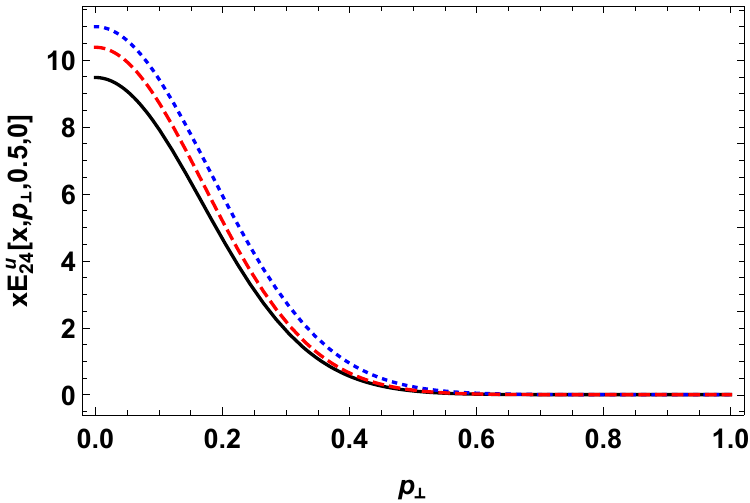}
				\hspace{0.05cm}
				(h)\includegraphics[width=7.3cm]{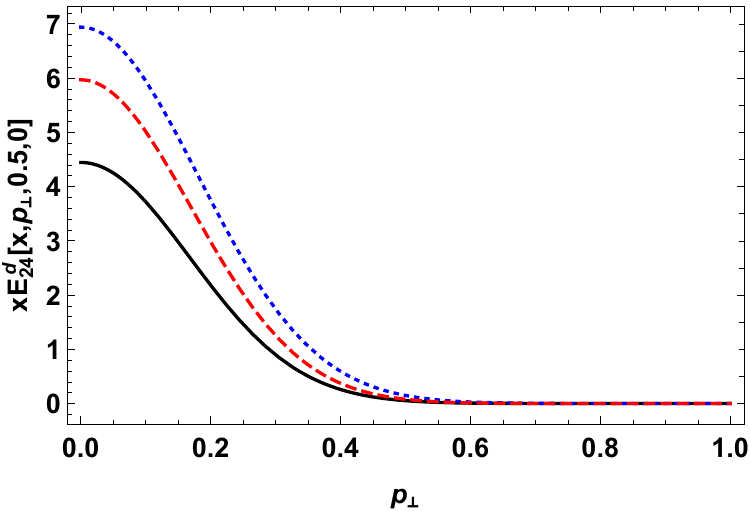}
				\hspace{0.05cm}\\
			\end{minipage}
		\caption{\label{fig3dXPE1} The sub-leading twist GTMDs 		
				$x E_{2,1}^{\nu}(x, p_{\perp},\Delta_{\perp},\theta)$,
				$x E_{2,2}^{\nu}(x, p_{\perp},\Delta_{\perp},\theta)$,
				$x E_{2,3}^{\nu}(x, p_{\perp},\Delta_{\perp},\theta)$, and
				$x E_{2,4}^{\nu}(x, p_{\perp},\Delta_{\perp},\theta)$
				are	plotted about ${{ p_\perp}}$ and $x=0.1,0.3,0.5$ keeping ${ \Delta_\perp}= 0.5~\mathrm{GeV}$ for ${\bfp} \parallel {\Dp}$.  In sequential order, $u$ and $d$ quarks are in the left and right columns.
			}
	\end{figure*}

\begin{figure*}
		\centering
		\begin{minipage}[c]{0.98\textwidth}
				(a)\includegraphics[width=7.3cm]{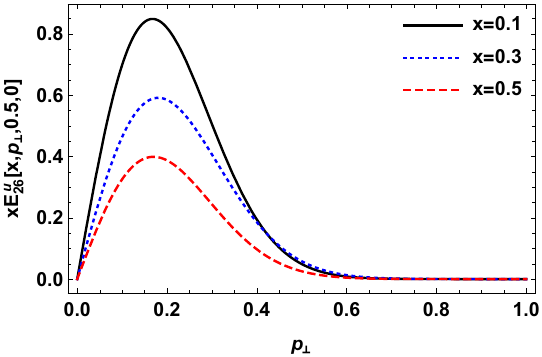}
				\hspace{0.05cm}
				(b)\includegraphics[width=7.3cm]{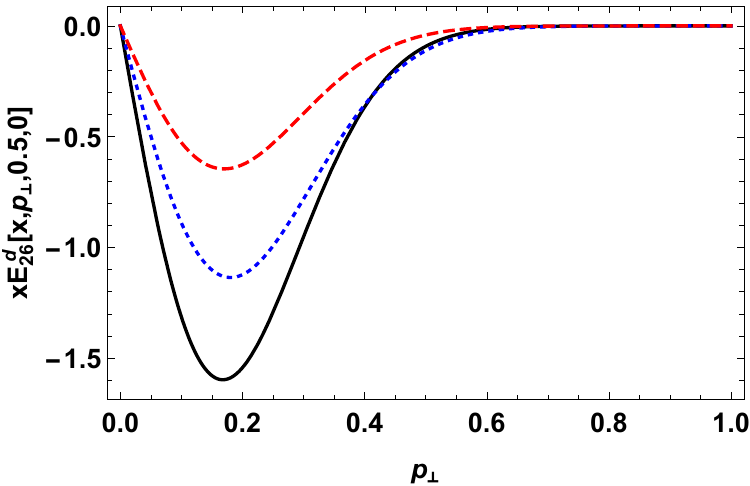}
				\hspace{0.05cm}
				(c)\includegraphics[width=7.3cm]{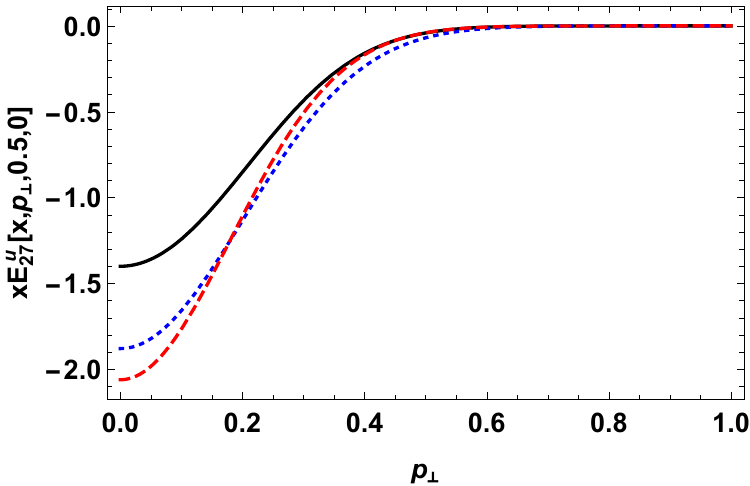}
				\hspace{0.05cm}
				(d)\includegraphics[width=7.3cm]{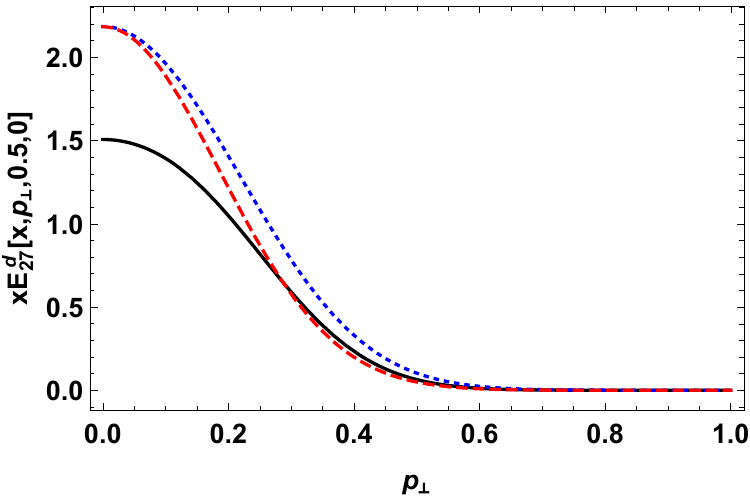}
				\hspace{0.05cm}
				(e)\includegraphics[width=7.3cm]{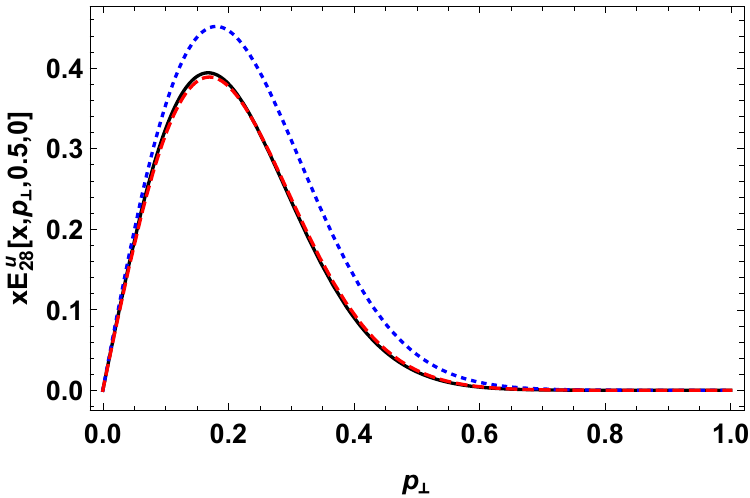}
				\hspace{0.05cm}
				(f)\includegraphics[width=7.3cm]{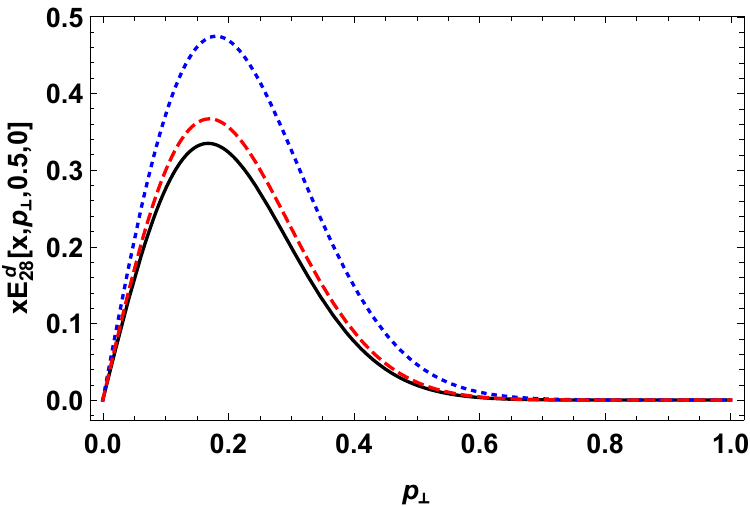}
				\hspace{0.05cm}
				\\
			\end{minipage}
		\caption{\label{fig3dXPE2} The sub-leading twist GTMDs 		
				$x E_{2,6}^{\nu}(x, p_{\perp},\Delta_{\perp},\theta)$,
				$x E_{2,7}^{\nu}(x, p_{\perp},\Delta_{\perp},\theta)$,
			 and
				$x E_{2,8}^{\nu}(x, p_{\perp},\Delta_{\perp},\theta)$
				are	plotted about ${{ p_\perp}}$ and $x=0.1,0.3,0.5$ keeping ${ \Delta_\perp}= 0.5~\mathrm{GeV}$ for ${\bfp} \parallel {\Dp}$.  In sequential order, $u$ and $d$ quarks are in the left and right columns.
			}
	\end{figure*}

\begin{figure*}
		\centering
		\begin{minipage}[c]{0.98\textwidth}
				(a)\includegraphics[width=7.3cm]{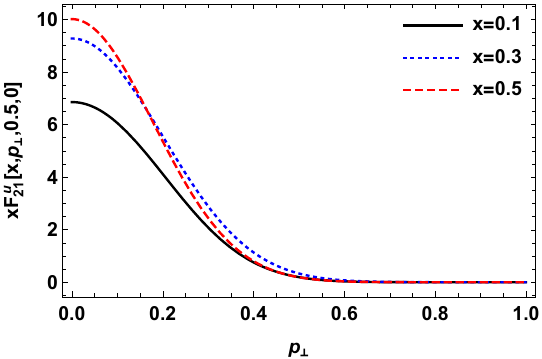}
				\hspace{0.05cm}
				(b)\includegraphics[width=7.3cm]{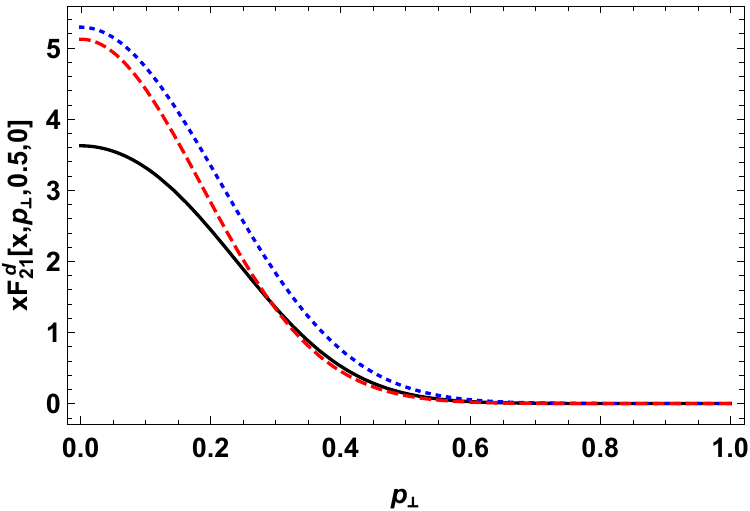}
				\hspace{0.05cm}
				(c)\includegraphics[width=7.3cm]{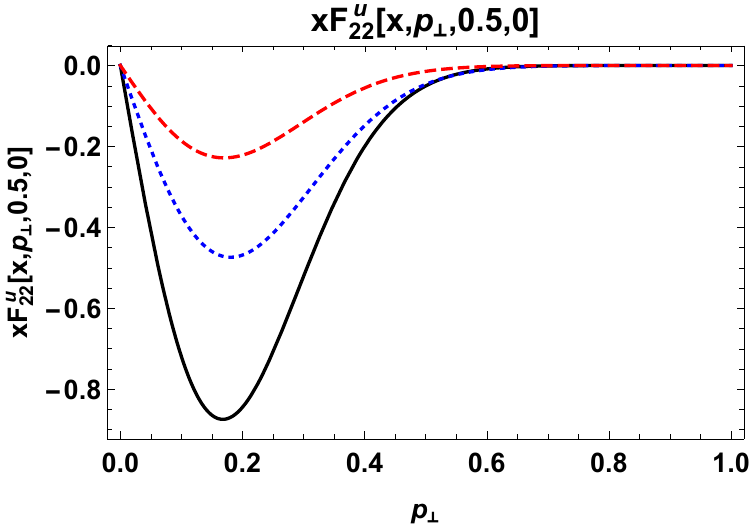}
				\hspace{0.05cm}
				(d)\includegraphics[width=7.3cm]{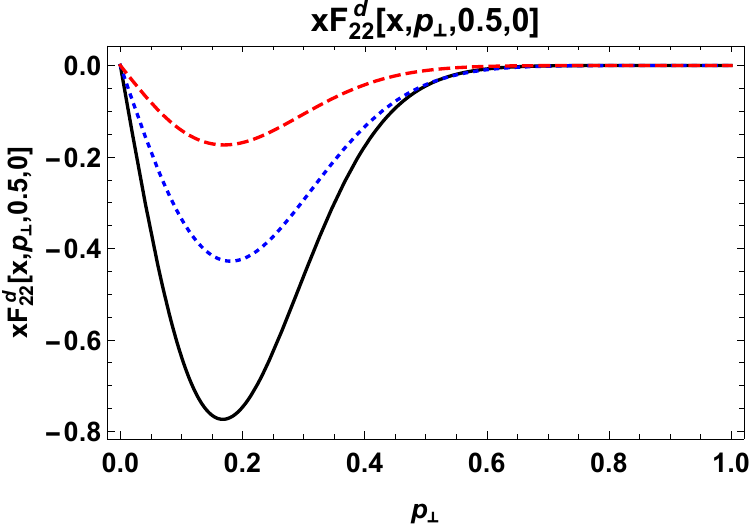}
				\hspace{0.05cm}
				(e)\includegraphics[width=7.3cm]{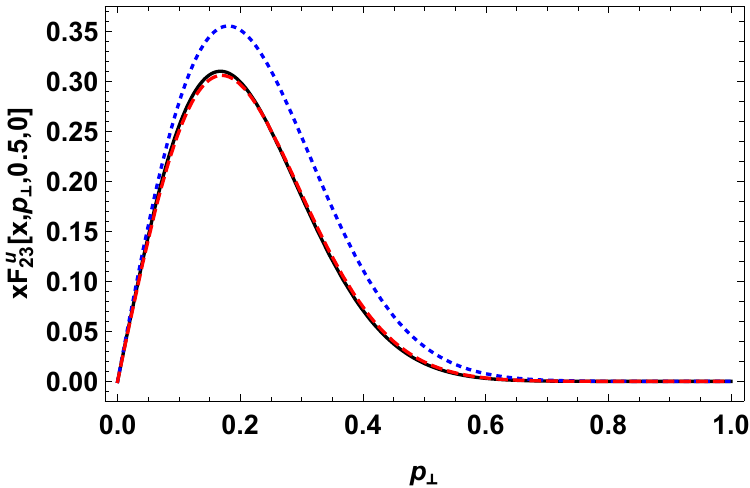}
				\hspace{0.05cm}
				(f)\includegraphics[width=7.3cm]{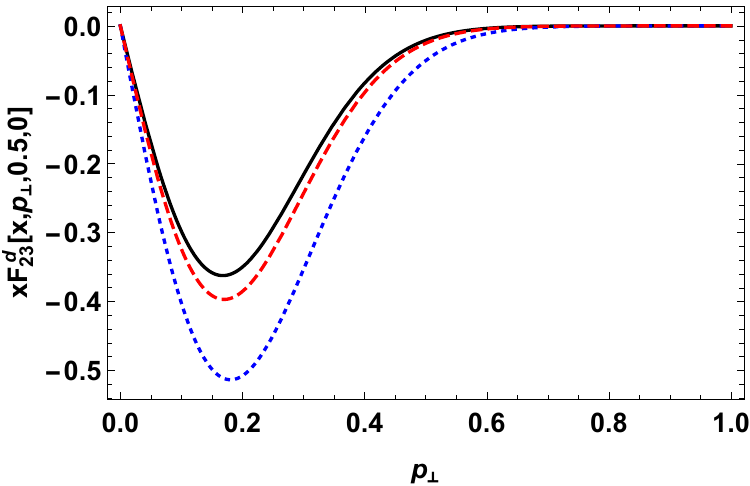}
				\hspace{0.05cm}
				(g)\includegraphics[width=7.3cm]{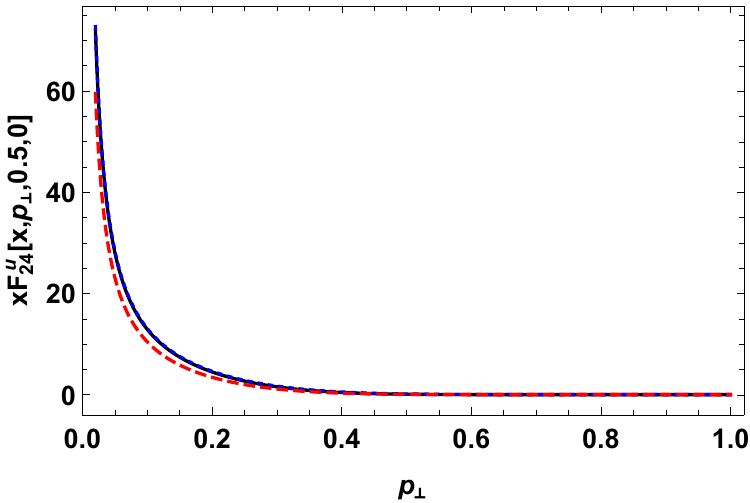}
				\hspace{0.05cm}
				(h)\includegraphics[width=7.3cm]{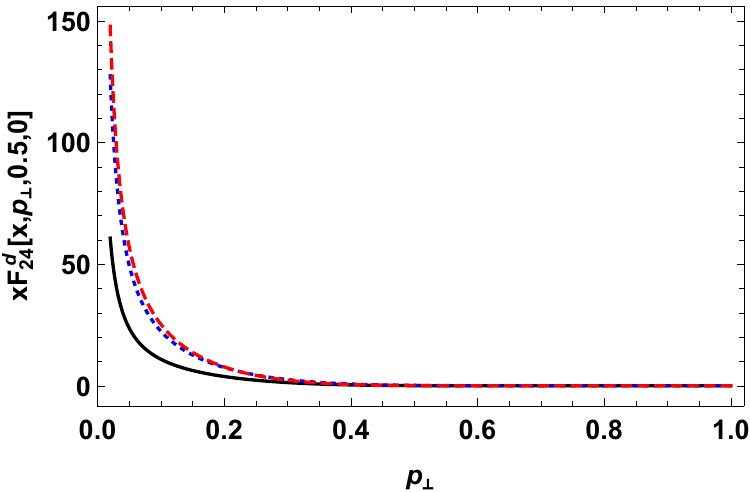}
				\hspace{0.05cm}\\
			\end{minipage}
		\caption{\label{fig3dXPF1} The sub-leading twist GTMDs 		
				$x F_{2,1}^{\nu}(x, p_{\perp},\Delta_{\perp},\theta)$,
				$x F_{2,2}^{\nu}(x, p_{\perp},\Delta_{\perp},\theta)$,
				$x F_{2,3}^{\nu}(x, p_{\perp},\Delta_{\perp},\theta)$, and
				$x F_{2,4}^{\nu}(x, p_{\perp},\Delta_{\perp},\theta)$
				are	plotted about ${{ p_\perp}}$ and $x=0.1,0.3,0.5$ keeping ${ \Delta_\perp}= 0.5~\mathrm{GeV}$ for ${\bfp} \parallel {\Dp}$.  In sequential order, $u$ and $d$ quarks are in the left and right columns.
			}
	\end{figure*}

\begin{figure*}
		\centering
		\begin{minipage}[c]{0.98\textwidth}
				(a)\includegraphics[width=7.3cm]{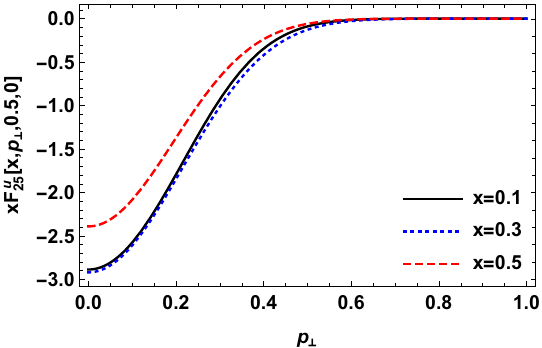}
				\hspace{0.05cm}
				(b)\includegraphics[width=7.3cm]{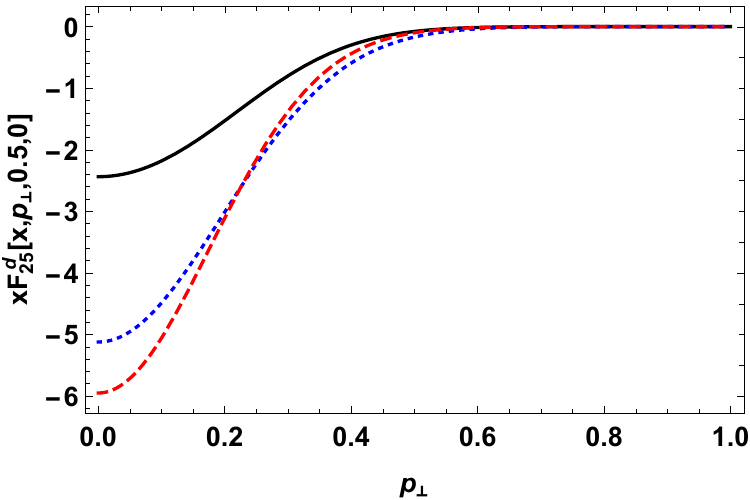}
				\hspace{0.05cm}
				(c)\includegraphics[width=7.3cm]{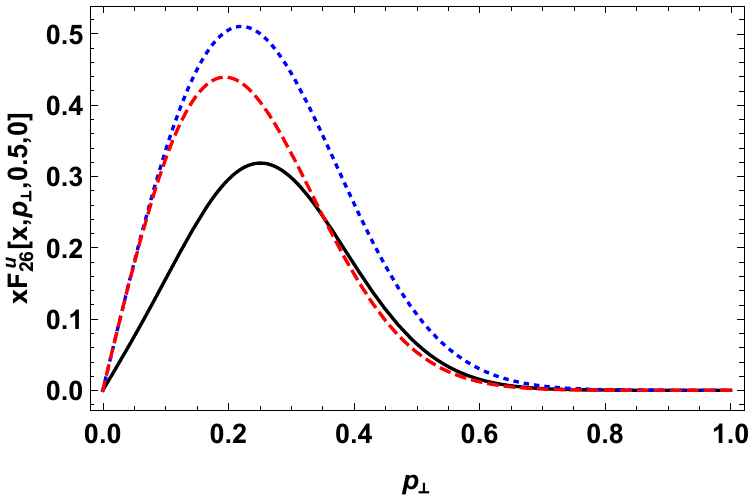}
				\hspace{0.05cm}
				(d)\includegraphics[width=7.3cm]{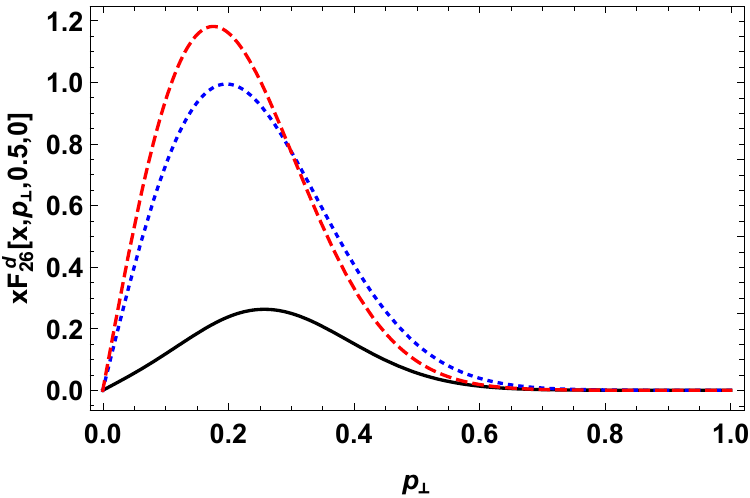}
				\hspace{0.05cm}
				(e)\includegraphics[width=7.3cm]{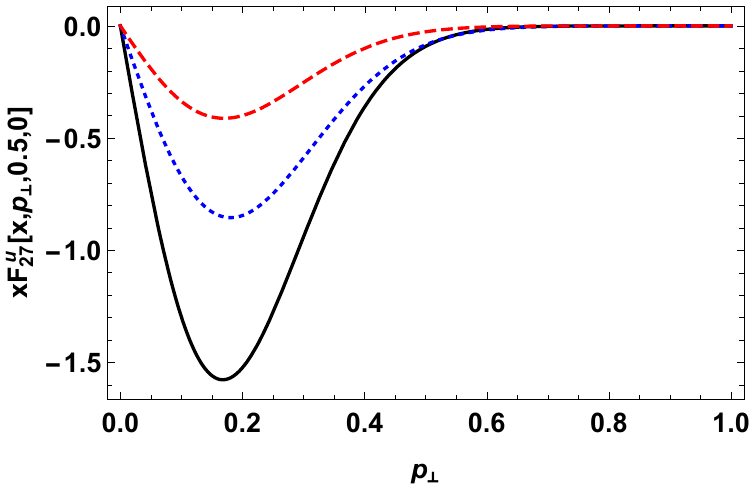}
				\hspace{0.05cm}
				(f)\includegraphics[width=7.3cm]{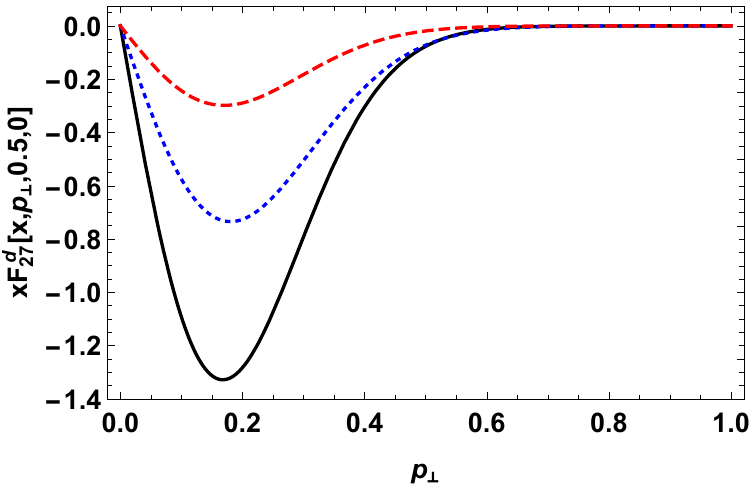}
				\hspace{0.05cm}
				(g)\includegraphics[width=7.3cm]{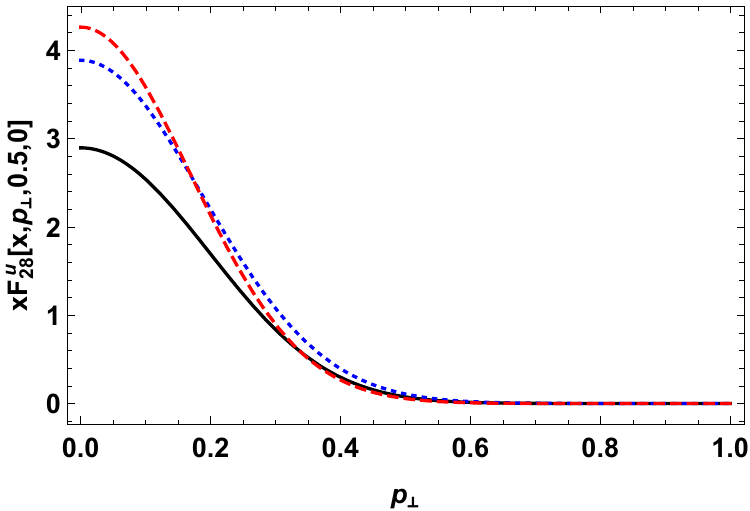}
				\hspace{0.05cm}
				(h)\includegraphics[width=7.3cm]{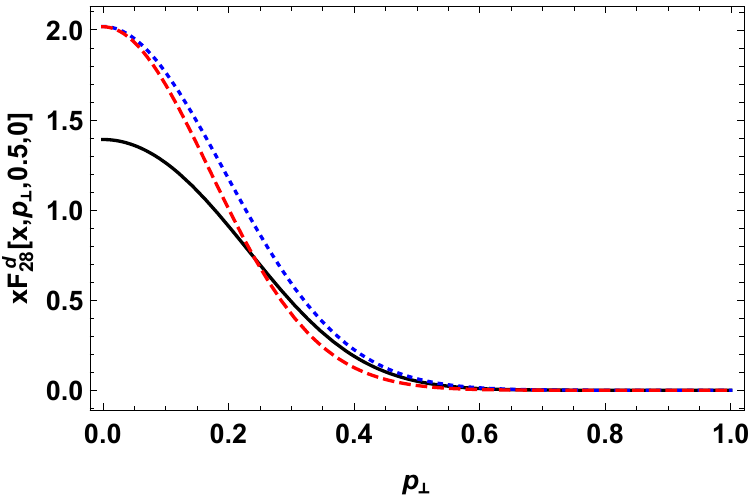}
				\hspace{0.05cm}\\
			\end{minipage}
		\caption{\label{fig3dXPF2} The sub-leading twist GTMDs 		
				$x F_{2,5}^{\nu}(x, p_{\perp},\Delta_{\perp},\theta)$,
				$x F_{2,6}^{\nu}(x, p_{\perp},\Delta_{\perp},\theta)$,
				$x F_{2,7}^{\nu}(x, p_{\perp},\Delta_{\perp},\theta)$, and
				$x F_{2,8}^{\nu}(x, p_{\perp},\Delta_{\perp},\theta)$
				are	plotted about ${{ p_\perp}}$ and $x=0.1,0.3,0.5$ keeping ${ \Delta_\perp}= 0.5~\mathrm{GeV}$ for ${\bfp} \parallel {\Dp}$.  In sequential order, $u$ and $d$ quarks are in the left and right columns.
			}
	\end{figure*}

\begin{figure*}
		\centering
		\begin{minipage}[c]{0.98\textwidth}
				(a)\includegraphics[width=7.3cm]{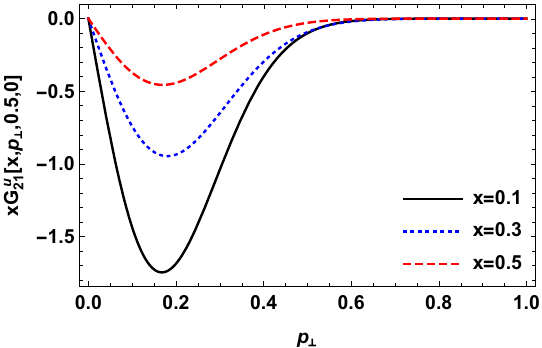}
				\hspace{0.05cm}
				(b)\includegraphics[width=7.3cm]{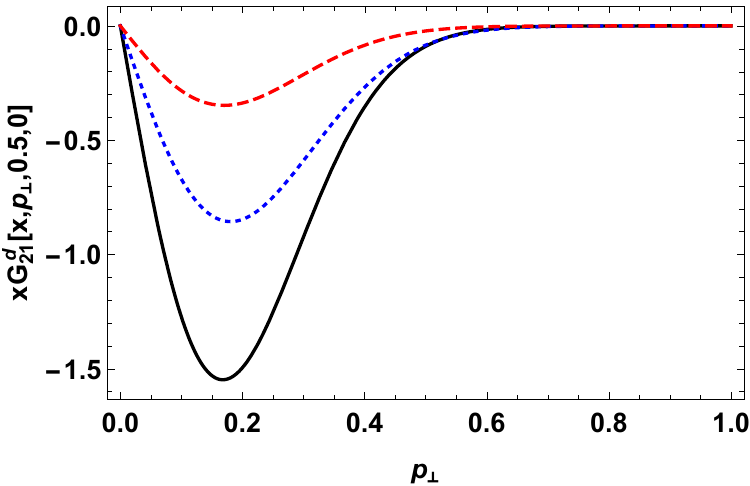}
				\hspace{0.05cm}
				(c)\includegraphics[width=7.3cm]{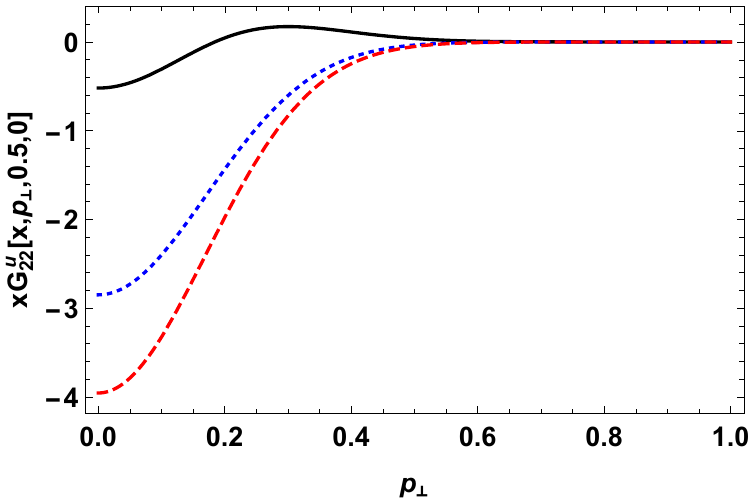}
				\hspace{0.05cm}
				(d)\includegraphics[width=7.3cm]{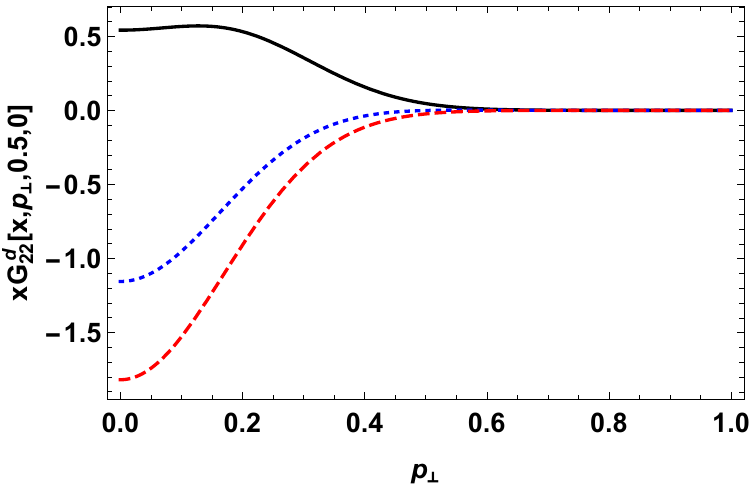}
				\hspace{0.05cm}
				(e)\includegraphics[width=7.3cm]{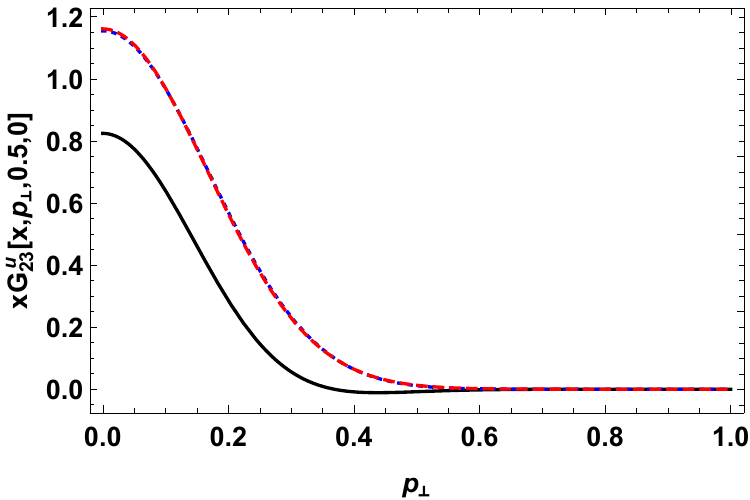}
				\hspace{0.05cm}
				(f)\includegraphics[width=7.3cm]{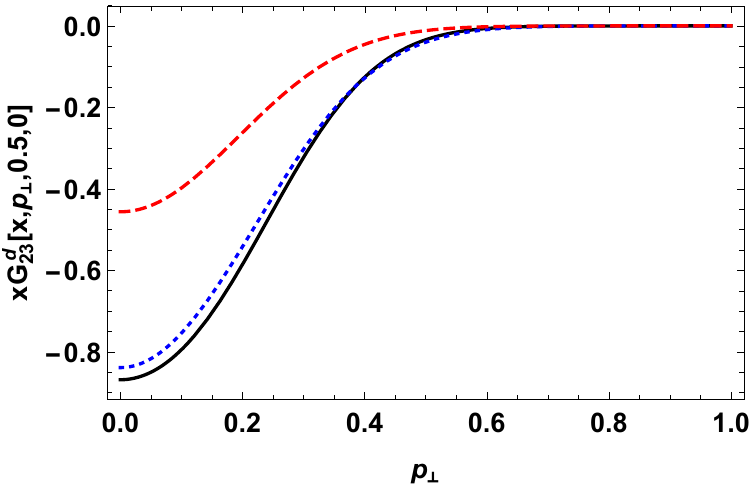}
				\hspace{0.05cm}
				(g)\includegraphics[width=7.3cm]{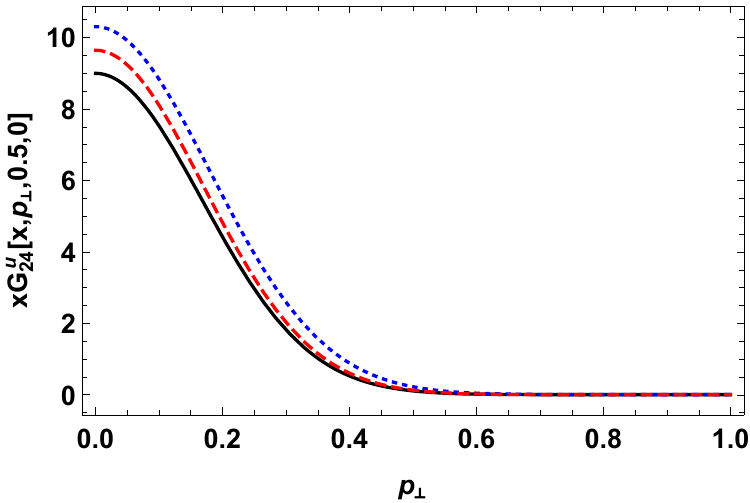}
				\hspace{0.05cm}
				(h)\includegraphics[width=7.3cm]{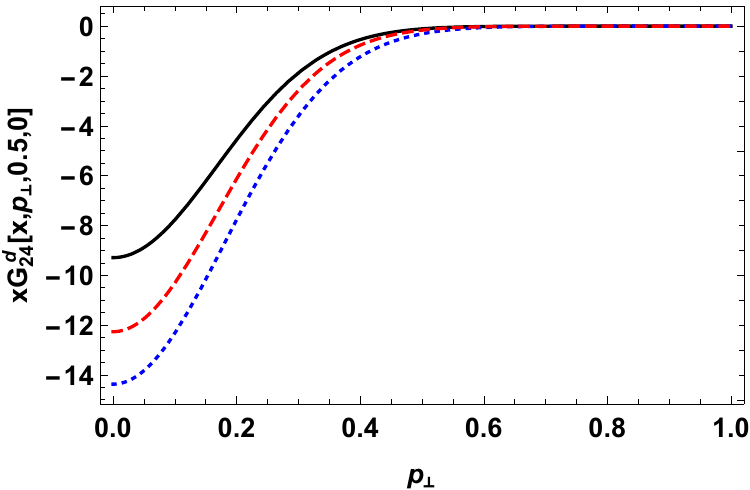}
				\hspace{0.05cm}\\
			\end{minipage}
		\caption{\label{fig3dXPG1} The sub-leading twist GTMDs 		
				$x G_{2,1}^{\nu}(x, p_{\perp},\Delta_{\perp},\theta)$,
				$x G_{2,2}^{\nu}(x, p_{\perp},\Delta_{\perp},\theta)$,
				$x G_{2,3}^{\nu}(x, p_{\perp},\Delta_{\perp},\theta)$, and
				$x G_{2,4}^{\nu}(x, p_{\perp},\Delta_{\perp},\theta)$
				are	plotted about ${{ p_\perp}}$ and $x=0.1,0.3,0.5$ keeping ${ \Delta_\perp}= 0.5~\mathrm{GeV}$ for ${\bfp} \parallel {\Dp}$.  In sequential order, $u$ and $d$ quarks are in the left and right columns.
			}
	\end{figure*}

\begin{figure*}
		\centering
		\begin{minipage}[c]{0.98\textwidth}
				(a)\includegraphics[width=7.3cm]{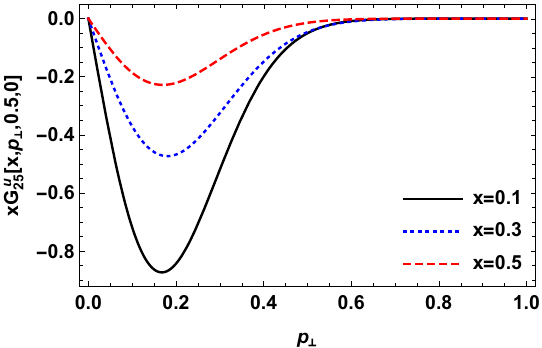}
				\hspace{0.05cm}
				(b)\includegraphics[width=7.3cm]{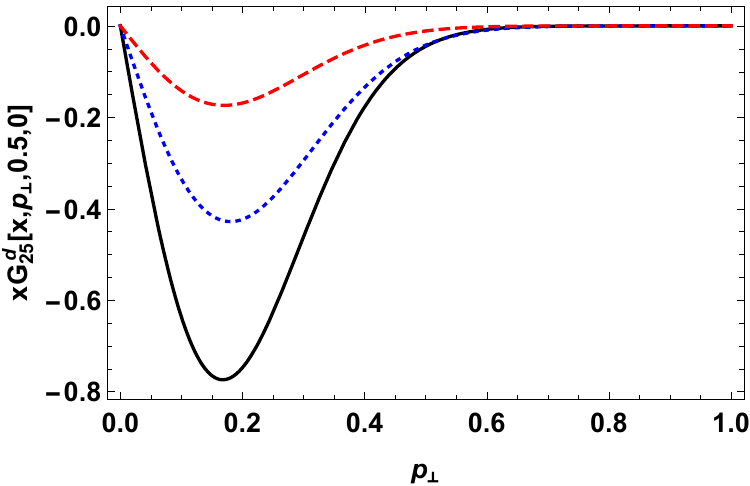}
				\hspace{0.05cm}
				(c)\includegraphics[width=7.3cm]{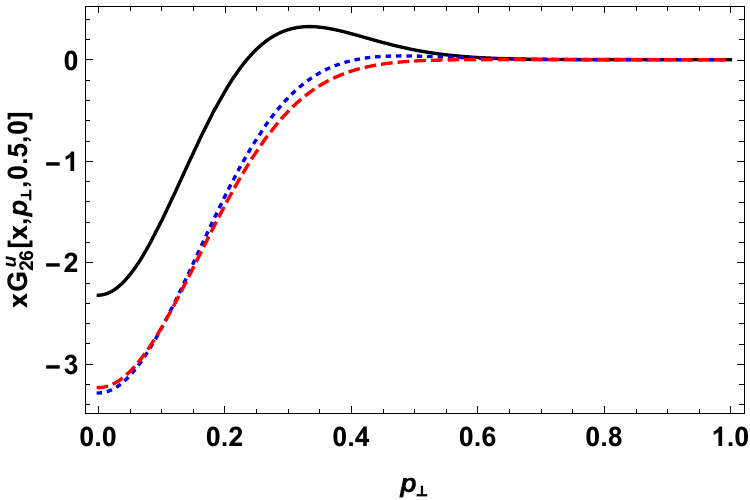}
				\hspace{0.05cm}
				(d)\includegraphics[width=7.3cm]{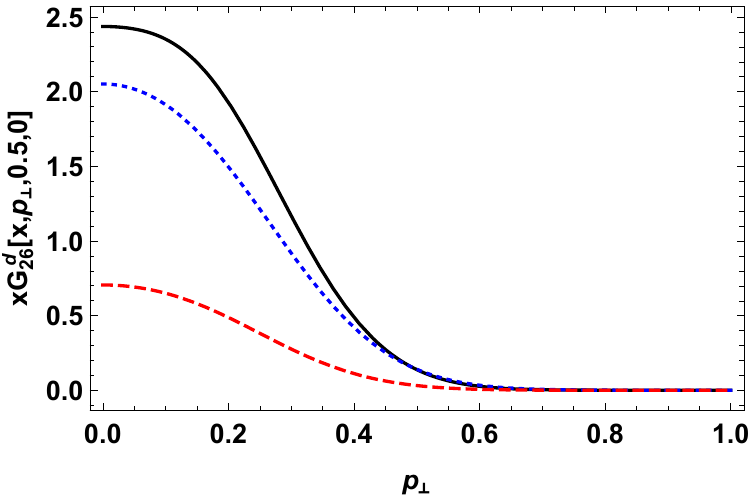}
				\hspace{0.05cm}
				(e)\includegraphics[width=7.3cm]{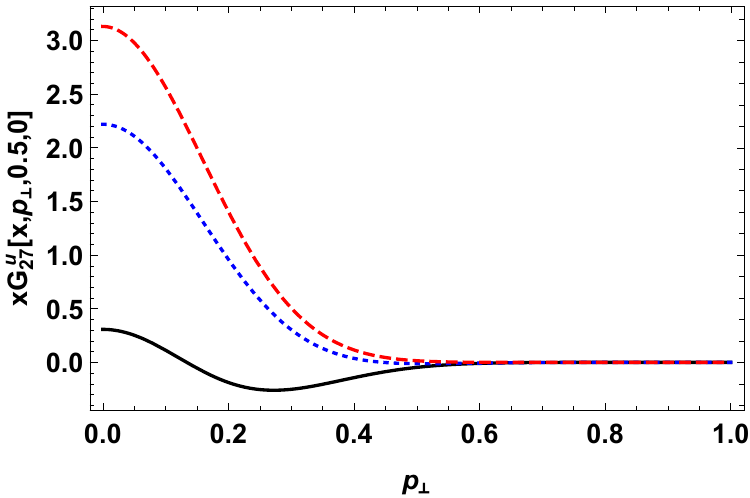}
				\hspace{0.05cm}
				(f)\includegraphics[width=7.3cm]{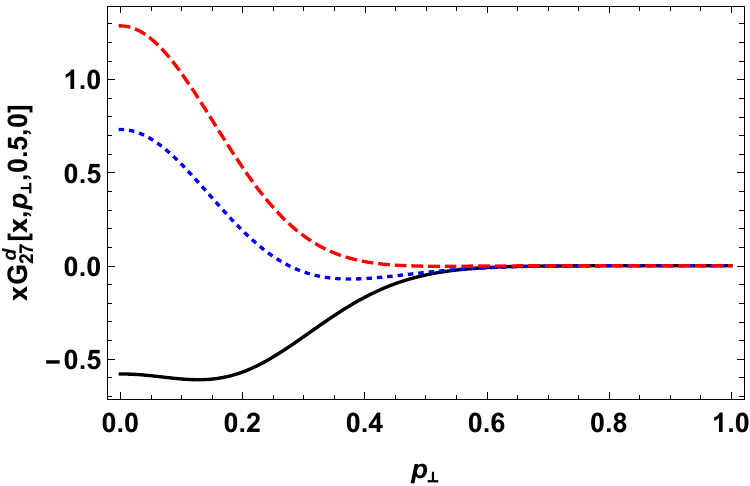}
				\hspace{0.05cm}
				(g)\includegraphics[width=7.3cm]{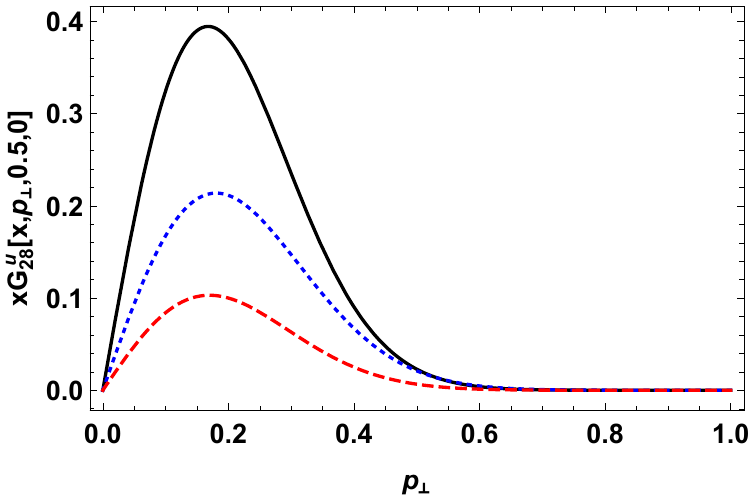}
				\hspace{0.05cm}
				(h)\includegraphics[width=7.3cm]{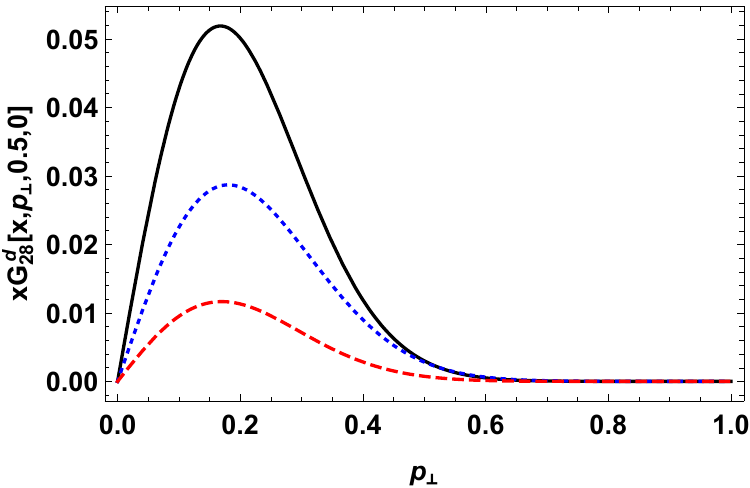}
				\hspace{0.05cm}\\
			\end{minipage}
		\caption{\label{fig3dXPG2} The sub-leading twist GTMDs 		
				$x G_{2,5}^{\nu}(x, p_{\perp},\Delta_{\perp},\theta)$,
				$x G_{2,6}^{\nu}(x, p_{\perp},\Delta_{\perp},\theta)$,
				$x G_{2,7}^{\nu}(x, p_{\perp},\Delta_{\perp},\theta)$, and
				$x G_{2,8}^{\nu}(x, p_{\perp},\Delta_{\perp},\theta)$
				are	plotted about ${{ p_\perp}}$ and $x=0.1,0.3,0.5$ keeping ${ \Delta_\perp}= 0.5~\mathrm{GeV}$ for ${\bfp} \parallel {\Dp}$.  In sequential order, $u$ and $d$ quarks are in the left and right columns.
			}
	\end{figure*}

\begin{figure*}
		\centering
		\begin{minipage}[c]{0.98\textwidth}
				(a)\includegraphics[width=7.3cm]{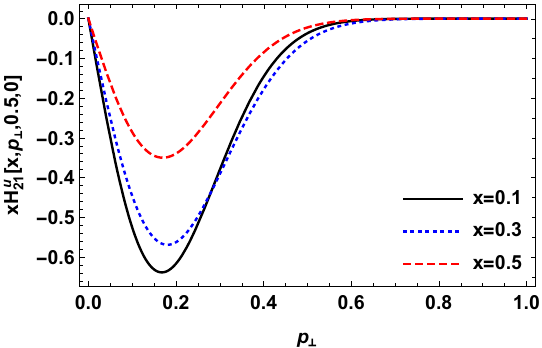}
				\hspace{0.05cm}
				(b)\includegraphics[width=7.3cm]{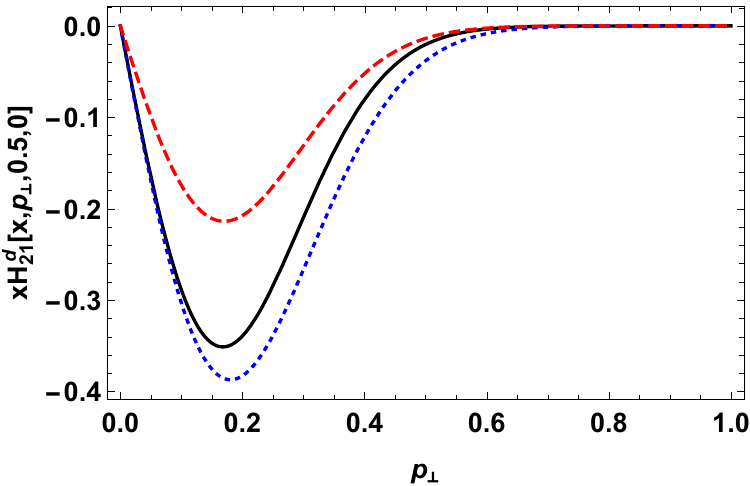}
				\hspace{0.05cm}
				(c)\includegraphics[width=7.3cm]{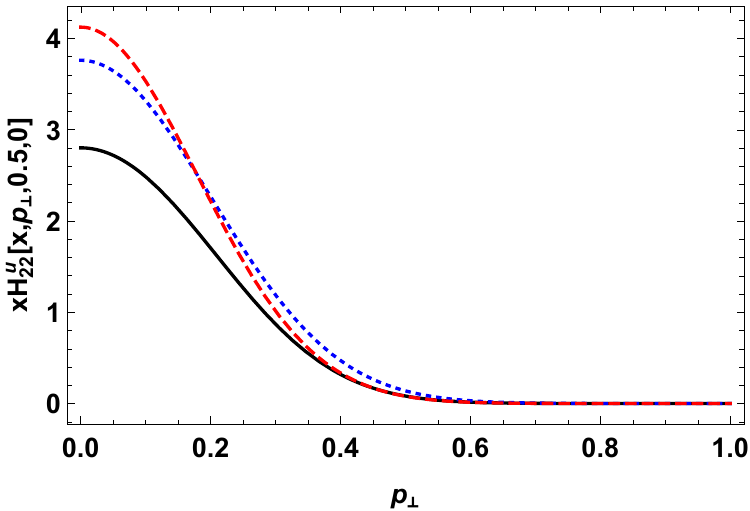}
				\hspace{0.05cm}
				(d)\includegraphics[width=7.3cm]{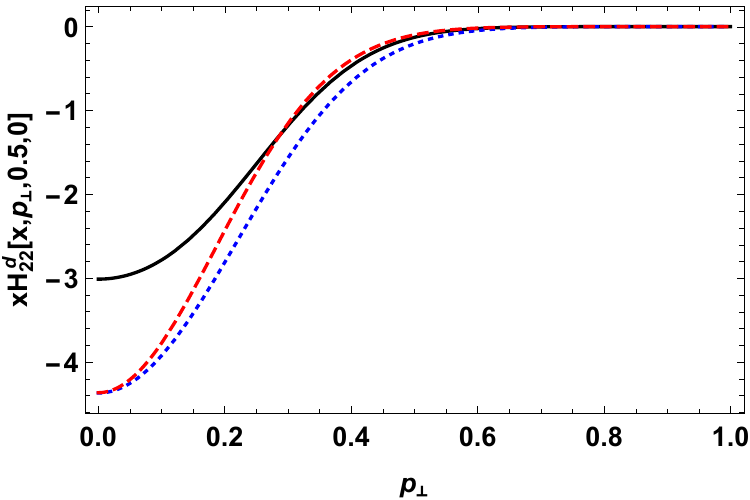}
				\hspace{0.05cm}
				(e)\includegraphics[width=7.3cm]{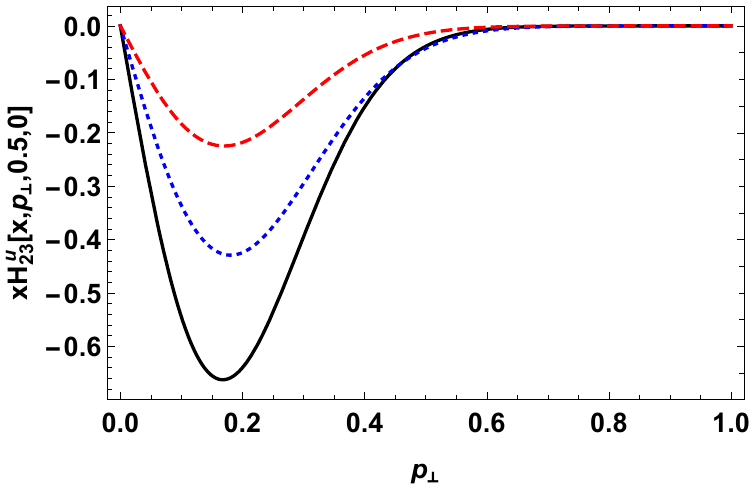}
				\hspace{0.05cm}
				(f)\includegraphics[width=7.3cm]{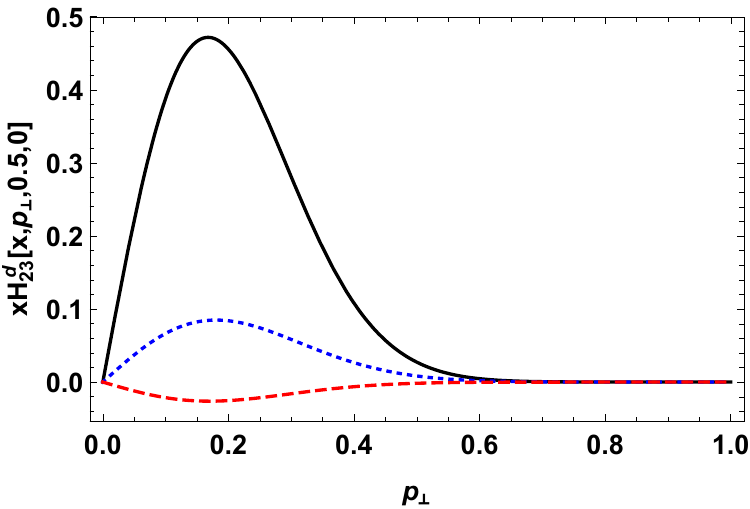}
				\hspace{0.05cm}
				\\
			\end{minipage}
		\caption{\label{fig3dXPH1} The sub-leading twist GTMDs 		
				$x H_{2,1}^{\nu}(x, p_{\perp},\Delta_{\perp},\theta)$,
				$x H_{2,2}^{\nu}(x, p_{\perp},\Delta_{\perp},\theta)$,
				and
				$x H_{2,3}^{\nu}(x, p_{\perp},\Delta_{\perp},\theta)$
				are	plotted about ${{ p_\perp}}$ and $x=0.1,0.3,0.5$ keeping ${ \Delta_\perp}= 0.5~\mathrm{GeV}$ for ${\bfp} \parallel {\Dp}$.  In sequential order, $u$ and $d$ quarks are in the left and right columns.
			}
	\end{figure*}

\begin{figure*}
		\centering
		\begin{minipage}[c]{0.98\textwidth}
				(a)\includegraphics[width=7.3cm]{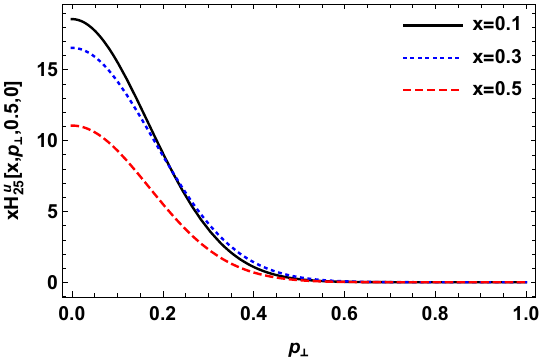}
				\hspace{0.05cm}
				(b)\includegraphics[width=7.3cm]{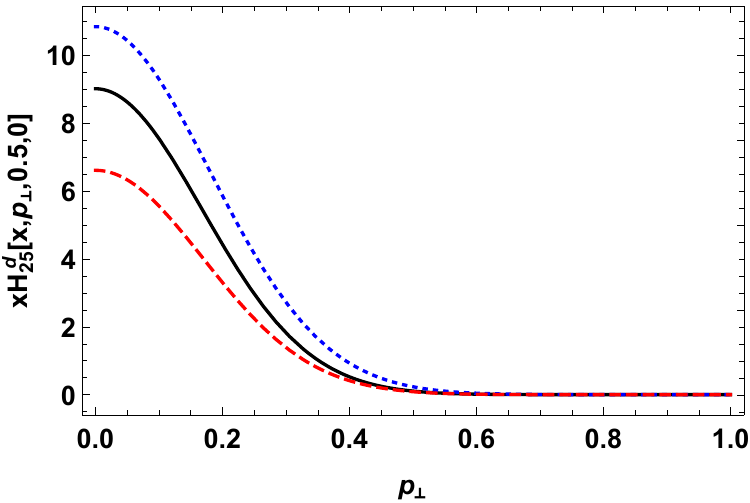}
				\hspace{0.05cm}
				(c)\includegraphics[width=7.3cm]{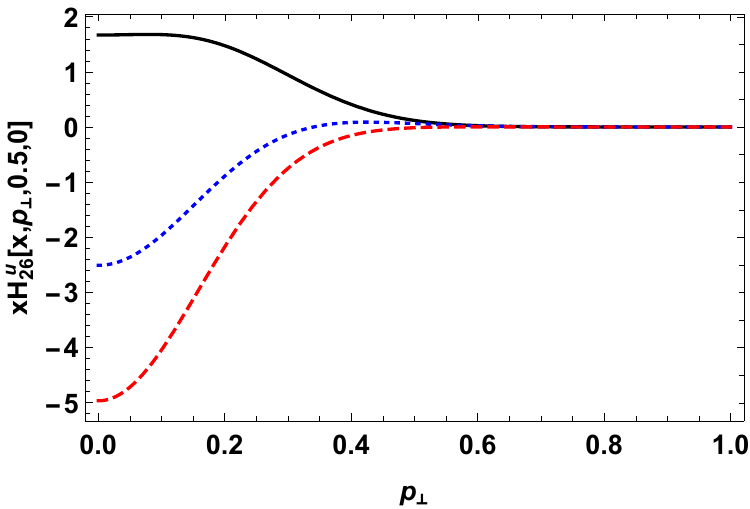}
				\hspace{0.05cm}
				(d)\includegraphics[width=7.3cm]{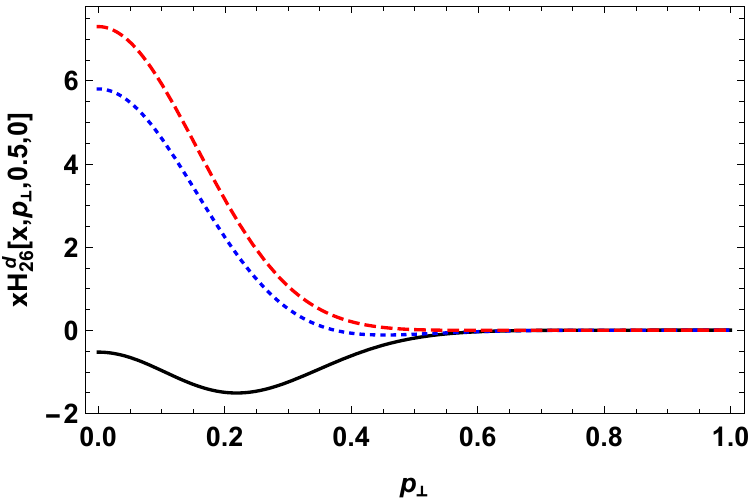}
				\hspace{0.05cm}
				(e)\includegraphics[width=7.3cm]{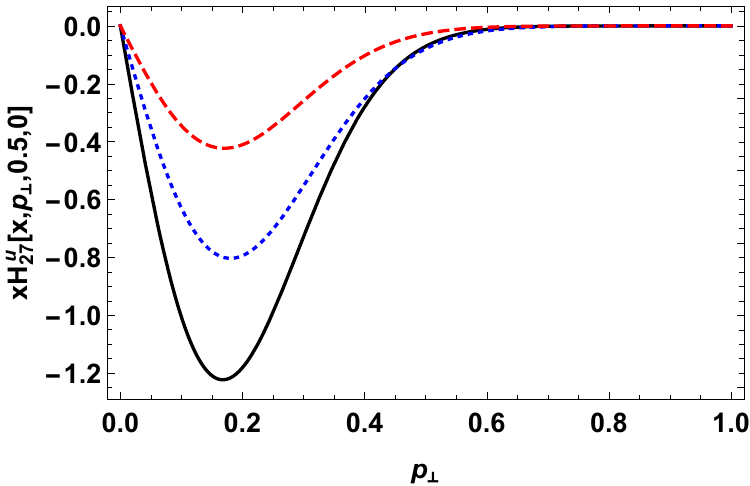}
				\hspace{0.05cm}
				(f)\includegraphics[width=7.3cm]{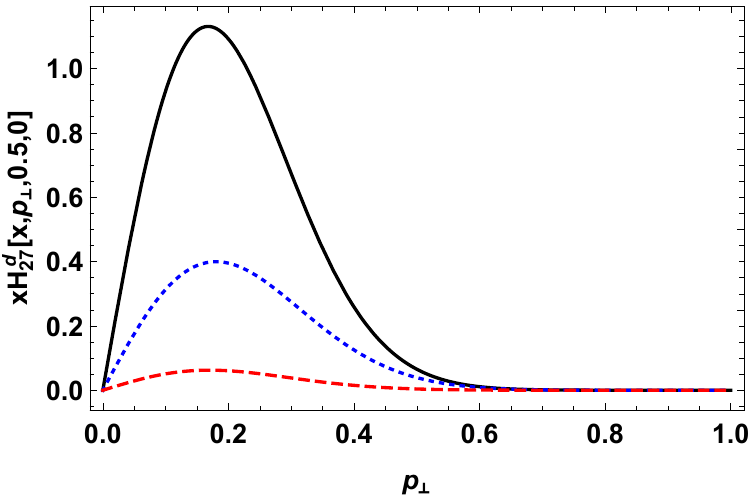}
				\hspace{0.05cm}
				(g)\includegraphics[width=7.3cm]{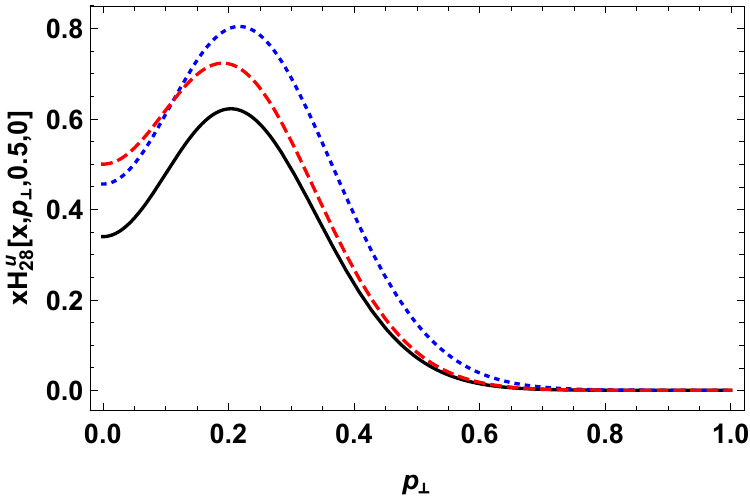}
				\hspace{0.05cm}
				(h)\includegraphics[width=7.3cm]{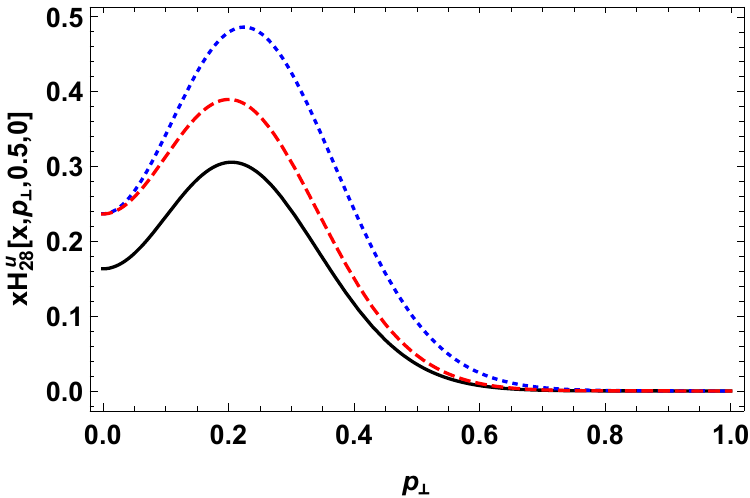}
				\hspace{0.05cm}\\
			\end{minipage}
		\caption{\label{fig3dXPH2} The sub-leading twist GTMDs 		
				$x H_{2,5}^{\nu}(x, p_{\perp},\Delta_{\perp},\theta)$,
				$x H_{2,6}^{\nu}(x, p_{\perp},\Delta_{\perp},\theta)$,
				$x H_{2,7}^{\nu}(x, p_{\perp},\Delta_{\perp},\theta)$, and
				$x H_{2,8}^{\nu}(x, p_{\perp},\Delta_{\perp},\theta)$
				are	plotted about ${{ p_\perp}}$ and $x=0.1,0.3,0.5$ keeping ${ \Delta_\perp}= 0.5~\mathrm{GeV}$ for ${\bfp} \parallel {\Dp}$. In sequential order, $u$ and $d$ quarks are in the left and right columns.
			}
	\end{figure*}

We can achieve the sub-leading twist TMDs by applying the TMD-limit to the sub-leading twist GTMDs. The relation between TMDs and GTMDs that follow are \cite{Meissner:2009ww}

	\begin{eqnarray}
		e^{\nu}(x,{ p_\perp}) & = & E_{2,1}^{\nu}(x, { p_\perp},0,\theta) \, \label{tmd1}, \\
		e^{\perp \nu}_{T}(x,{ p_\perp}) & = & -E^{\nu}_{2,2}(x, { p_\perp},0,\theta) \,, \label{tmd2}\\
		e_{T}^{\nu}(x,{ p_\perp}) & = & -E_{2,6}^{\nu}(x, { p_\perp},0,\theta) \,, \label{tmd3} \\
		e_{L}^{\nu}(x,{ p_\perp}) & = & -E_{2,8}^{\nu}(x, { p_\perp},0,\theta) \,, \label{tmd4}\\
		f^{\perp \nu}(x,{ p_\perp}) & = & F_{2,1}^{\nu}(x, { p_\perp},0,\theta) \,, \label{tmd5}\\
		f^{' \nu}_{T}(x,{ p_\perp}) & = & F_{2,3}^{\nu}(x, { p_\perp},0,\theta) \,, \label{tmd6} \\
		f^{\perp \nu}_{T}(x,{ p_\perp}) & = & F_{2,4}^{\nu}(x, { p_\perp},0,\theta) \,, \label{tmd7}\\
		f^{\perp \nu}_{L}(x,{ p_\perp}) & = & F_{2,7}^{\nu}(x, { p_\perp},0,\theta) \,, \label{tmd8}		\\
		g^{\perp \nu}(x,{ p_\perp}) & = & -G_{2,1}^{\nu}(x, { p_\perp},0,\theta) \,, \label{tmd9}\\
		g^{' \nu}_{T}(x,{ p_\perp}) & = & G_{2,3}^{\nu}(x, { p_\perp},0,\theta) \,, \label{tmd10}\\
		g^{\perp \nu}_{T}(x,{ p_\perp}) & = & G_{2,4}^{\nu}(x, { p_\perp},0,\theta) \,, \label{tmd11} 		\\
		g^{\perp \nu}_{L}(x,{ p_\perp}) & = & G_{2,7}^{\nu}(x, { p_\perp},0,\theta) \,, \label{tmd12}		\\ 
		h(x,{ p_\perp}) & = & -H_{2,1}^{\nu}(x, { p_\perp},0,\theta) \,, \label{tmd13}	\\
		h^{\perp \nu}_{T}(x,{ p_\perp}) & = & H_{2,2}^{\nu}(x, { p_\perp},0,\theta) \,, \label{tmd14}\\
		h^{\nu}_{T}(x,{ p_\perp}) & = & H_{2,6}^{\nu}(x, { p_\perp},0,\theta) \,, \label{tmd15}\\
		h^{ \nu}_{L}(x,{ p_\perp}) & = & H_{2,8}^{\nu}(x, { p_\perp},0,\theta) \,. \label{tmd16} 
	\end{eqnarray}
In Figs. (\ref{fig3dXPE1}) to (\ref{fig3dXPH2}), the GTMDs $xE_{2,1},~xE_{2,2},~xE_{2,3},~xE_{2,4},~xE_{2,6},~xE_{2,7},~xE_{2,8},~xF_{2,1},~xF_{2,2},~xF_{2,3},$ $xF_{2,4},~xF_{2,5},~xF_{2,6},~xF_{2,7},~xF_{2,8},~xG_{2,1},~xG_{2,2},~xG_{2,3},~xG_{2,4},~xG_{2,5},~xG_{2,6},~xG_{2,7},~xG_{2,8},~xH_{2,1},$
$~xH_{2,2},~xH_{2,3},~xH_{2,5},~xH_{2,6},~xH_{2,7},$ and $xH_{2,8}$ corresponding to sub-leading twist Dirac matrix structures have been plotted, which are given by Eqs. {\eqref{e21s}} to {\eqref{h28a}}. To analyze the characteristics of sub-leading twist GTMDs with respect to longitudinal momentum fraction $x$ and transverse momentum $p_{\perp}$, we have plotted the relevant GTMDs with respect to $p_{\perp}$, keeping $x=0.1, 0.3,$ and $0.5$ at a fixed value of momentum transfer to the proton ${\Delta_\perp}=0.5~\mathrm{GeV}$ for $\bfp$ parallel to $\Dp$.
\par
In Fig. (\ref{fig3dXPE1}), the GTMDs $xE_{2,1},~xE_{2,2},~xE_{2,3}$, and $~xE_{2,4}$, relating to the Dirac matrix structure $\Gamma=1$, have been plotted. We start our discussion with the GTMD $xE_{2,1}$, linked with the unpolarized TMD $xe^{\perp \nu}_{T}(x,{ p_\perp})$ via Eq. \eqref{tmd1} and is plotted in Figs. \ref{fig3dXPE1} $(a)$ and \ref{fig3dXPE1} $(b)$ for $u$ and $d$ quarks, respectively. Since this GTMD corresponds to an unpolarized TMD, quark flavor symmetry is identified. It has been noted that due to the occurrence of non-zero momentum transfer in the GTMD $xE_{2,1}$, the likelihood of an active quark possessing a low value of longitudinal momentum is heightened. The GTMD $xE_{2,2}$ has been plotted for each quark $u$ and $d$  in Figs. \ref{fig3dXPE1} $(c)$ and \ref{fig3dXPE1} $(d)$, respectively. As evident from Eq. \eqref{tmd2}, it is related to the T-odd TMD $xe^{\perp \nu}_{T}(x,{ p_\perp})$ and does not survive at the TMD-limit ${\Delta_\perp}=0$ in the LFQDM. For a non-zero transverse momentum transfer $\Delta_{\perp}$, its behavior is observed to be anti-symmetric over the quark flavor, as shown in Figs. \ref{fig3dXPE1} $(c)$ and \ref{fig3dXPE1} $(d)$. It can be observed that the quark and nucleon polarization associated with the GTMD $xE_{2,2}$ are largely expected at a minimal value of longitudinal momentum fraction $x$. Now let us discuss the GTMD $xE_{2,3}$, which has been plotted in Figs. \ref{fig3dXPE1} $(e)$ and \ref{fig3dXPE1} $(f)$ for the quarks $u$ and $d$, sequentially. In contrast to the previously discussed GTMDs, non-existence of quark flavor symmetry has been noticed. This property can be explained by the occurrence of an additional term in the vector diquark expression given in Eq. \eqref{e23a}. In Figs. \ref{fig3dXPE1} $(g)$ and \ref{fig3dXPE1} $(h)$, the GTMD $xE_{2,4}$ has been plotted for the quarks $u$ and $d$ , sequentially. It is found to be symmetric over the quark flavor $\nu$, which can also be inferred from Eqs. \eqref{e24s} and \eqref{e24a}. This suggests that an active $u$ or active $d$ quark will contribute similarly to the corresponding scattering process.
\par
Now, in analyzing GTMDs pertaining with the Dirac matrix structure $\Gamma=\gamma_5$ (i.e., $xE_{2,6},~xE_{2,7}$, and $xE_{2,8}$), we have plotted these GTMDs about ${ p_\perp}$ for $x =0.1,~0.3$, and $0.5$ in Fig. (\ref{fig3dXPE2}), while ${\Delta_\perp}$ and $\theta$ were kept fixed at $0.5~\mathrm{GeV}$ and $0$ sequentially. In Figs. \ref{fig3dXPE2} $(a)$ and \ref{fig3dXPE2} $(b)$, the GTMD $xE_{2,6}$, corresponding with the TMD $xe_{T}^{\nu}(x,{ p_\perp})$, has been plotted for $u$ and $d$ quarks, sequentially. This GTMD is equal to $xE_{2,2}$ in magnitude but with opposite polarity. In Figs. \ref{fig3dXPE2} $(c)$ and \ref{fig3dXPE2} $(d)$, the GTMD $xE_{2,7}$ has been plotted for the active quarks $u$ and $d$, accodrdingly. Aside from magnitude, it has been noted that the GTMD amplitude changes from negative to positive when the active quark's flavor changes from $u$ to $d$. This observation implies that the active quark's polarity shifts along with its flavor. In Figs. \ref{fig3dXPE2} $(e)$ and \ref{fig3dXPE2} $(f)$, the GTMD $xE_{2,8}$ is plotted for $u$ and $d$ quarks, accordingly. This GTMD corresponds with the T-odd TMD $xe_{L}^{\nu}(x,{ p_\perp})$, and thus, applying the TMD-limit, the distribution vanishes.
\par
Proceeding to the GTMDs ($xF_{2,1}, xF_{2,3}, xF_{2,4}$, and $xF_{2,7}$), which have been plotted in Figs. (\ref{fig3dXPF1}) and (\ref{fig3dXPF2}), they correspond to the Dirac matrix structure $\Gamma=\gamma^j$. In Figs. \ref{fig3dXPF1} $(a)$ and \ref{fig3dXPF1} $(b)$, the GTMD $xF_{2,1}^{\nu}$ has been plotted for both quark flavors. According to Eq. \eqref{tmd5}, this GTMD is connected to the unpolarized TMD $xf^{\perp \nu}(x,{ p_\perp})$. Furthermore, at the TMD-limit, this GTMD is connected to the unpolarized TMDs at leading twist as well as at twist-$4$ level, as described in Refs. \cite{sstwist3,sstwist4}. In Figs. \ref{fig3dXPF1} $(e)$ to \ref{fig3dXPF1} $(h)$, GTMDs $xF_{2,3}^{\nu}$ and $xF_{2,4}^{\nu}$ have been plotted for each quark flavors $u$ and $d$ and are linked with T-odd TMDs $xf^{\prime \nu}_{T}(x,{ p_\perp})$ and $xf^{\perp \nu}_{T}(x,{ p_\perp})$ via Eqs. \eqref{tmd6} and \eqref{tmd7}, respectively. For GTMD $xF_{2,4}^{\nu}$, it is surprising to notice that the GTMD falls to zero more rapidly with a slight increase in the transverse momentum $ p_\perp$. In Figs. \ref{fig3dXPF2} $(a)$ and \ref{fig3dXPF2} $(b)$, GTMD $xF_{2,7}^{\nu}$ is plotted for $u$ and $d$ quark, respectively, which corresponds to the longitudinally polarized T-odd TMD $xf^{\perp \nu}_{L}(x,{ p_\perp})$ through Eq. \eqref{tmd8}. Since it is a T-odd TMD, it vanishes in the zero momentum transfer limit.
\par
Now, we move towards the GTMDs corresponding with the Dirac matrix structure $\Gamma=\gamma^j \gamma_5$, which are $xG_{2,1},~xG_{2,2},~xG_{2,3},~xG_{2,4},~xG_{2,5},~xG_{2,6},~xG_{2,7}$, and $xG_{2,8}$. These GTMDs have been plotted in Figs. (\ref{fig3dXPG1}) and (\ref{fig3dXPG2}). GTMD $xG_{2,1}$ related to the unpolarized T-odd TMD $xg^{\perp \nu}(x,{ p_\perp}) $ via Eq. \eqref{tmd9} is plotted in Figs. \ref{fig3dXPG1} $(a)$ and \ref{fig3dXPG1} $(b)$. Eqs. \eqref{g21s} and \eqref{g21a} clearly suggest that the only difference between the GTMDs for active $u$ or active $d$ quarks is the magnitude due to their respective normalization constants. This GTMD also follows the generic condition for GTMDs related to T-odd TMDs, vanishing in the zero momentum transfer limit. GTMDs $xG_{2,3}$ and $xG_{2,4}$ yield the transversely polarized T-even TMDs $xg^{' \nu}_{T}(x,{ p_\perp})$ and $xg^{\perp \nu}_{T}(x,{ p_\perp})$ in the TMD limit.
	
These GTMDs have been plotted in Fig. \ref{fig3dXPG1}. After examining the GTMD $xG_{2,3}$, it is noted that when the momentum transfer ${\Delta_\perp}$ is increased to $0.5~\mathrm{GeV}$, the GTMD's amplitude increases and the maxima of the DF for active $u$ quark shift to the larger value of the longitudinal momentum fraction $x$, while for the active $d$ quark it shifts towards the lower values of the longitudinal momentum fraction $x$. In contrast, for the GTMD $xG_{2,4}$, the opposite behavior is observed i.e., the amplitude of the GTMD decreases as the momentum transfer ${\Delta_\perp}$ is increased from zero to $0.5~\mathrm{GeV}$. In Figs. \ref{fig3dXPG2} $(c)$ and \ref{fig3dXPG2} $(d)$, the GTMD $xG_{2,7}$, which relates to the longitudinally polarized T-even TMD $xg^{\perp \nu}_{L}(x,{ p_\perp})$ via Eq. \eqref{tmd12}, is plotted. It is found that the GTMD $xG_{2,7}$ has a negative amplitude for lower values of the longitudinal momentum fraction $x$ and transverse momentum $p_\perp$. When compared with the TMD limit, i.e., the ${\Delta_\perp}=0$ distribution, it is observed that due to the momentum transfer not being equal to zero, the maxima of the distribution shift towards the lower value of the longitudinal momentum fraction $x$.
\par 
Now, we move towards the GTMDs associated with the Dirac matrix structure $\Gamma=i \sigma^{ij} \gamma_5$. The GTMDs, $xH_{2,1}$ and $xH_{2,2}$, are plotted in Fig. (\ref{fig3dXPH1}) and are related to the unpolarized T-odd TMD $xh(x,{ p_\perp})$ and $xh^{\perp \nu}_{T}(x,{ p_\perp})$ via Eqs. \eqref{tmd13} and \eqref{tmd14}, respectively. In the case of GTMD $xH_{2,1}$, significant quark flavor symmetry is observed, and as evident from Eq. \eqref{tmd13}, the distribution vanishes in the zero momentum transfer limit. Conversely, for GTMD $xH_{2,2}$, quark flavor anti-symmetry is observed. When comparing with the TMD-limit distribution, a small shift in the maxima of GTMD $xH_{2,2}$ towards the higher (lower) value of the longitudinal momentum fraction $x$ for $u$ ($d$) quarks is observed. Finally, we discuss the GTMDs $xH_{2,6}$ and $xH_{2,8}$ concerning the Dirac matrix structure $\Gamma=i \sigma^{+-} \gamma_5$, which are plotted in Fig. \ref{fig3dXPH2}. For the GTMD $xH_{2,6}$, related to the transversely polarized T-even TMD $xh^{\nu}_{T}(x,{ p_\perp})$ via Eq. \eqref{tmd15}, it is observed that the maximum amplitude increases (decreases) as the non-zero momentum transfer is introduced in the expression for the $u$ ($d$) quark. For the GTMD $xH_{2,8}$, related to the longitudinally polarized T-even TMD $xh^{\nu}_{L}(x,{ p_\perp})$, some quark flavor symmetry is observed. Compared to the corresponding TMD, these GTMDs show a decrease in maximum amplitude for both quark flavors.
%
	\subsubsection{Variation with $x$ and ${ \Delta_\perp}$}\label{ssxd}
	%
	%
		\begin{figure*}
				\centering
				\begin{minipage}[c]{0.98\textwidth}
						(a)\includegraphics[width=7.3cm]{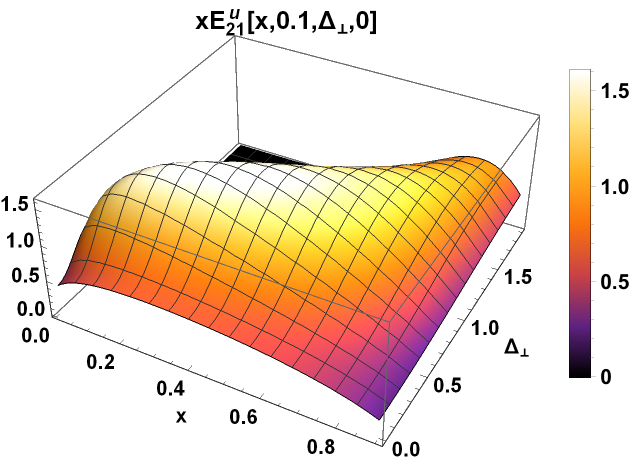}
						\hspace{0.05cm}
						(b)\includegraphics[width=7.3cm]{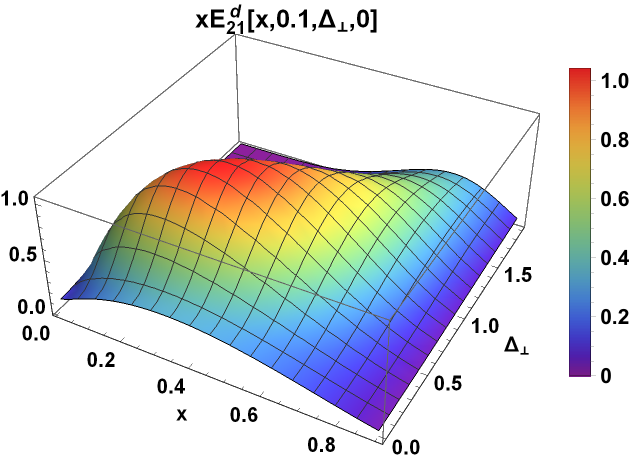}
						\hspace{0.05cm}
						(c)\includegraphics[width=7.3cm]{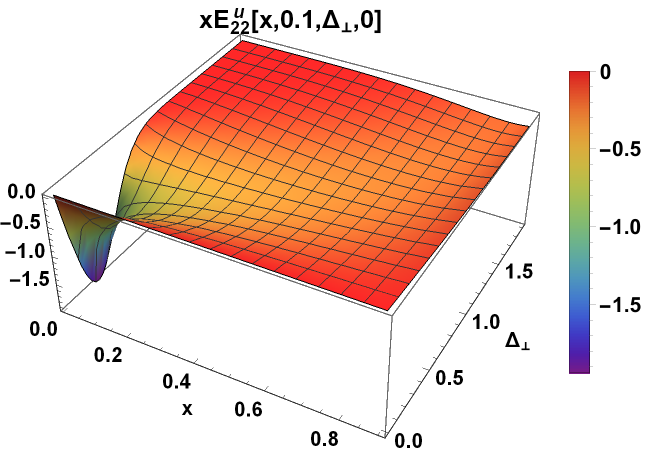}
						\hspace{0.05cm}
						(d)\includegraphics[width=7.3cm]{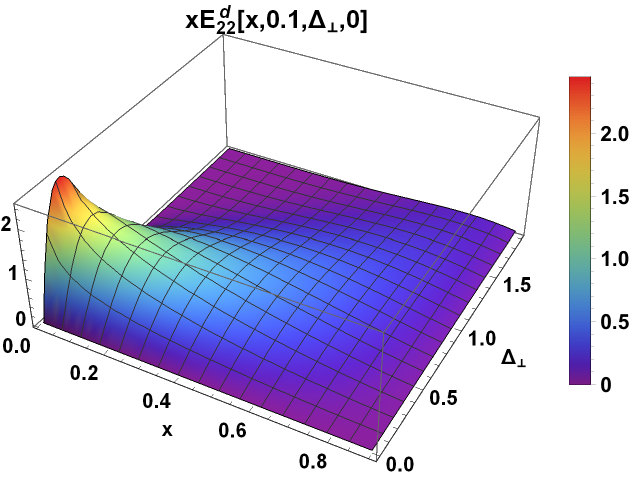}
						\hspace{0.05cm}
						(e)\includegraphics[width=7.3cm]{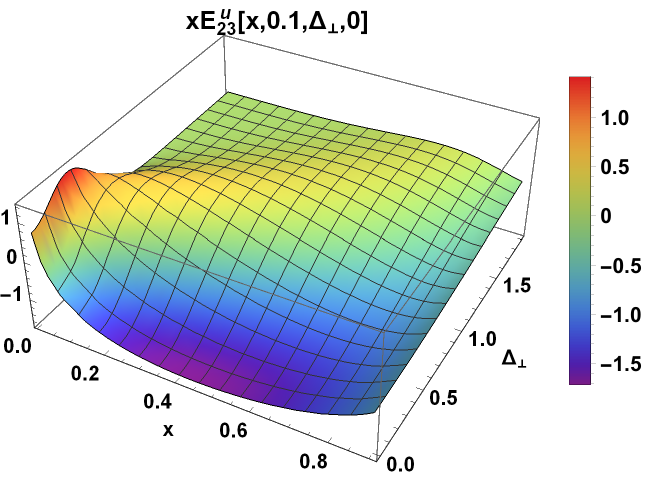}
						\hspace{0.05cm}
						(f)\includegraphics[width=7.3cm]{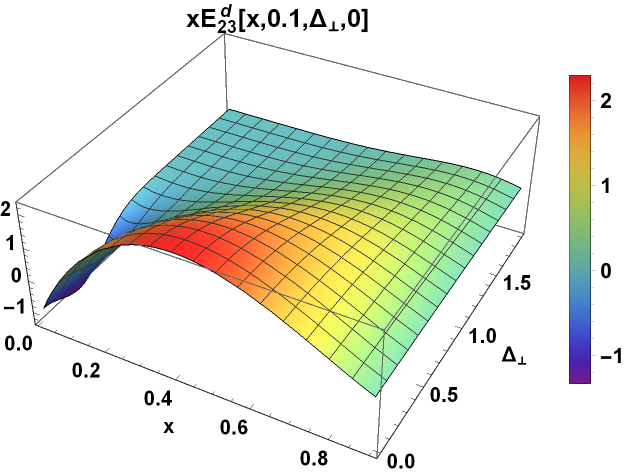}
						\hspace{0.05cm}
						(g)\includegraphics[width=7.3cm]{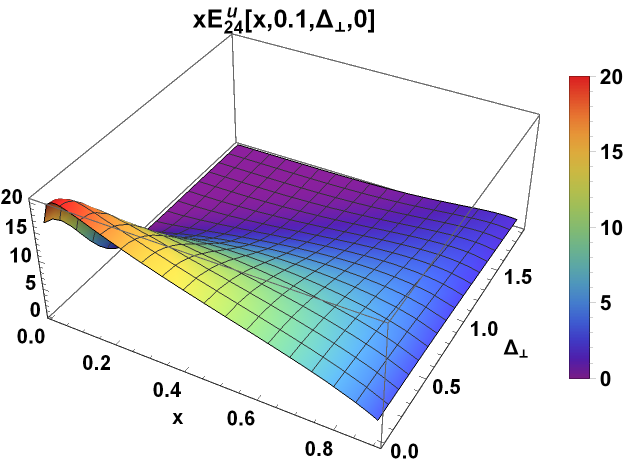}
						\hspace{0.05cm}
						(h)\includegraphics[width=7.3cm]{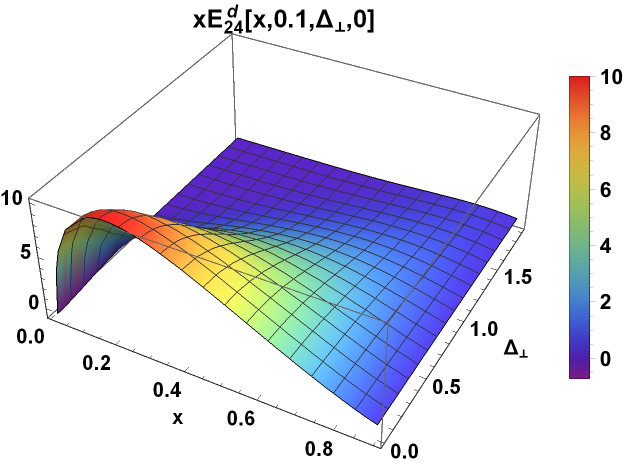}
						\hspace{0.05cm}\\
					\end{minipage}
				\caption{\label{fig3dXDE1} The sub-leading twist GTMDs 		
						$x E_{2,1}^{\nu}(x, p_{\perp},\Delta_{\perp},\theta)$,
						$x E_{2,2}^{\nu}(x, p_{\perp},\Delta_{\perp},\theta)$,
						$x E_{2,3}^{\nu}(x, p_{\perp},\Delta_{\perp},\theta)$, and
						$x E_{2,4}^{\nu}(x, p_{\perp},\Delta_{\perp},\theta)$
						are	plotted about $x$ and ${{ \Delta_\perp}}$ keeping ${ p_\perp}= 0.1~\mathrm{GeV}$ for ${\bfp} \parallel {\Dp}$. In sequential order, $u$ and $d$ quarks are in the left and right columns.
					}
			\end{figure*}
		\begin{figure*}
				\centering
				\begin{minipage}[c]{0.98\textwidth}
						(a)\includegraphics[width=7.3cm]{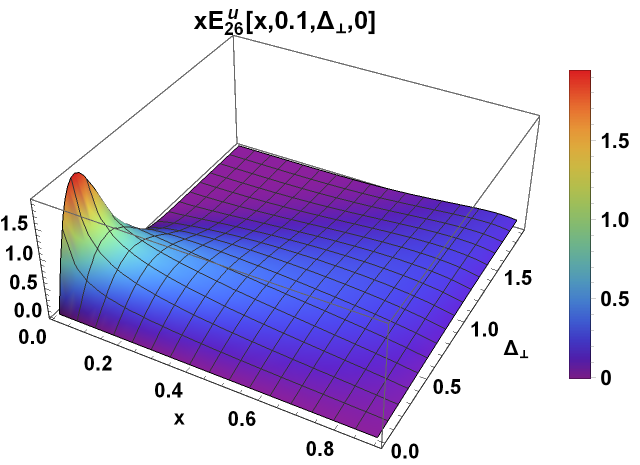}
						\hspace{0.05cm}
						(b)\includegraphics[width=7.3cm]{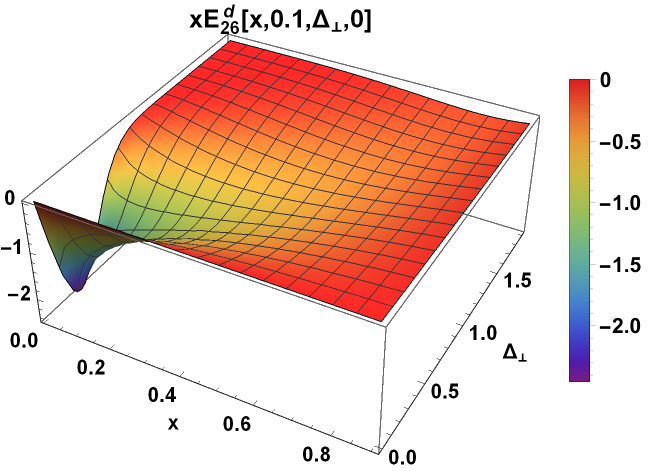}
						\hspace{0.05cm}
						(c)\includegraphics[width=7.3cm]{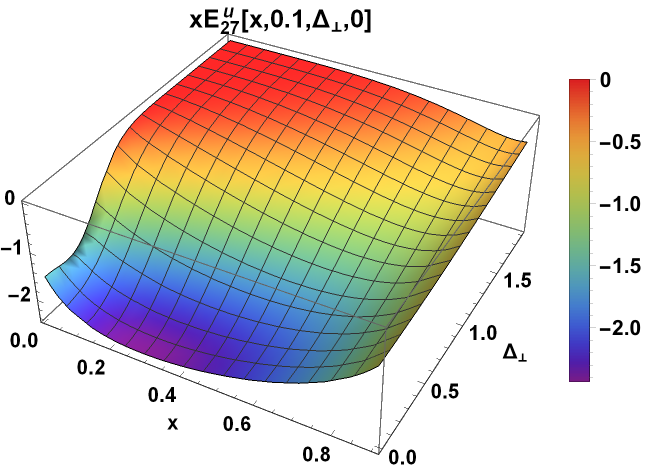}
						\hspace{0.05cm}
						(d)\includegraphics[width=7.3cm]{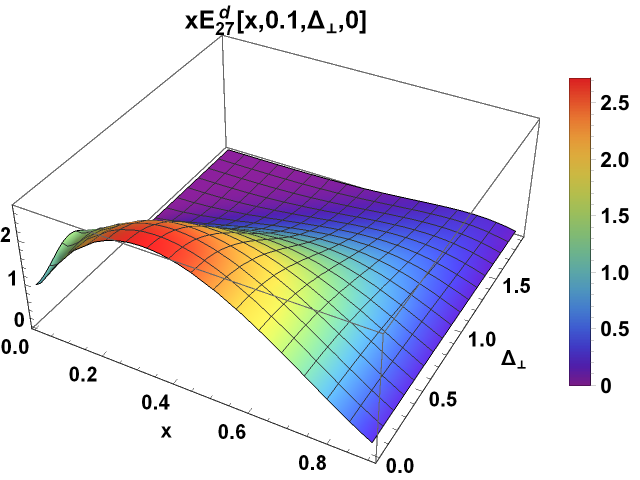}
						\hspace{0.05cm}
						(e)\includegraphics[width=7.3cm]{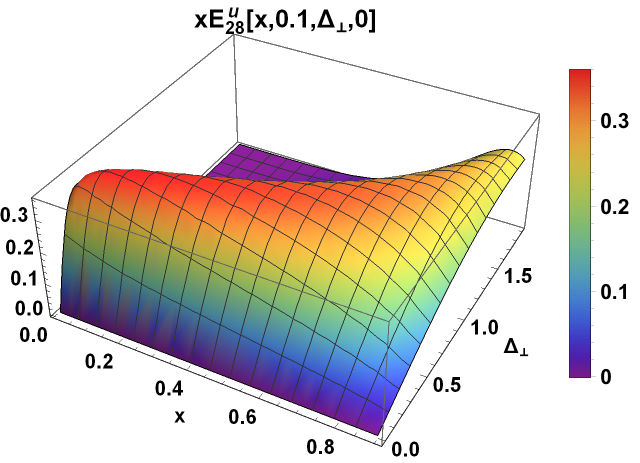}
						\hspace{0.05cm}
						(f)\includegraphics[width=7.3cm]{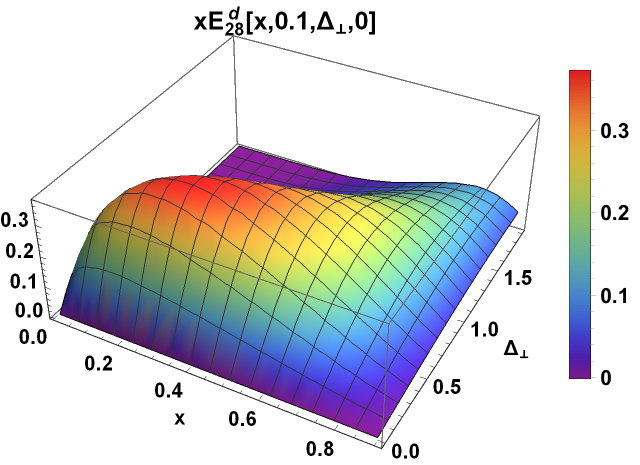}
						\hspace{0.05cm}
						\\
					\end{minipage}
				\caption{\label{fig3dXDE2} The sub-leading twist GTMDs 		
						$x E_{2,6}^{\nu}(x, p_{\perp},\Delta_{\perp},\theta)$,
						$x E_{2,7}^{\nu}(x, p_{\perp},\Delta_{\perp},\theta)$,
						and
						$x E_{2,8}^{\nu}(x, p_{\perp},\Delta_{\perp},\theta)$
							are	plotted about $x$ and ${{ \Delta_\perp}}$ keeping ${ p_\perp}= 0.1~\mathrm{GeV}$ for ${\bfp} \parallel {\Dp}$. In sequential order, $u$ and $d$ quarks are in the left and right columns.
					}
			\end{figure*}
		\begin{figure*}
				\centering
				\begin{minipage}[c]{0.98\textwidth}
						(a)\includegraphics[width=7.3cm]{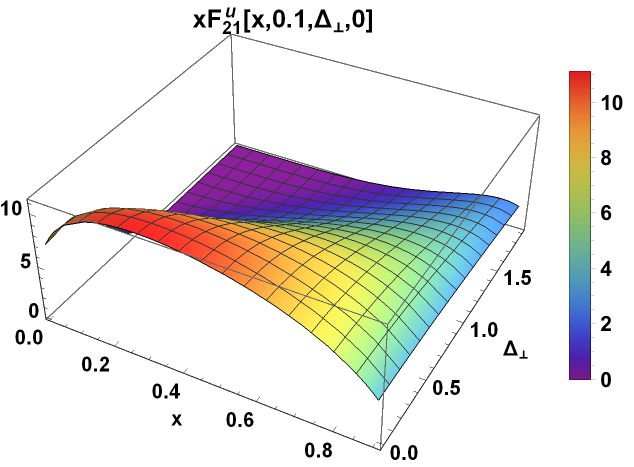}
						\hspace{0.05cm}
						(b)\includegraphics[width=7.3cm]{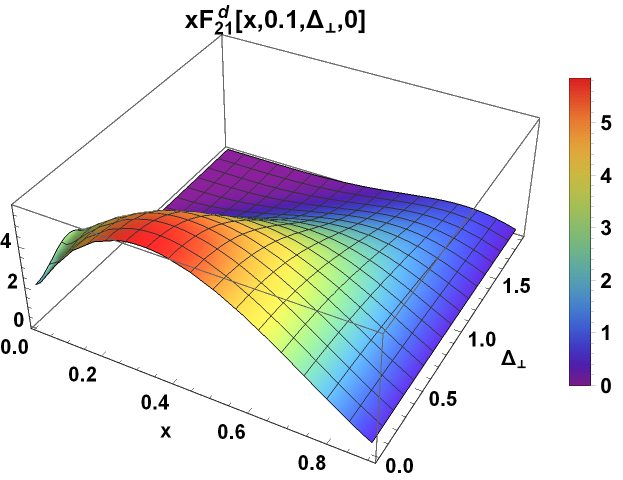}
						\hspace{0.05cm}
						(c)\includegraphics[width=7.3cm]{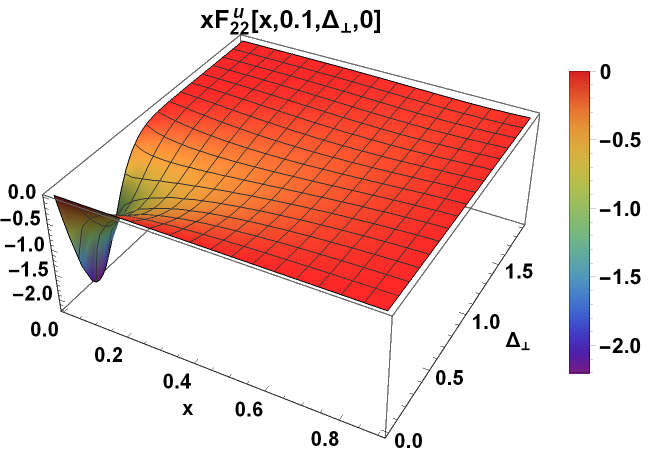}
						\hspace{0.05cm}
						(d)\includegraphics[width=7.3cm]{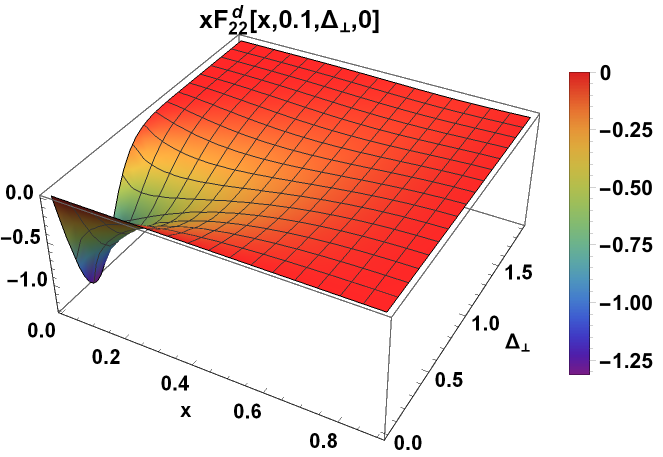}
						\hspace{0.05cm}
						(e)\includegraphics[width=7.3cm]{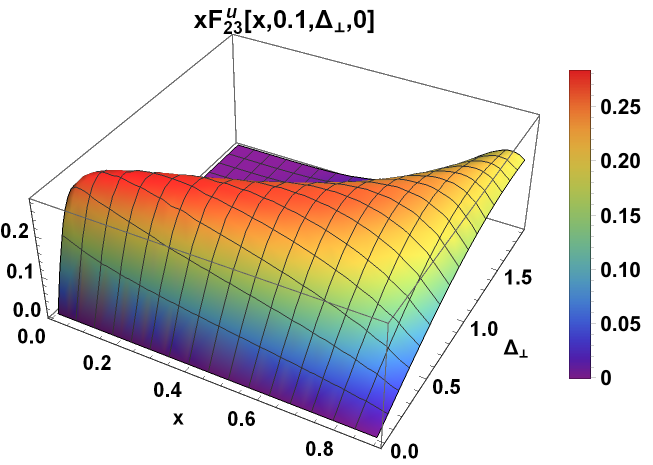}
						\hspace{0.05cm}
						(f)\includegraphics[width=7.3cm]{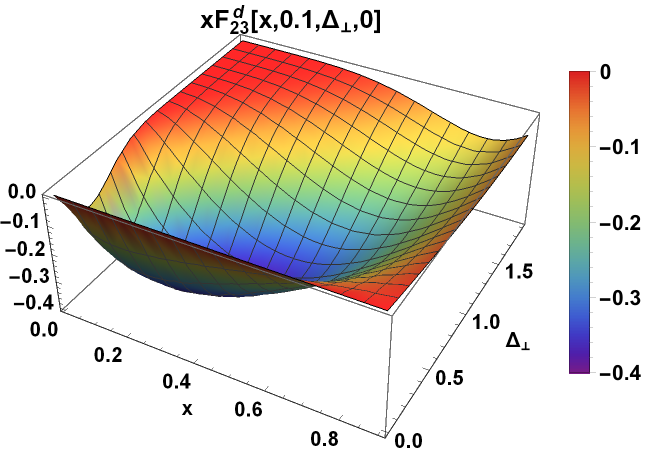}
						\hspace{0.05cm}
						(g)\includegraphics[width=7.3cm]{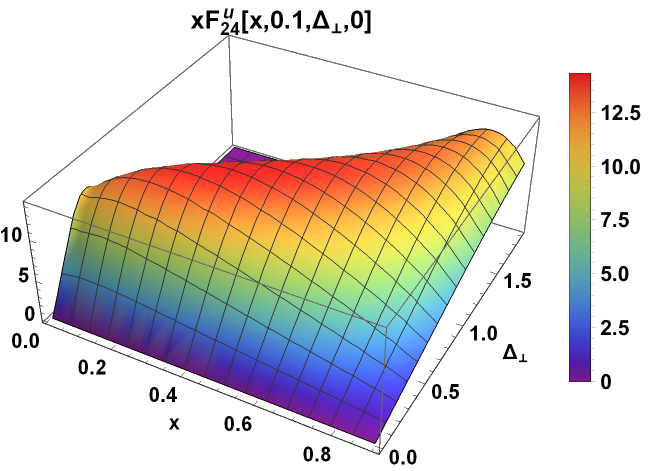}
						\hspace{0.05cm}
						(h)\includegraphics[width=7.3cm]{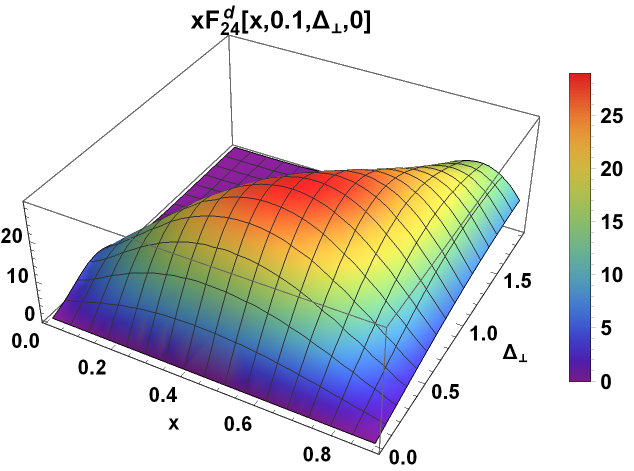}
						\hspace{0.05cm}\\
					\end{minipage}
				\caption{\label{fig3dXDF1} The sub-leading twist GTMDs 		
						$x F_{2,1}^{\nu}(x, p_{\perp},\Delta_{\perp},\theta)$,
						$x F_{2,2}^{\nu}(x, p_{\perp},\Delta_{\perp},\theta)$,
						$x F_{2,3}^{\nu}(x, p_{\perp},\Delta_{\perp},\theta)$, and
						$x F_{2,4}^{\nu}(x, p_{\perp},\Delta_{\perp},\theta)$
						are	plotted about $x$ and ${{ \Delta_\perp}}$ keeping ${ p_\perp}= 0.1~\mathrm{GeV}$ for ${\bfp} \parallel {\Dp}$. In sequential order, $u$ and $d$ quarks are in the left and right columns.
					}
			\end{figure*}
		\begin{figure*}
				\centering
				\begin{minipage}[c]{0.98\textwidth}
						(a)\includegraphics[width=7.3cm]{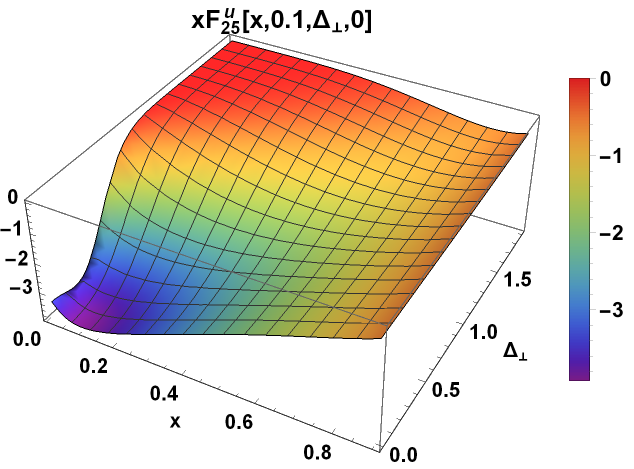}
						\hspace{0.05cm}
						(b)\includegraphics[width=7.3cm]{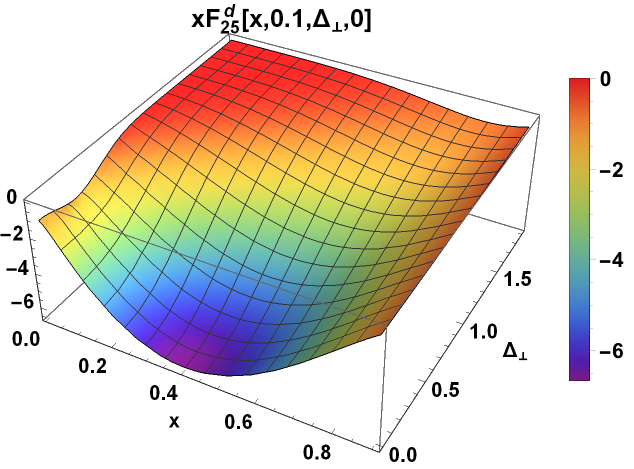}
						\hspace{0.05cm}
						(c)\includegraphics[width=7.3cm]{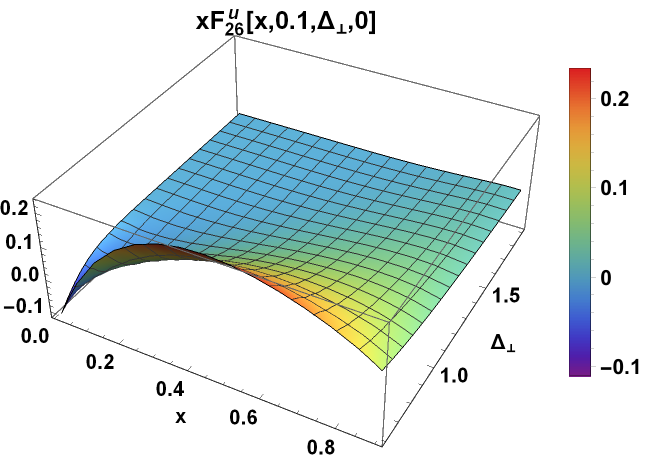}
						\hspace{0.05cm}
					    (d)\includegraphics[width=7.3cm]{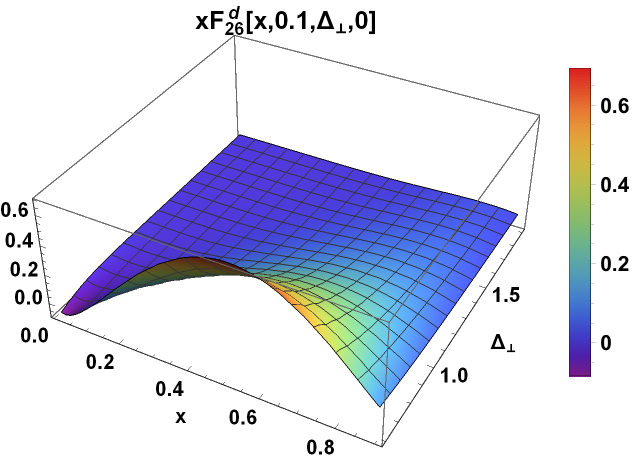}
						\hspace{0.05cm}
						(e)\includegraphics[width=7.3cm]{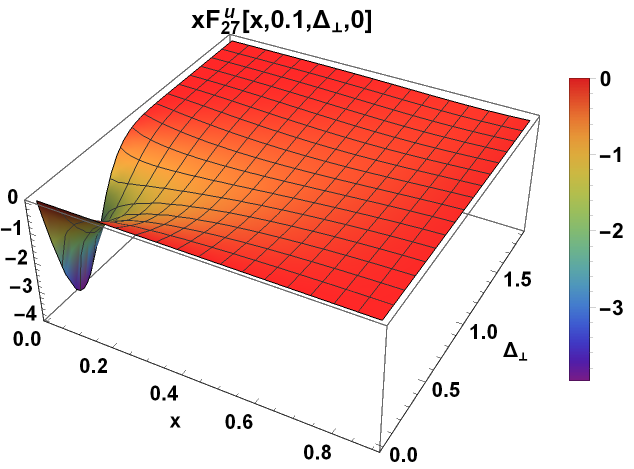}
						\hspace{0.05cm}
						(f)\includegraphics[width=7.3cm]{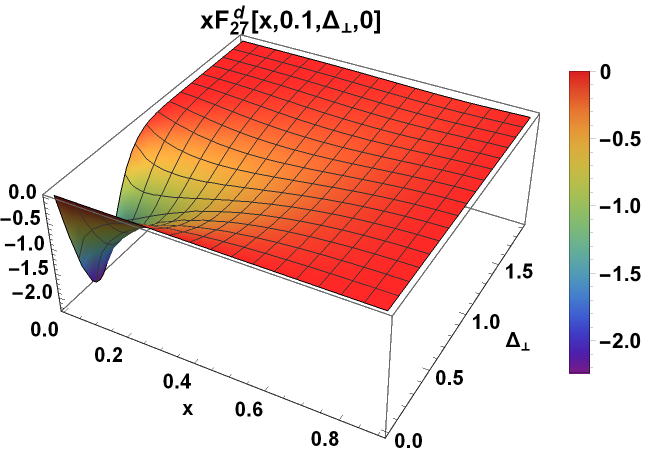}
						\hspace{0.05cm}
						(g)\includegraphics[width=7.3cm]{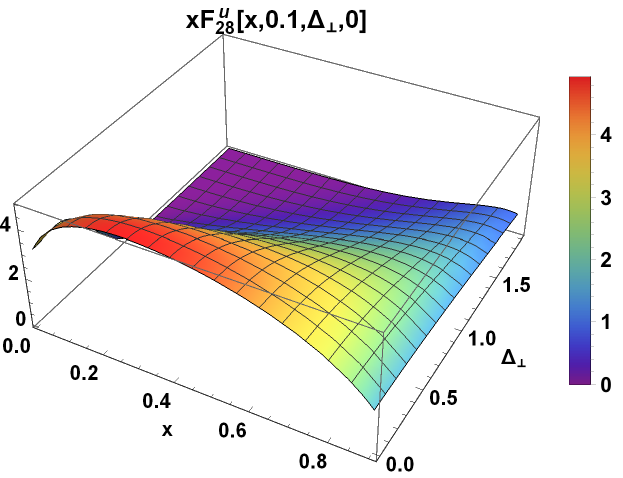}
						\hspace{0.05cm}
						(h)\includegraphics[width=7.3cm]{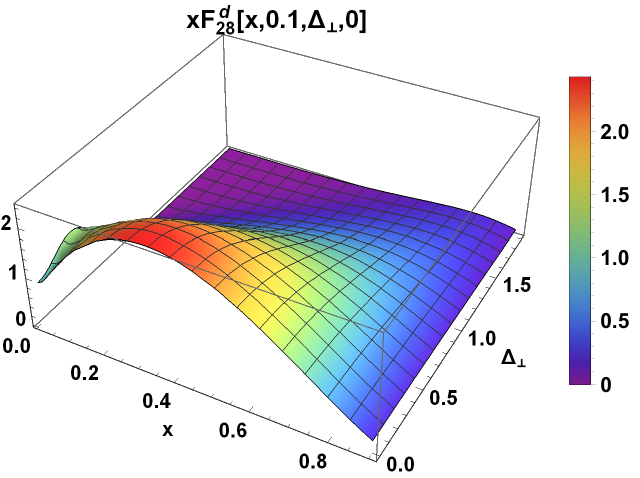}
						\hspace{0.05cm}\\
					\end{minipage}
				\caption{\label{fig3dXDF2} The sub-leading twist GTMDs 		
						$x F_{2,5}^{\nu}(x, p_{\perp},\Delta_{\perp},\theta)$,
						$x F_{2,6}^{\nu}(x, p_{\perp},\Delta_{\perp},\theta)$,
						$x F_{2,7}^{\nu}(x, p_{\perp},\Delta_{\perp},\theta)$, and
						$x F_{2,8}^{\nu}(x, p_{\perp},\Delta_{\perp},\theta)$
						are	plotted about $x$ and ${{ \Delta_\perp}}$ keeping ${ p_\perp}= 0.1~\mathrm{GeV}$ for ${\bfp} \parallel {\Dp}$. In sequential order, $u$ and $d$ quarks are in the left and right columns.
					}
			\end{figure*}
		\begin{figure*}
				\centering
				\begin{minipage}[c]{0.98\textwidth}
						(a)\includegraphics[width=7.3cm]{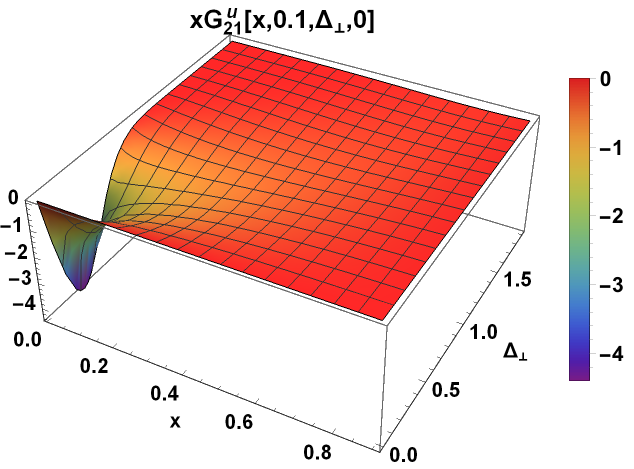}
						\hspace{0.05cm}
						(b)\includegraphics[width=7.3cm]{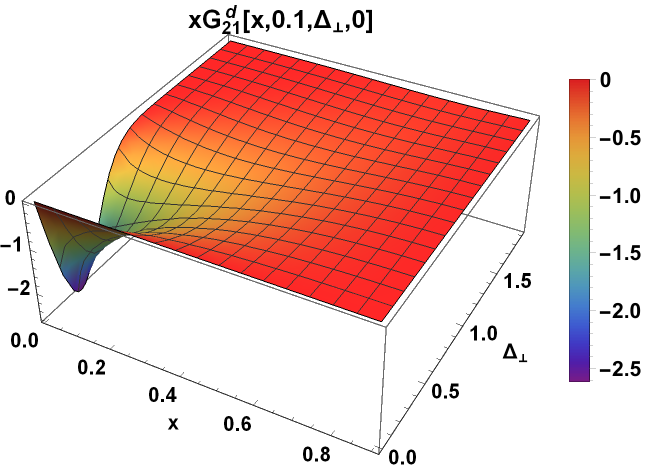}
						\hspace{0.05cm}
						(c)\includegraphics[width=7.3cm]{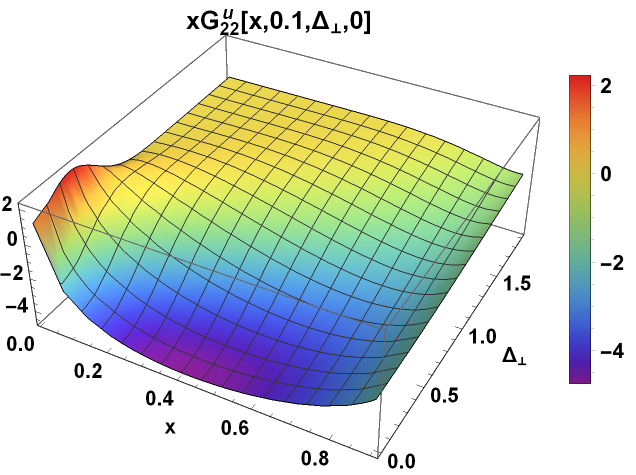}
						\hspace{0.05cm}
						(d)\includegraphics[width=7.3cm]{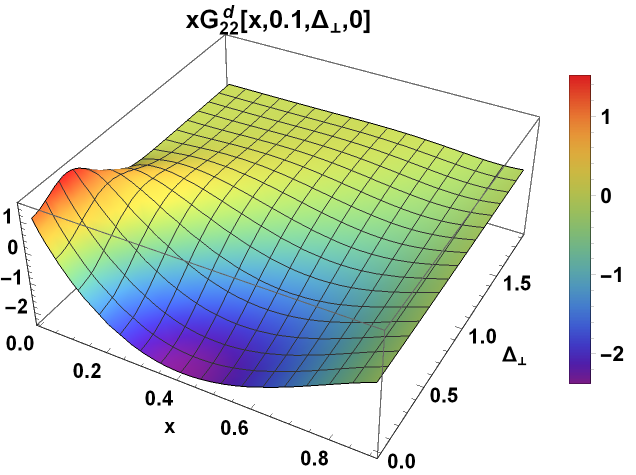}
						\hspace{0.05cm}
						(e)\includegraphics[width=7.3cm]{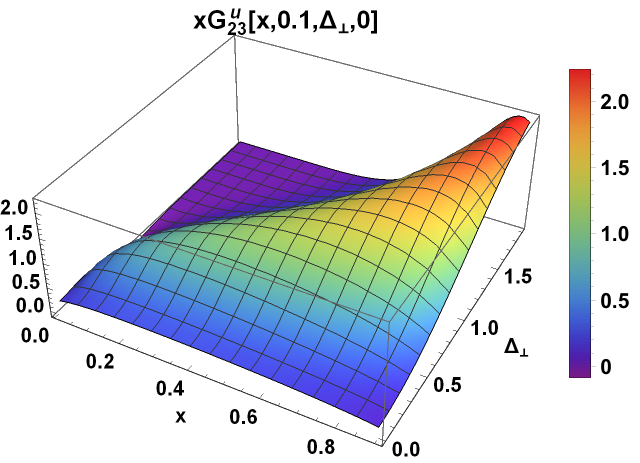}
						\hspace{0.05cm}
						(f)\includegraphics[width=7.3cm]{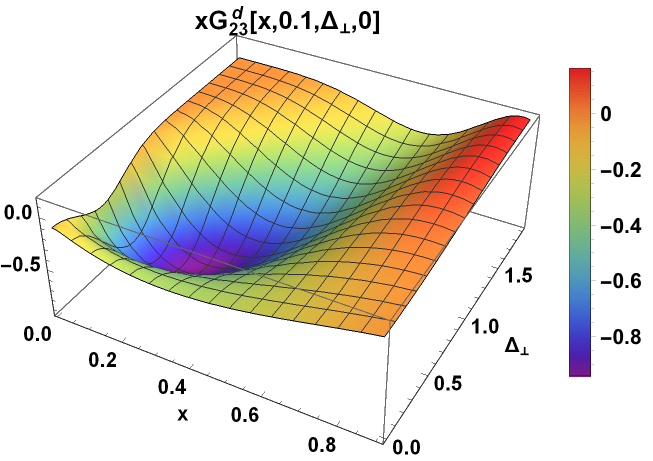}
						\hspace{0.05cm}
						(g)\includegraphics[width=7.3cm]{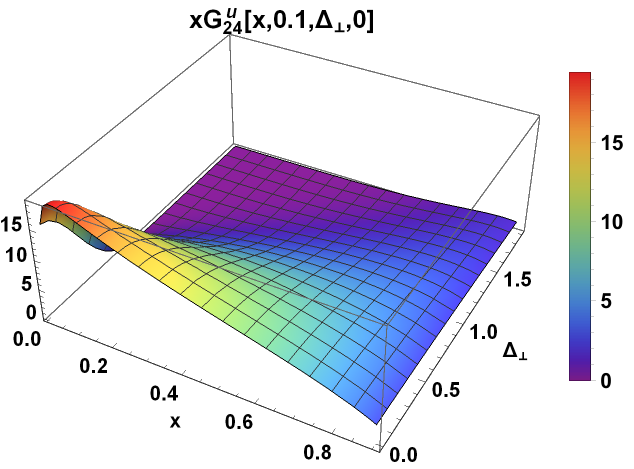}
						\hspace{0.05cm}
						(h)\includegraphics[width=7.3cm]{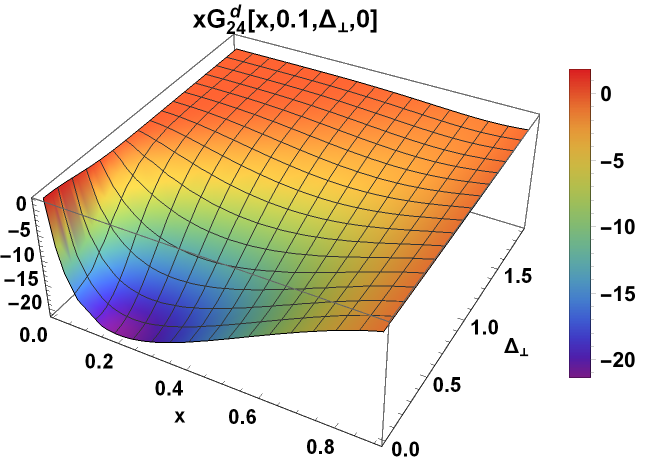}
						\hspace{0.05cm}\\
					\end{minipage}
				\caption{\label{fig3dXDG1} The sub-leading twist GTMDs 		
						$x G_{2,1}^{\nu}(x, p_{\perp},\Delta_{\perp},\theta)$,
						$x G_{2,2}^{\nu}(x, p_{\perp},\Delta_{\perp},\theta)$,
						$x G_{2,3}^{\nu}(x, p_{\perp},\Delta_{\perp},\theta)$, and
						$x G_{2,4}^{\nu}(x, p_{\perp},\Delta_{\perp},\theta)$
						are	plotted about $x$ and ${{ \Delta_\perp}}$ keeping ${ p_\perp}= 0.1~\mathrm{GeV}$ for ${\bfp} \parallel {\Dp}$. In sequential order, $u$ and $d$ quarks are in the left and right columns.
					}
			\end{figure*}
		\begin{figure*}
				\centering
				\begin{minipage}[c]{0.98\textwidth}
						(a)\includegraphics[width=7.3cm]{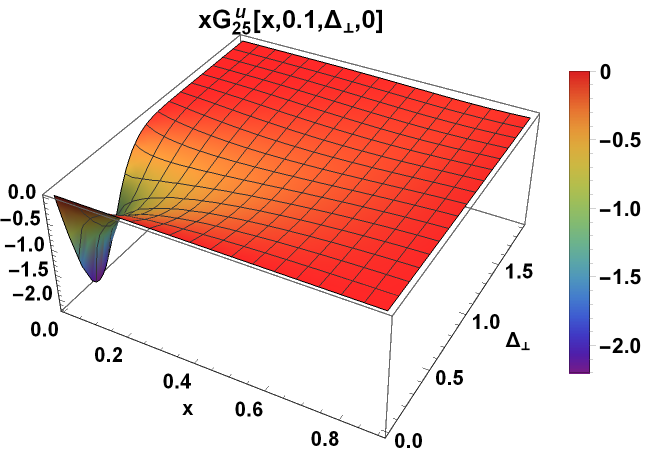}
						\hspace{0.05cm}
						(b)\includegraphics[width=7.3cm]{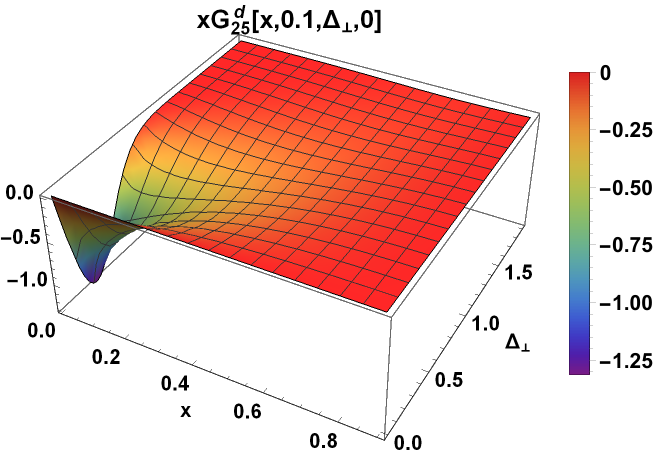}
						\hspace{0.05cm}
						(c)\includegraphics[width=7.3cm]{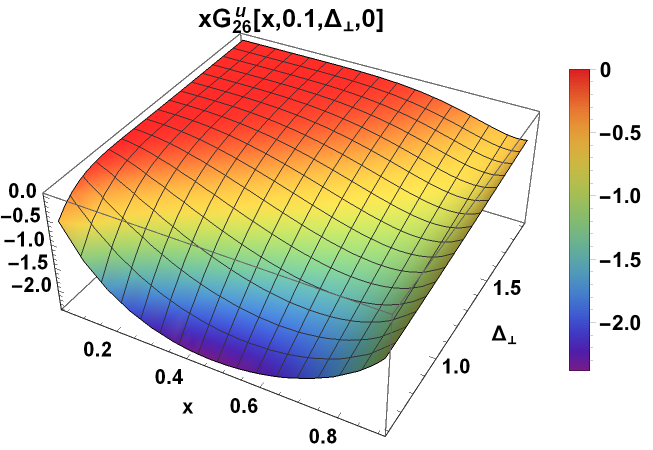}
						\hspace{0.05cm}
						(d)\includegraphics[width=7.3cm]{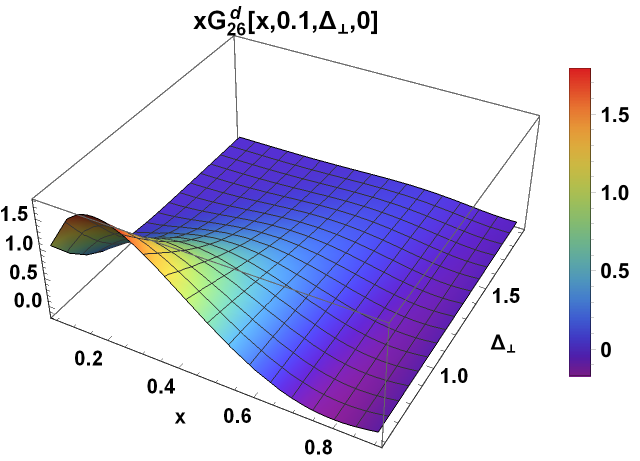}
						\hspace{0.05cm}
						(e)\includegraphics[width=7.3cm]{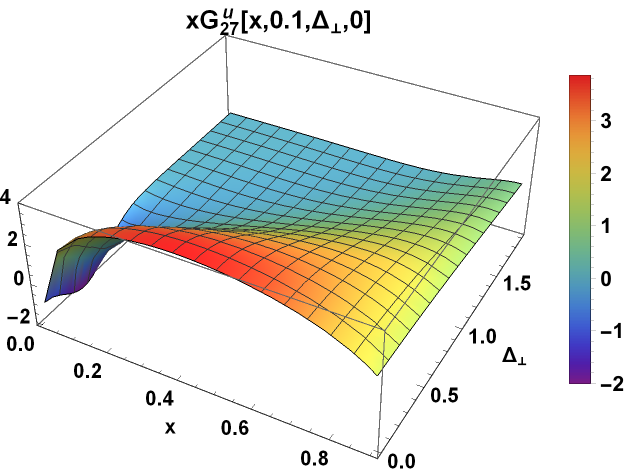}
						\hspace{0.05cm}
						(f)\includegraphics[width=7.3cm]{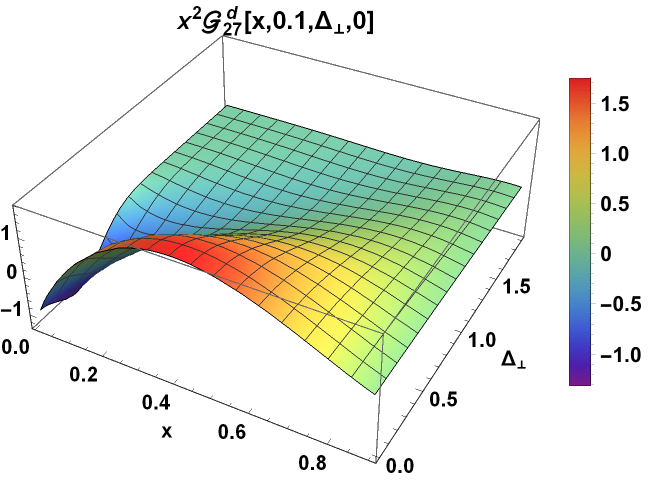}
						\hspace{0.05cm}
						(g)\includegraphics[width=7.3cm]{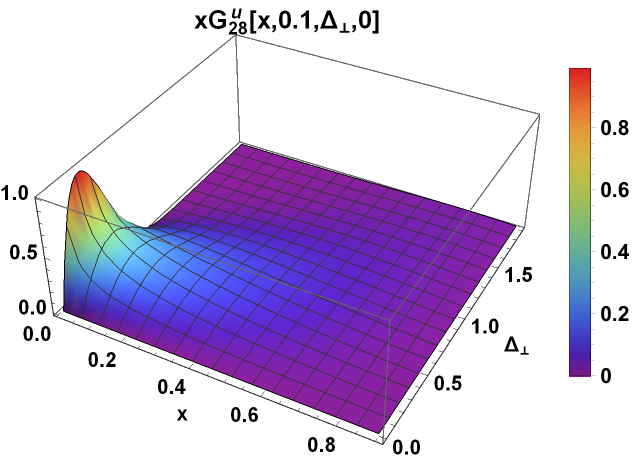}
						\hspace{0.05cm}
						(h)\includegraphics[width=7.3cm]{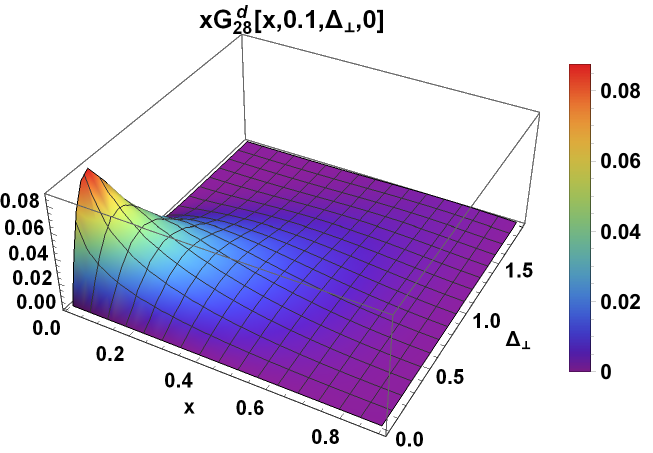}
						\hspace{0.05cm}\\
					\end{minipage}
				\caption{\label{fig3dXDG2} The sub-leading twist GTMDs 		
						$x G_{2,5}^{\nu}(x, p_{\perp},\Delta_{\perp},\theta)$,
						$x G_{2,6}^{\nu}(x, p_{\perp},\Delta_{\perp},\theta)$,
						$x G_{2,7}^{\nu}(x, p_{\perp},\Delta_{\perp},\theta)$, and
						$x G_{2,8}^{\nu}(x, p_{\perp},\Delta_{\perp},\theta)$
							are	plotted about $x$ and ${{ \Delta_\perp}}$ keeping ${ p_\perp}= 0.1~\mathrm{GeV}$ for ${\bfp} \parallel {\Dp}$. In sequential order, $u$ and $d$ quarks are in the left and right columns.
					}
			\end{figure*}
		\begin{figure*}
				\centering
				\begin{minipage}[c]{0.98\textwidth}
						(a)\includegraphics[width=7.3cm]{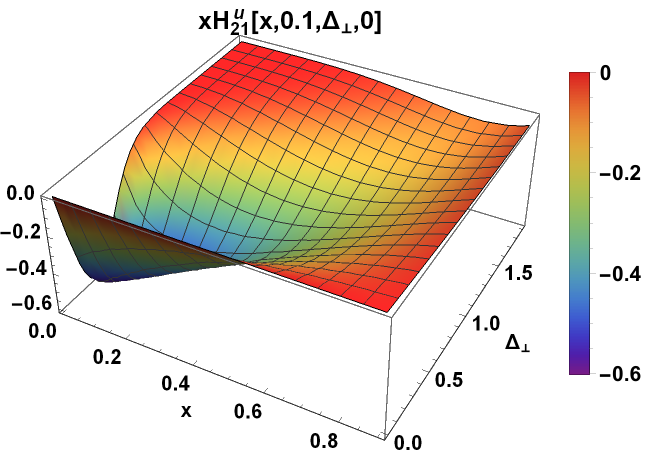}
						\hspace{0.05cm}
						(b)\includegraphics[width=7.3cm]{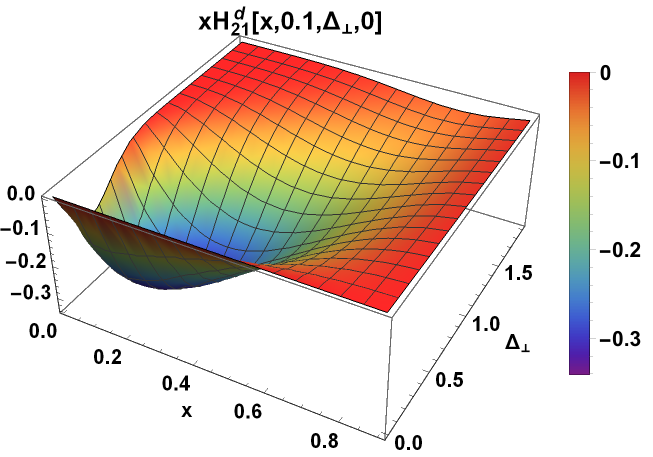}
						\hspace{0.05cm}
						(c)\includegraphics[width=7.3cm]{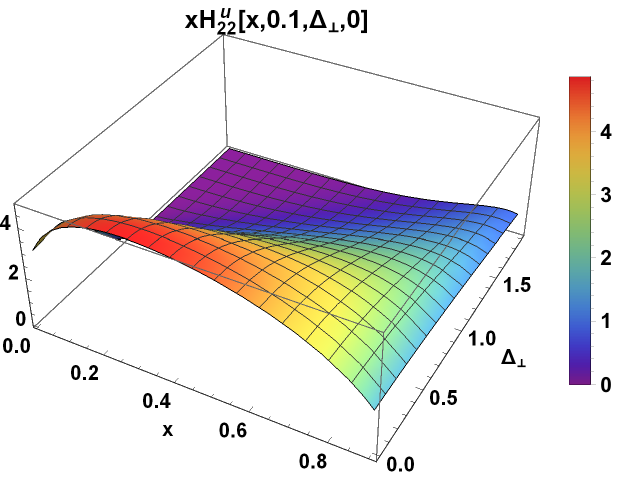}
						\hspace{0.05cm}
						(d)\includegraphics[width=7.3cm]{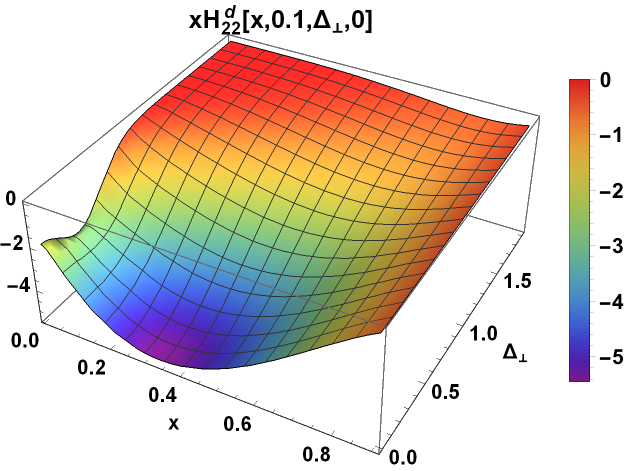}
						\hspace{0.05cm}
						\hspace{0.05cm}
						(c)\includegraphics[width=7.3cm]{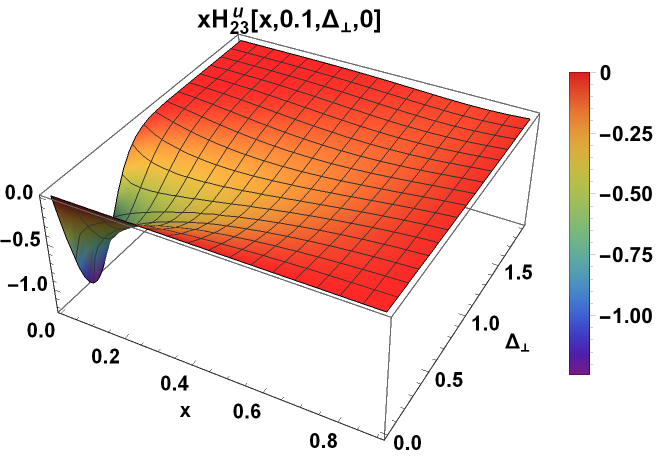}
						\hspace{0.05cm}
						(d)\includegraphics[width=7.3cm]{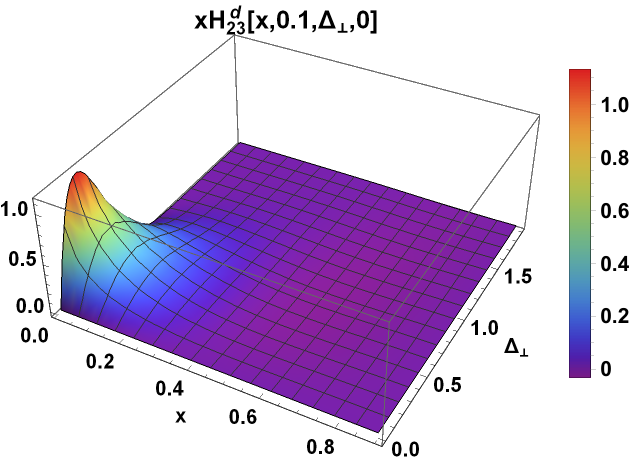}
						\hspace{0.05cm}
						\\
					\end{minipage}
				\caption{\label{fig3dXDH1} The sub-leading twist GTMDs 		
						$x H_{2,1}^{\nu}(x, p_{\perp},\Delta_{\perp},\theta)$,
						$x H_{2,2}^{\nu}(x, p_{\perp},\Delta_{\perp},\theta)$,
						and
						$x H_{2,3}^{\nu}(x, p_{\perp},\Delta_{\perp},\theta)$
						are	plotted about $x$ and ${{ \Delta_\perp}}$ keeping ${ p_\perp}= 0.1~\mathrm{GeV}$ for ${\bfp} \parallel {\Dp}$. In sequential order, $u$ and $d$ quarks are in the left and right columns.
					}
			\end{figure*}
		\begin{figure*}
				\centering
				\begin{minipage}[c]{0.98\textwidth}
						(a)\includegraphics[width=7.3cm]{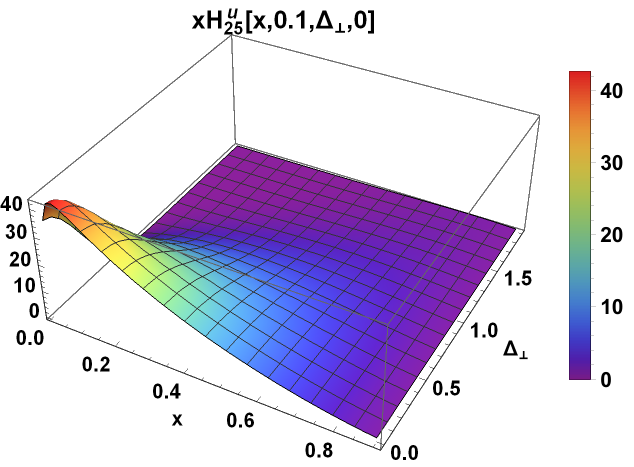}
						\hspace{0.05cm}
						(b)\includegraphics[width=7.3cm]{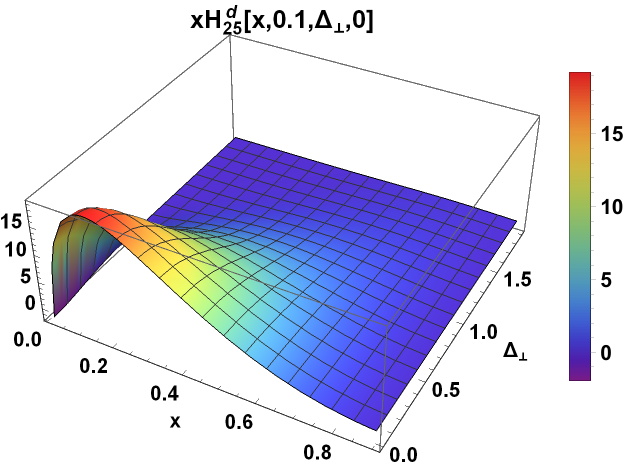}
						\hspace{0.05cm}
						(c)\includegraphics[width=7.3cm]{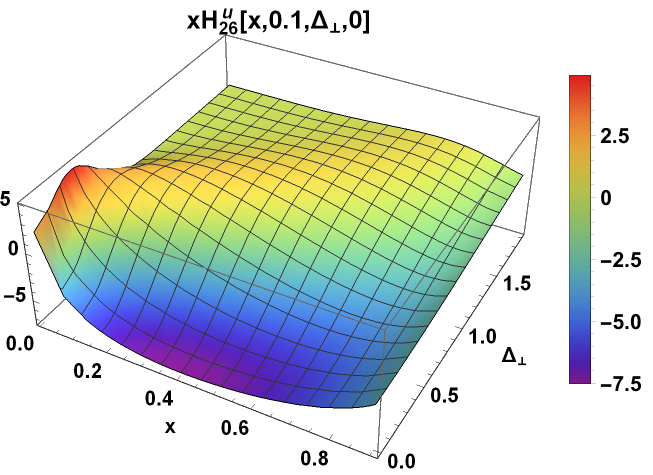}
						\hspace{0.05cm}
						(d)\includegraphics[width=7.3cm]{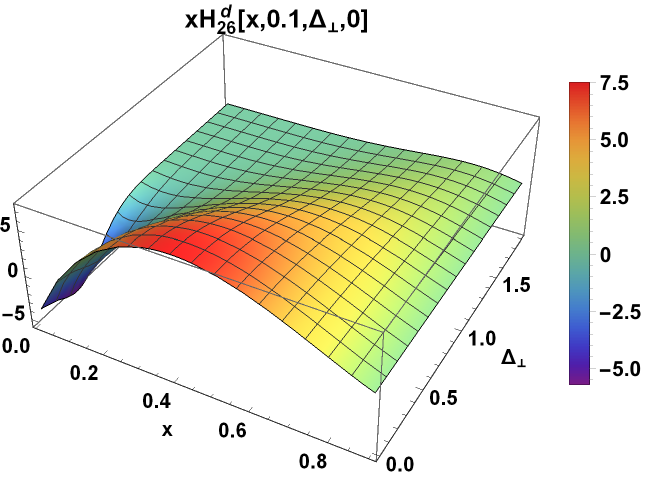}
						\hspace{0.05cm}
						(e)\includegraphics[width=7.3cm]{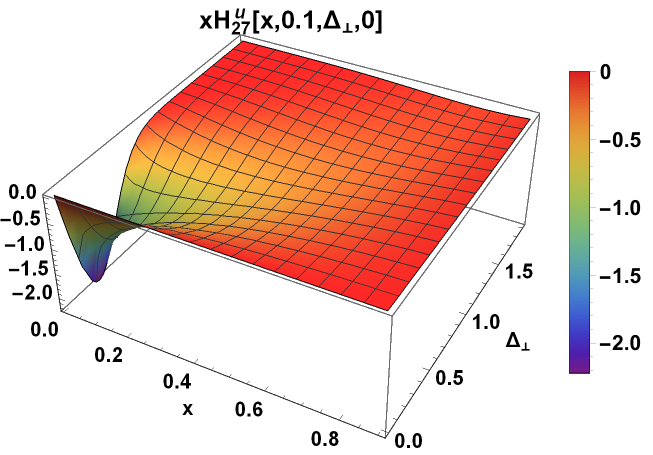}
						\hspace{0.05cm}
						(f)\includegraphics[width=7.3cm]{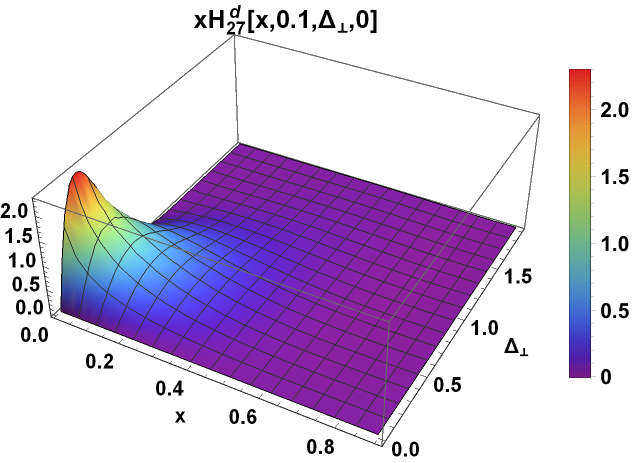}
						\hspace{0.05cm}
						(g)\includegraphics[width=7.3cm]{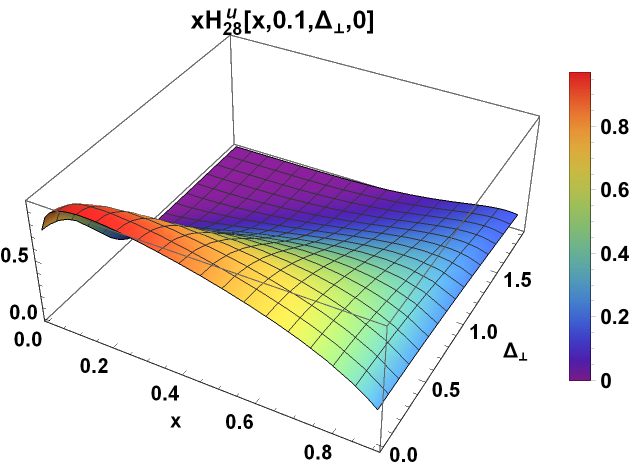}
						\hspace{0.05cm}
						(h)\includegraphics[width=7.3cm]{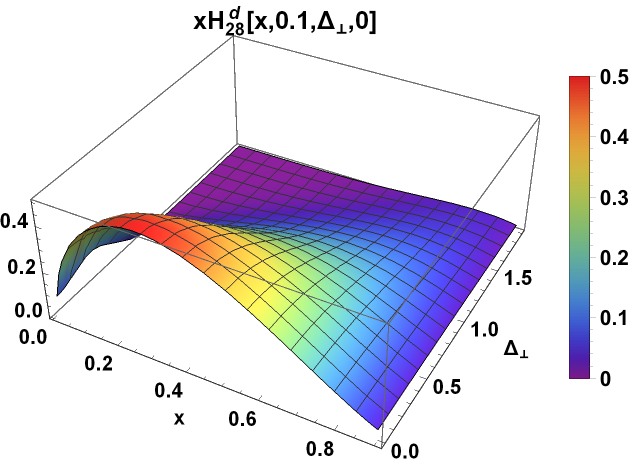}
						\hspace{0.05cm}\\
					\end{minipage}
				\caption{\label{fig3dXDH2} The sub-leading twist GTMDs 		
						$x H_{2,5}^{\nu}(x, p_{\perp},\Delta_{\perp},\theta)$,
						$x H_{2,6}^{\nu}(x, p_{\perp},\Delta_{\perp},\theta)$,
						$x H_{2,7}^{\nu}(x, p_{\perp},\Delta_{\perp},\theta)$, and
						$x H_{2,8}^{\nu}(x, p_{\perp},\Delta_{\perp},\theta)$
						are	plotted plotted about $x$ and ${{ \Delta_\perp}}$ keeping ${ p_\perp}= 0.1~\mathrm{GeV}$ for ${\bfp} \parallel {\Dp}$. In sequential order, $u$ and $d$ quarks are in the left and right columns.
					}
			\end{figure*}
%
In this subsubsection, we study the variation of different GTMDs with respect to the variables $x$ and ${\Delta_\perp}$ while keeping the value of transverse momentum ${p_\perp}=0.1~\mathrm{GeV}$, and here we have taken momentum transfer ${\Dp}$ to be parallel to the transverse momentum $\bfp$. $3$-dimensional plots of GTMDs $xE_{2,1},~xE_{2,2},~xE_{2,3}$, and $xE_{2,4}$ concerning the Dirac matrix structure $\Gamma=1$ are given in Fig. (\ref{fig3dXDE1}). From the plots represented in Fig. \ref{fig3dXDE1}, it is clear that the GTMD $xE_{2,1}$ shows a similar pattern for each quark flavor $u$ and $d$, changing only in magnitude. GTMD $xE_{2,2}$ has been plotted for the active quarks $u$ and $d$  in Figs. \ref{fig3dXDE1} $(c)$ and \ref{fig3dXDE1} $(d)$ sequenitially. As the plots show, the maximum amplitude of these distributions is fixed towards the low values of longitudinal momentum fraction $x$ and transverse momentum transfer ${\Dp}$. For the active quark flavors $u$ and $d$, GTMD $xE_{2,3}$ has been plotted in Figs. \ref{fig3dXDE1} $(e)$ and \ref{fig3dXDE1} $(f)$, respectively. This GTMD is observed to possess negative values for low values of the longitudinal momentum fraction $x$ and positive values for high longitudinal momentum fraction $x$, which can be explained due to the multipolar behavior of this GTMD. In Figs. \ref{fig3dXDE1} $(g)$ and \ref{fig3dXDE1} $(h)$, GTMD $xE_{2,4}$ has been plotted for the active quarks $u$ and $d$ accordingly. As expected from the Eqs. \eqref{e24s} and \eqref{e24a}, the GTMD $xE_{2,4}$ shows similar distributions for the active quarks $u$ and $d$ varying only in magnitude, this property is in line with the previously observed GTMD $xE_{2,1}$.
\par 
Now, we discuss the GTMDs $xE_{2,6},~xE_{2,7},$ and $xE_{2,8}$ corresponding with the Dirac matrix structure $\Gamma=\gamma_5$. In Figs. \ref{fig3dXDE2} $(a)$ and \ref{fig3dXDE2} $(b)$, GTMD $xE_{2,6}$ has been plotted for both quark flavors, respectively. As the expression of this GTMD is precisely the same as that of GTMD $xE_{2,2}$, but it has opposite polarity, therefore it exhibits similar behaviors to GTMD $xE_{2,2}$. GTMD $xE_{2,7}$ has been plotted in Figs. \ref{fig3dXDE2} $(c)$ and \ref{fig3dXDE2} $(d)$ for $u$ and $d$ quarks, respectively. Although this GTMD does not exhibit the exact quark flavor symmetry, the behavior of the distribution for active $u$ quark is very similar to that of the active $d$ quark but with opposite polarity. In Figs. \ref{fig3dXDE2} $(e)$ and \ref{fig3dXDE2} $(f)$, GTMD $xE_{2,8}$ has been plotted for active quarks $u$ and $d$, sequentially. The maximum amplitude for each quark flavor is found to be around $0.3$, suggesting the equal allocation of longitudinal momentum fraction $x$ for all three valence quarks, and this GTMD is found to be positive for all ranges of longitudinal momentum fraction $x$ and transverse momentum ${p_\perp}$.
\par 
Now, we move towards the discussion of GTMDs corresponding with the Dirac matrix structure $\Gamma=\gamma^j$. The GTMDs $xF_{2,1}, xF_{2,3}, xF_{2,4}$, and $xF_{2,7}$ are plotted in Fig. (\ref{fig3dXDF1}) for the quarks $u$ and $d$ accordingly. In Figs. \ref{fig3dXDF1} $(a)$ and \ref{fig3dXDF1} $(b)$, GTMD $xF_{2,1}$ has been plotted for the quarks $u$ and $d$ accordingly. As evident from the plots, this GTMD is found to be flavor symmetric. $3$-dimensional plots of GTMD $xF_{2,3}$ and $xF_{2,4}$ are given in Figs. \ref{fig3dXDF1} $(e)$ to \ref{fig3dXDF1} $(h)$ for the quark flavor $u$ and $d$ accordingly. Since GTMD $xF_{2,5}$ is a part of expressions for both GTMDs $xF_{2,3}$ and $xF_{2,4}$, the GTMD $xF_{2,3}$ is found to have no quark flavor symmetry while GTMD $xF_{2,4}$ is positive over the entire range for both quark flavors. GTMD $xF_{2,7}$ has been plotted in Figs. \ref{fig3dXDF1} $(e)$ and \ref{fig3dXDF1} $(f)$ for $u$ and $d$ quark respectively. Since the expressions for the scalar and vector parts of GTMD $xF_{2,7}$ are similar to those of GTMD $xF_{2,2}$, the behavior of the distribution remains the same.
\par 
Here, we discuss the GTMDs $xG_{2,1},~xG_{2,3},~xG_{2,4}$ and $xG_{2,7}$ corresponding with the Dirac matrix structure $\Gamma=\gamma^j \gamma_5$, plotted in Figs. (\ref{fig3dXDG1}) and (\ref{fig3dXDG2}) for $u$ and $d$ quark accordingly. The only difference between the GTMDs $xF_{2,2}$ and $xF_{2,7}$ and the GTMD $xG_{2,1}$ expressions is in the coefficients. In addition, it is obvious that the $u$ and $d$ active quark Figs. \ref{fig3dXDG2} $(a)$ and \ref{fig3dXDG2} $(b)$ are comparable to those of $xF_{2,2}$ and $xF_{2,7}$, which differ only in magnitude. GTMDs $xG_{2,3}$ and $xG_{2,4}$ have been plotted in Fig. \ref{fig3dXDG2} $(e)$ to \ref{fig3dXDG2}$(h)$ for the active quarks $u$ and $d$ accordingly. Considering that the expression of GTMD $xG_{2,3}$ is constructed of several other GTMDs, plots of GTMD $xG_{2,3}$ for $u$ and $d$ quark are found to be significantly distinct since the scalar and vector parts of different expressions incorporate differently for constructing the resultant GTMD. Even though the GTMD $xG_{2,4}$ expression is comparatively simpler, it results in a bit of an anti-symmetric plot for the $u$ and $d$ quark. Plots of the GTMD $xG_{2,7}$ are given in Figs. \ref{fig3dXDG2} $(e)$ and \ref{fig3dXDG2} $(f)$ for $u$ and $d$ quark accordingly, and the specified GTMD shows significant quark flavor symmetry.
\par 
Now, we discuss the GTMDs $xH_{2,1},~xH_{2,2},~xH_{2,6}$, and $xH_{2,8}$ corresponding with the Dirac matrix structure $\Gamma=i\sigma^{ij} \gamma_5$ and $\Gamma=i\sigma^{+-} \gamma_5$, which have been plotted in Figs. \ref{fig3dXDH1} and \ref{fig3dXDH2} for $u$ and $d$ quark accordingly. When the quark flavor is changed, GTMD $xH_{2,1}$ is observed to keep its polarity unchanged, whereas GTMD $xH_{2,2}$ exhibits the opposite polarity for active $u$ or $d$ quarks. In Figs. \ref{fig3dXPH2} $(c)$ and \ref{fig3dXPH2} $(d)$, the GTMD $xH_{2,6}$ has been plotted for each quark flavors $u$ and $d$, this GTMD is found to be anti-symmetric for the   quark flavors $u$ and $d$. In Figs. \ref{fig3dXPH2} $(g)$ and \ref{fig3dXPH2} $(h)$, the GTMD $xH_{2,8}$ has been plotted for the quarks $u$ and $d$ sequentially, and is found to be positive for the complete range of the longitudinal momentum fraction $x$.
%
\subsubsection{Variation with ${ p_\perp}$ and ${ \Delta_\perp}$}\label{sspd}
\begin{figure*}
		\centering
		\begin{minipage}[c]{0.98\textwidth}
			(a)\includegraphics[width=7.3cm]{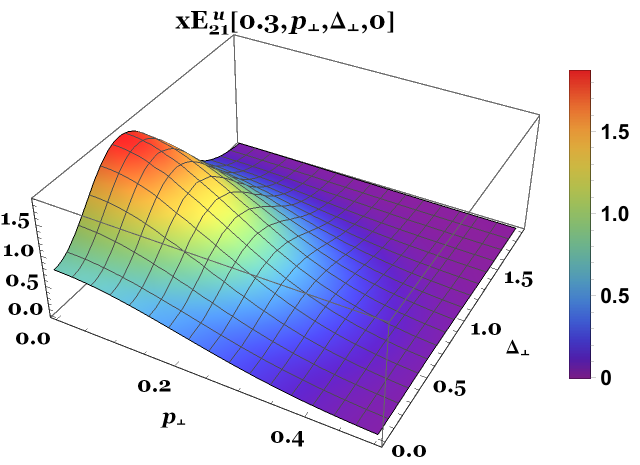}
			\hspace{0.05cm}
			(b)\includegraphics[width=7.3cm]{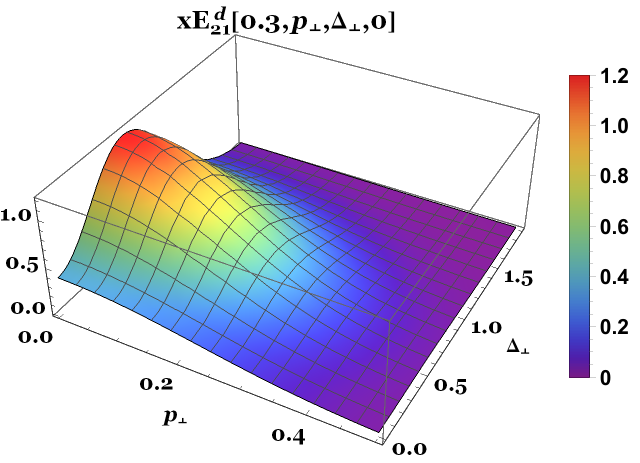}
			\hspace{0.05cm}
			(c)\includegraphics[width=7.3cm]{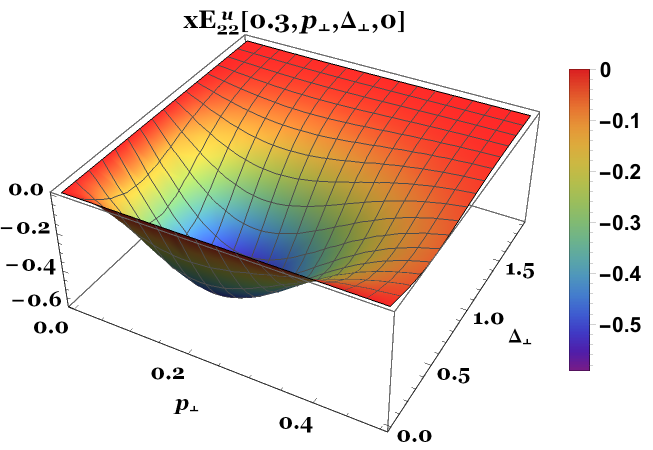}
			\hspace{0.05cm}
			(d)\includegraphics[width=7.3cm]{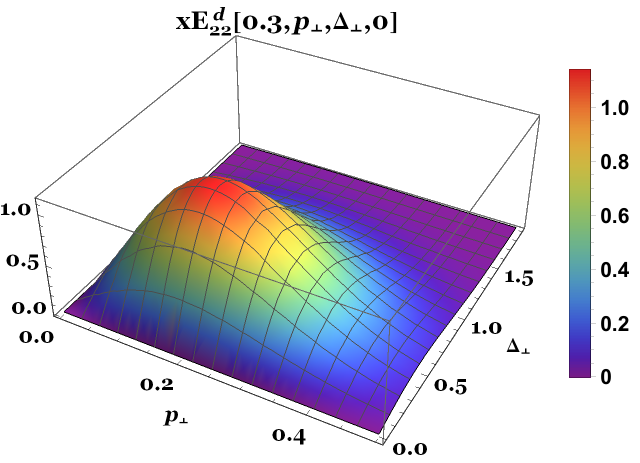}
			\hspace{0.05cm}
			(e)\includegraphics[width=7.3cm]{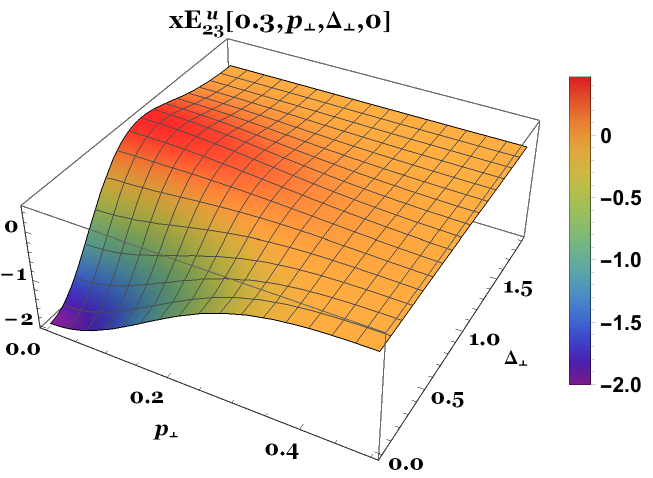}
			\hspace{0.05cm}
			(f)\includegraphics[width=7.3cm]{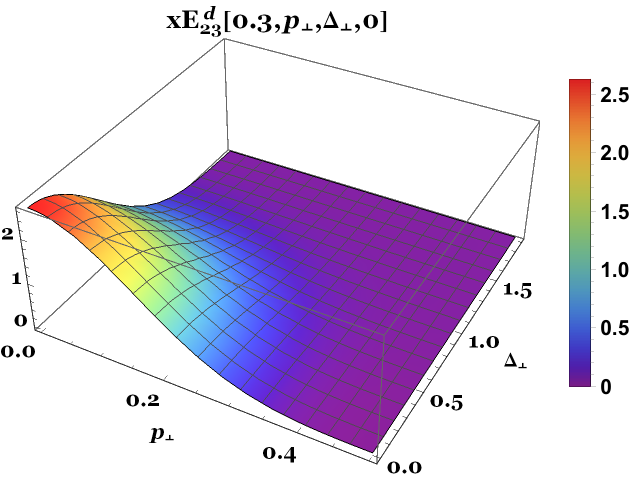}
			\hspace{0.05cm}
			(g)\includegraphics[width=7.3cm]{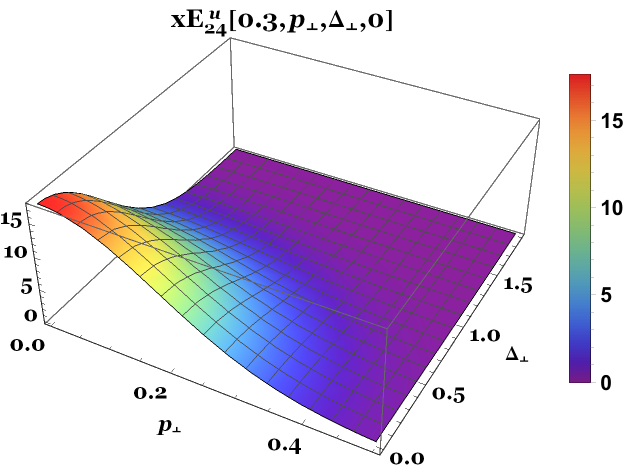}
			\hspace{0.05cm}
			(h)\includegraphics[width=7.3cm]{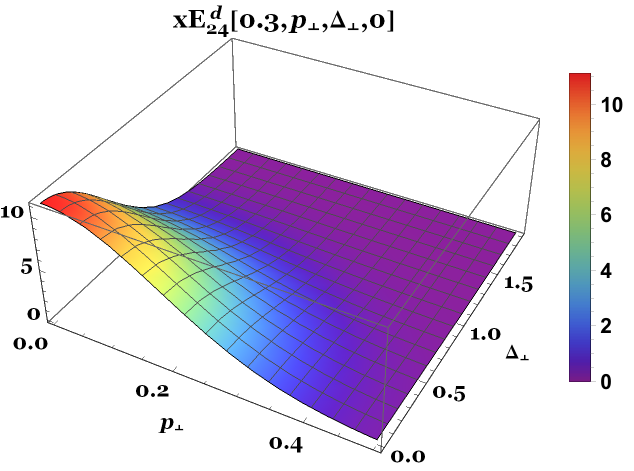}
			\hspace{0.05cm}\\
		\end{minipage}
		\caption{\label{fig3dPDE1} The sub-leading twist GTMDs 		
			$x E_{2,1}^{\nu}(x, p_{\perp},\Delta_{\perp},\theta)$,
			$x E_{2,2}^{\nu}(x, p_{\perp},\Delta_{\perp},\theta)$,
			$x E_{2,3}^{\nu}(x, p_{\perp},\Delta_{\perp},\theta)$, and
			$x E_{2,4}^{\nu}(x, p_{\perp},\Delta_{\perp},\theta)$
	are	plotted about ${ p_\perp}$ and ${{ \Delta_\perp}}$ for $x= 0.3$ keeping ${\bfp} \parallel {\Dp}$. In sequential order, $u$ and $d$ quarks are in the left and right columns.
		}
	\end{figure*}

\begin{figure*}
		\centering
		\begin{minipage}[c]{0.98\textwidth}
			(a)\includegraphics[width=7.3cm]{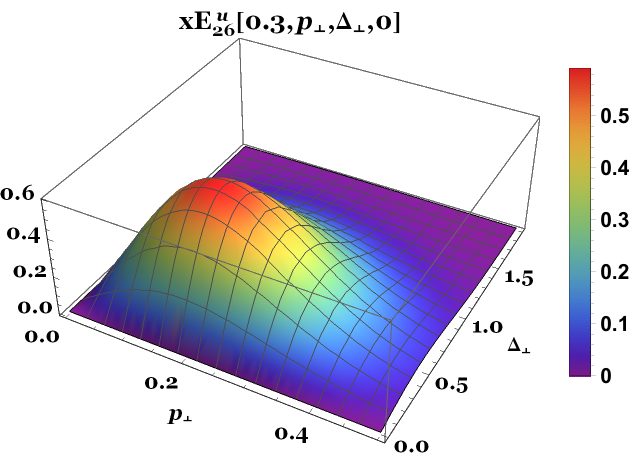}
			\hspace{0.05cm}
			(b)\includegraphics[width=7.3cm]{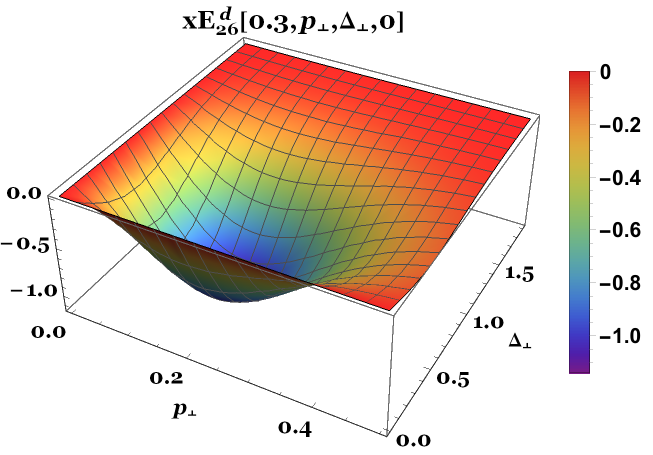}
			\hspace{0.05cm}
			(c)\includegraphics[width=7.3cm]{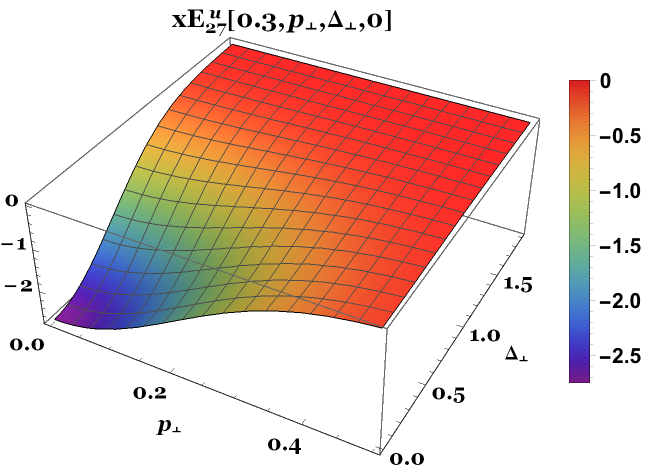}
			\hspace{0.05cm}
			(d)\includegraphics[width=7.3cm]{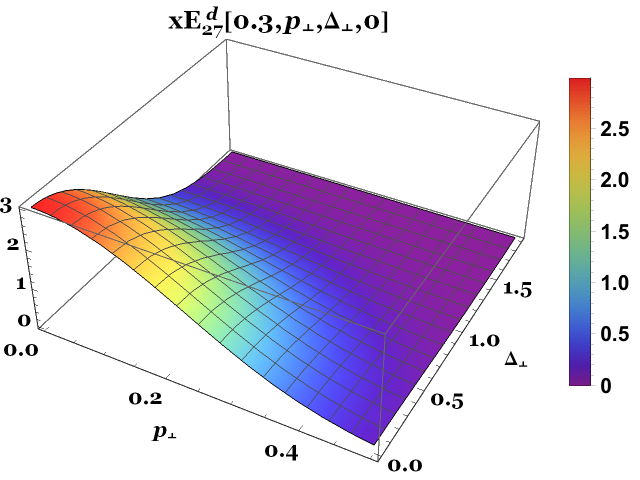}
			\hspace{0.05cm}
			(e)\includegraphics[width=7.3cm]{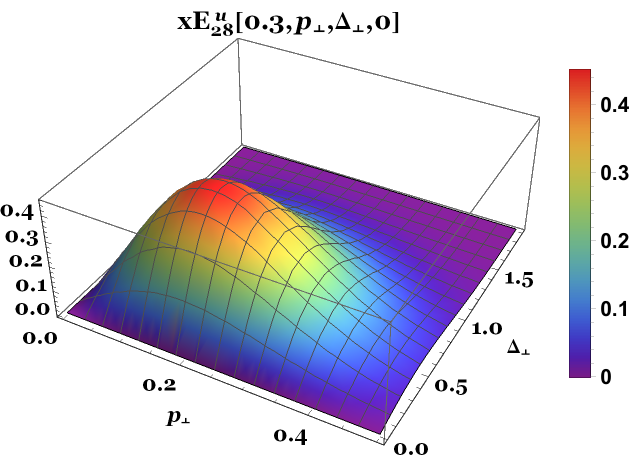}
			\hspace{0.05cm}
			(f)\includegraphics[width=7.3cm]{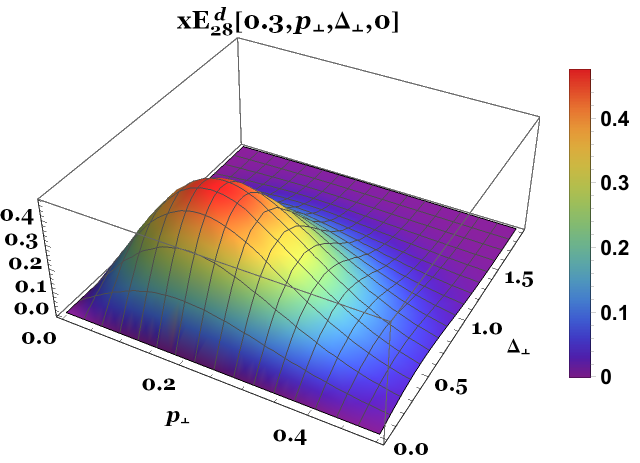}
			\hspace{0.05cm}
			\\
		\end{minipage}
		\caption{\label{fig3dPDE2} The sub-leading twist GTMDs 		
			$x E_{2,6}^{\nu}(x, p_{\perp},\Delta_{\perp},\theta)$,
			$x E_{2,7}^{\nu}(x, p_{\perp},\Delta_{\perp},\theta)$,
			and
			$x E_{2,8}^{\nu}(x, p_{\perp},\Delta_{\perp},\theta)$
			are	plotted about ${ p_\perp}$ and ${{ \Delta_\perp}}$ for $x= 0.3$ keeping ${\bfp} \parallel {\Dp}$. In sequential order, $u$ and $d$ quarks are in the left and right columns.
		}
	\end{figure*}

\begin{figure*}
		\centering
		\begin{minipage}[c]{0.98\textwidth}
			(a)\includegraphics[width=7.3cm]{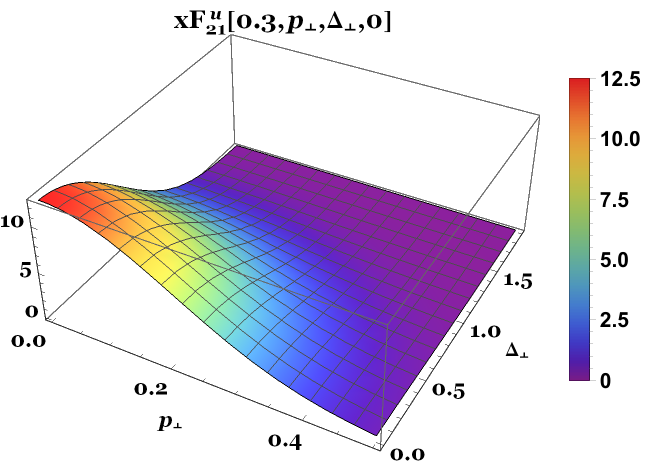}
			\hspace{0.05cm}
			(b)\includegraphics[width=7.3cm]{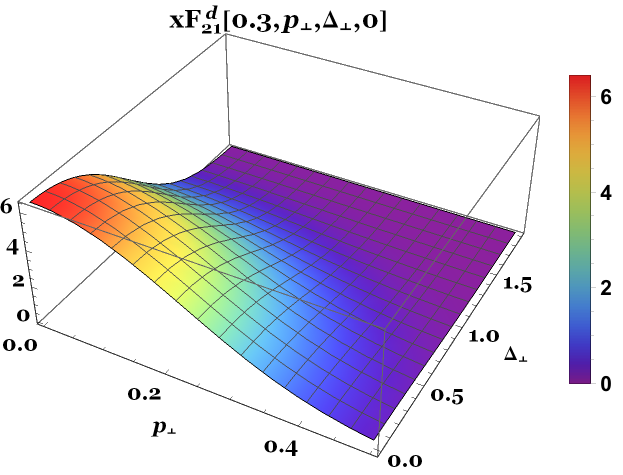}
			\hspace{0.05cm}
			(c)\includegraphics[width=7.3cm]{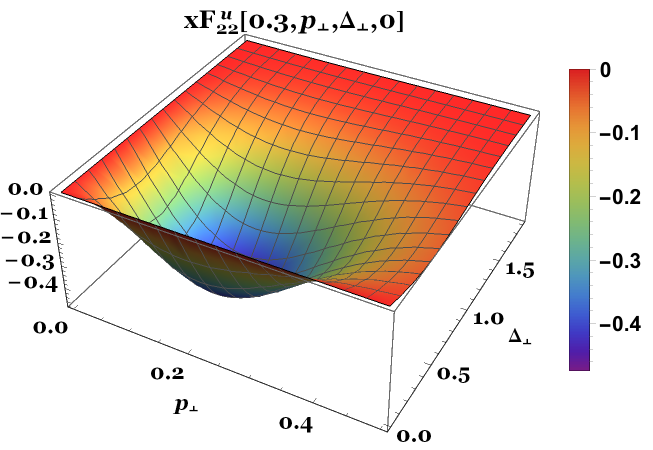}
			\hspace{0.05cm}
			(d)\includegraphics[width=7.3cm]{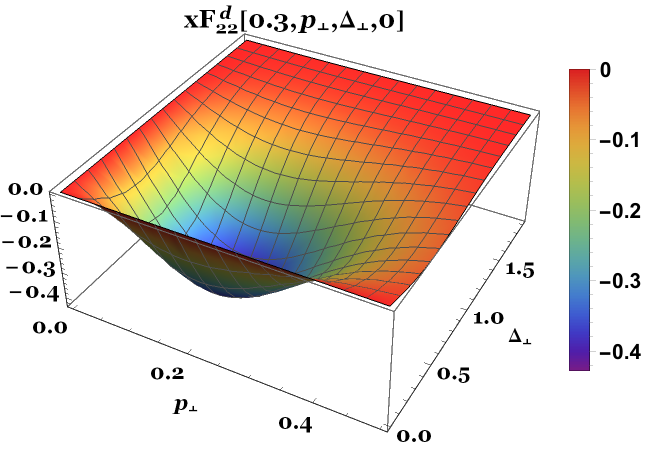}
			\hspace{0.05cm}
			(e)\includegraphics[width=7.3cm]{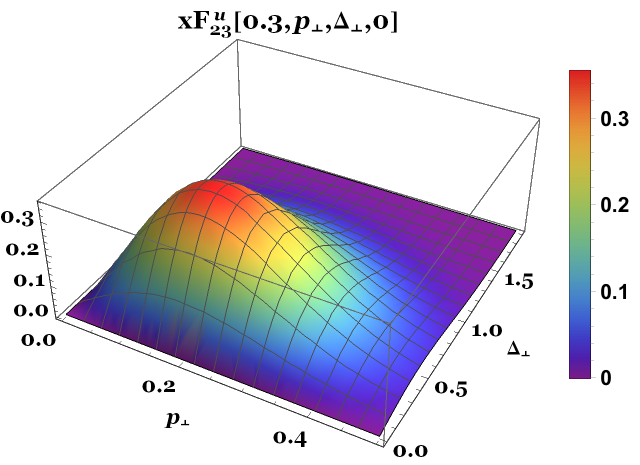}
			\hspace{0.05cm}
			(f)\includegraphics[width=7.3cm]{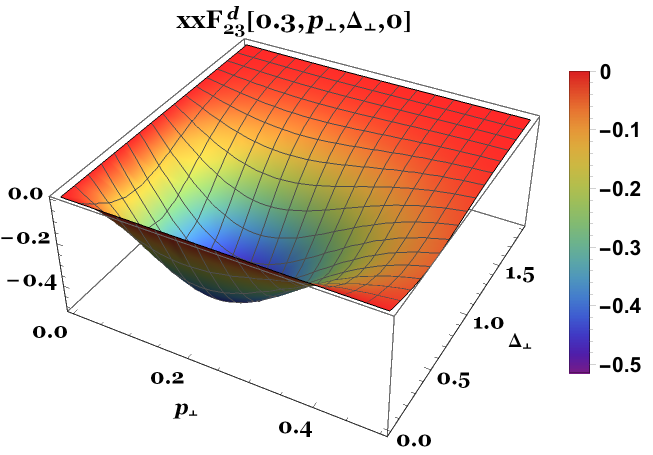}
			\hspace{0.05cm}
			(g)\includegraphics[width=7.3cm]{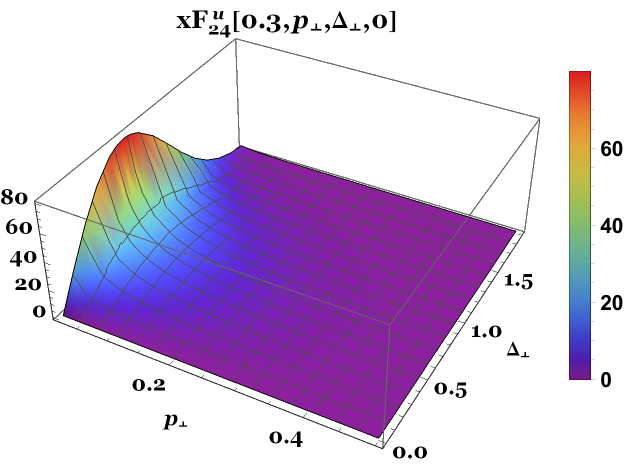}
			\hspace{0.05cm}
			(h)\includegraphics[width=7.3cm]{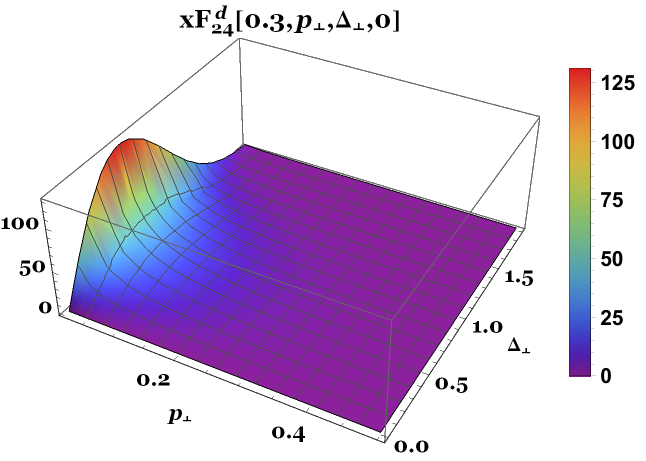}
			\hspace{0.05cm}\\
		\end{minipage}
		\caption{\label{fig3dPDF1} The sub-leading twist GTMDs 		
			$x F_{2,1}^{\nu}(x, p_{\perp},\Delta_{\perp},\theta)$,
			$x F_{2,2}^{\nu}(x, p_{\perp},\Delta_{\perp},\theta)$,
			$x F_{2,3}^{\nu}(x, p_{\perp},\Delta_{\perp},\theta)$, and
			$x F_{2,4}^{\nu}(x, p_{\perp},\Delta_{\perp},\theta)$
			are	plotted about ${ p_\perp}$ and ${{ \Delta_\perp}}$ for $x= 0.3$ keeping ${\bfp} \parallel {\Dp}$. In sequential order, $u$ and $d$ quarks are in the left and right columns.
		}
	\end{figure*}

\begin{figure*}
		\centering
		\begin{minipage}[c]{0.98\textwidth}
				(a)\includegraphics[width=7.3cm]{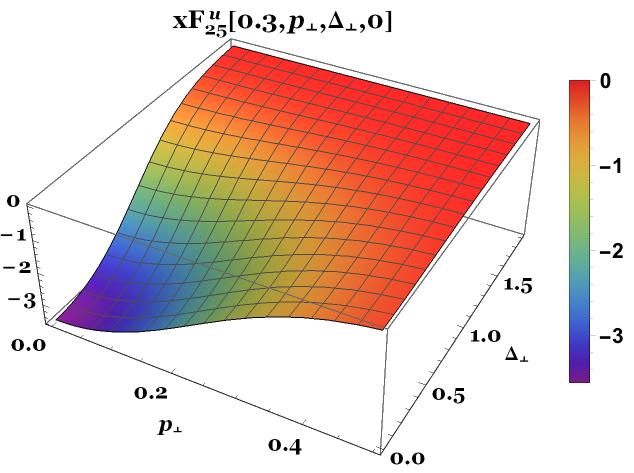}
			\hspace{0.05cm}
				(b)\includegraphics[width=7.3cm]{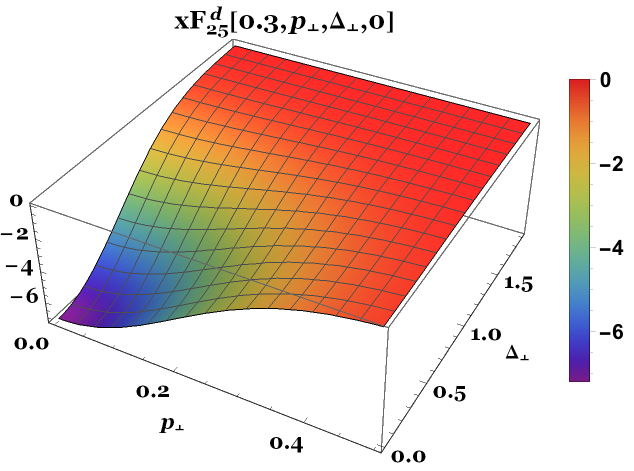}
			\hspace{0.05cm}
			(c)\includegraphics[width=7.3cm]{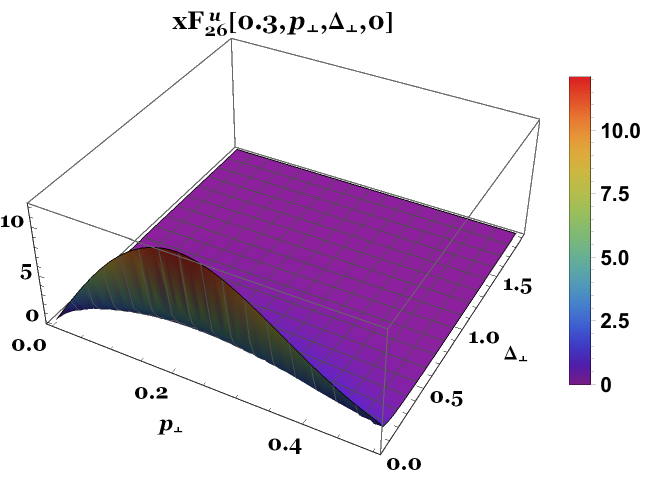}
			\hspace{0.05cm}
			(d)\includegraphics[width=7.3cm]{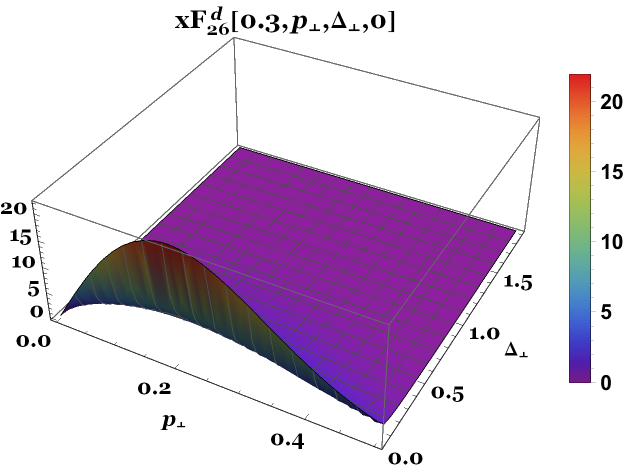}
			\hspace{0.05cm}
			(e)\includegraphics[width=7.3cm]{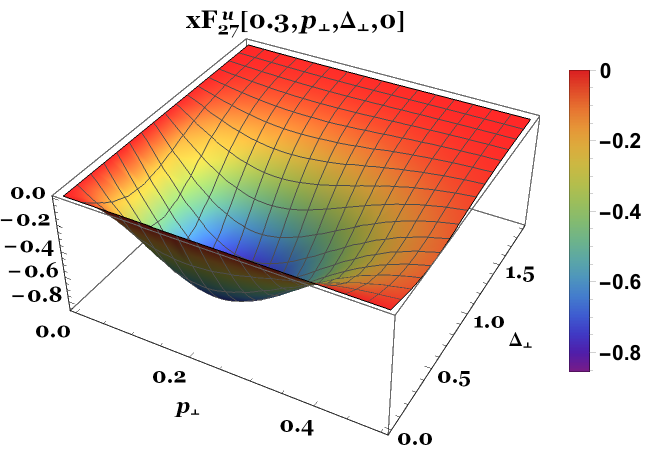}
			\hspace{0.05cm}
			(f)\includegraphics[width=7.3cm]{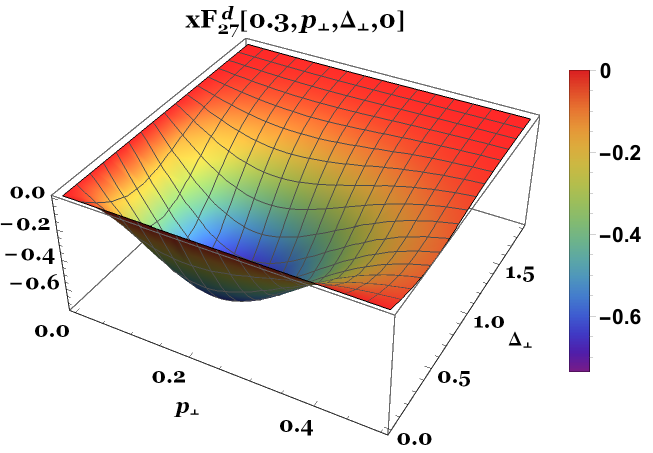}
			\hspace{0.05cm}
			(g)\includegraphics[width=7.3cm]{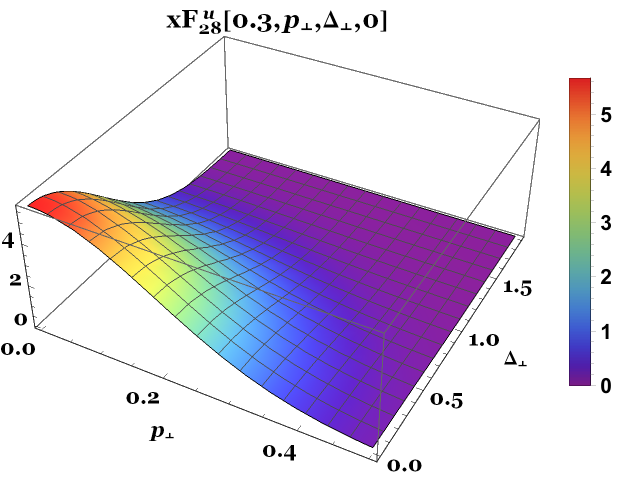}
			\hspace{0.05cm}
			(h)\includegraphics[width=7.3cm]{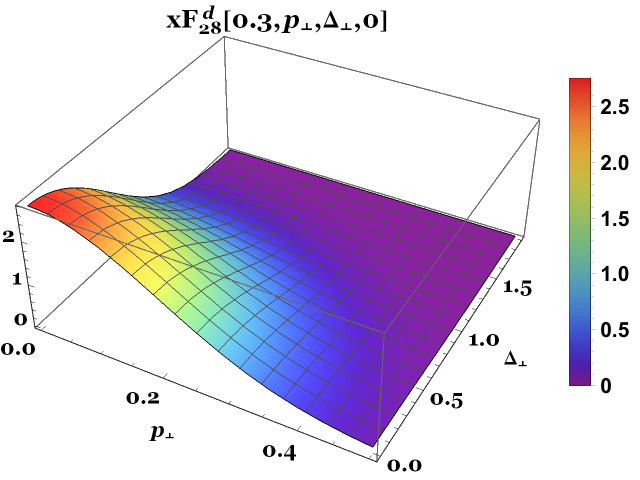}
			\hspace{0.05cm}\\
		\end{minipage}
		\caption{\label{fig3dPDF2} The sub-leading twist GTMDs 		
			$x F_{2,5}^{\nu}(x, p_{\perp},\Delta_{\perp},\theta)$,
			$x F_{2,6}^{\nu}(x, p_{\perp},\Delta_{\perp},\theta)$,
			$x F_{2,7}^{\nu}(x, p_{\perp},\Delta_{\perp},\theta)$, and
			$x F_{2,8}^{\nu}(x, p_{\perp},\Delta_{\perp},\theta)$
			are	plotted about ${ p_\perp}$ and ${{ \Delta_\perp}}$ for $x= 0.3$ keeping ${\bfp} \parallel {\Dp}$. In sequential order, $u$ and $d$ quarks are in the left and right columns.
		}
	\end{figure*}

\begin{figure*}
		\centering
		\begin{minipage}[c]{0.98\textwidth}
			(a)\includegraphics[width=7.3cm]{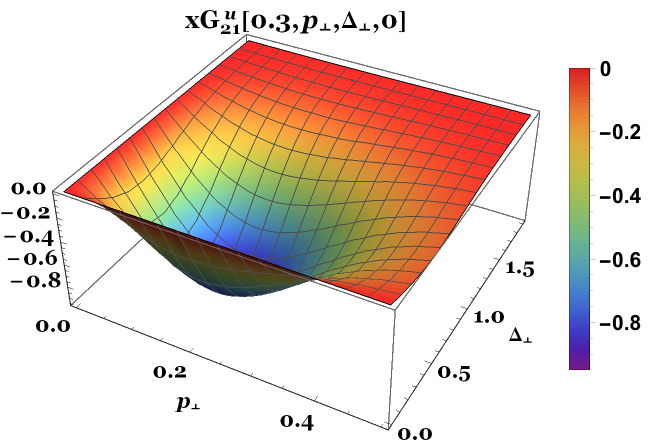}
			\hspace{0.05cm}
			(b)\includegraphics[width=7.3cm]{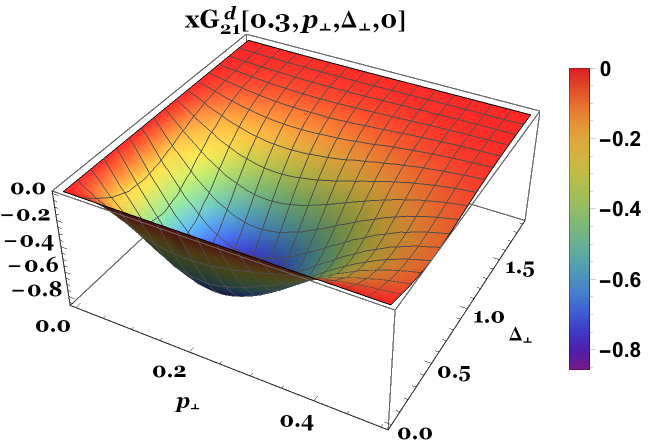}
			\hspace{0.05cm}
			(c)\includegraphics[width=7.3cm]{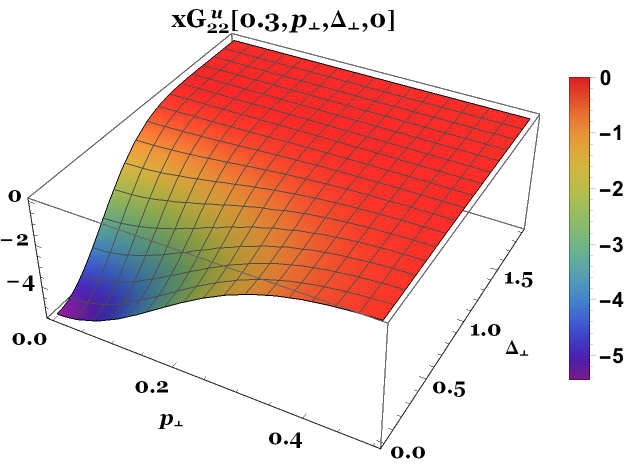}
			\hspace{0.05cm}
			(d)\includegraphics[width=7.3cm]{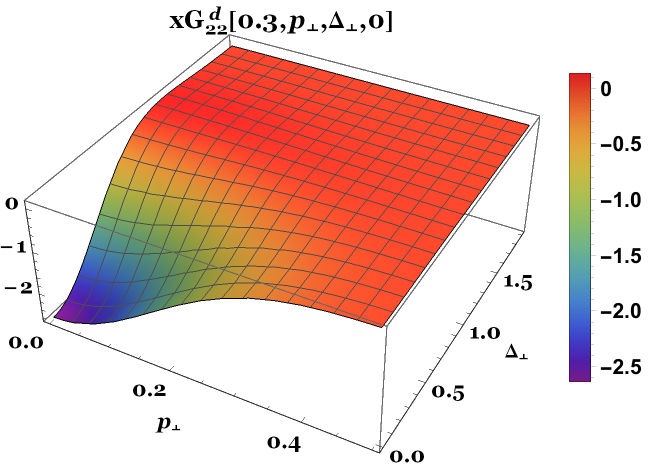}
			\hspace{0.05cm}
			(e)\includegraphics[width=7.3cm]{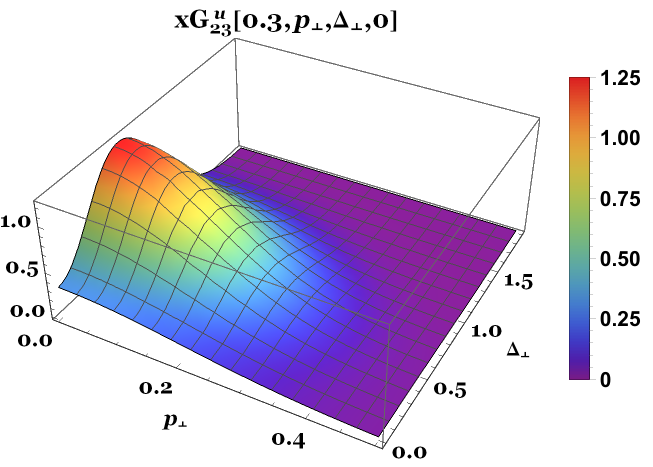}
			\hspace{0.05cm}
			(f)\includegraphics[width=7.3cm]{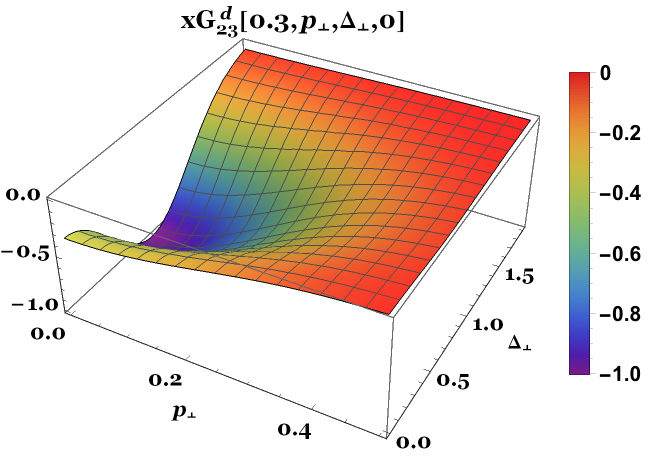}
			\hspace{0.05cm}
			(g)\includegraphics[width=7.3cm]{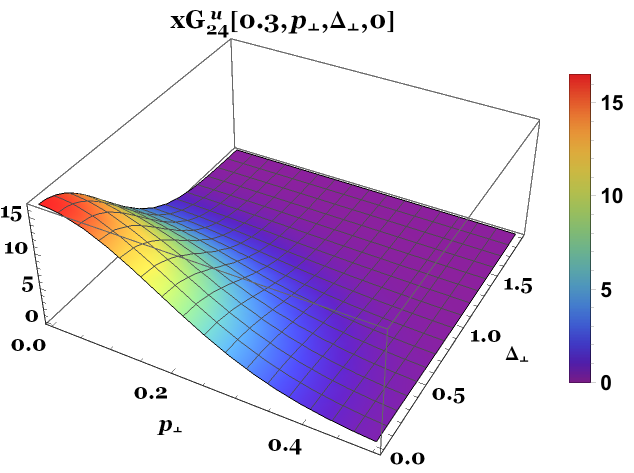}
			\hspace{0.05cm}
			(h)\includegraphics[width=7.3cm]{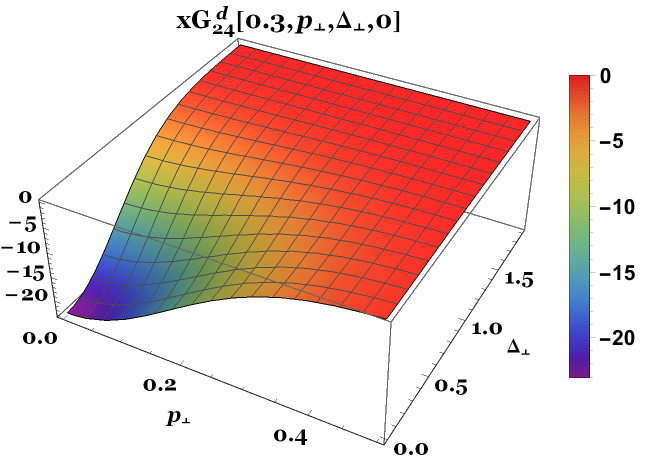}
			\hspace{0.05cm}\\
		\end{minipage}
		\caption{\label{fig3dPDG1} The sub-leading twist GTMDs 		
			$x G_{2,1}^{\nu}(x, p_{\perp},\Delta_{\perp},\theta)$,
			$x G_{2,2}^{\nu}(x, p_{\perp},\Delta_{\perp},\theta)$,
			$x G_{2,3}^{\nu}(x, p_{\perp},\Delta_{\perp},\theta)$, and
			$x G_{2,4}^{\nu}(x, p_{\perp},\Delta_{\perp},\theta)$
			are	plotted about ${ p_\perp}$ and ${{ \Delta_\perp}}$ for $x= 0.3$ keeping ${\bfp} \parallel {\Dp}$. In sequential order, $u$ and $d$ quarks are in the left and right columns.
		}
	\end{figure*}

\begin{figure*}
		\centering
		\begin{minipage}[c]{0.98\textwidth}
			(a)\includegraphics[width=7.3cm]{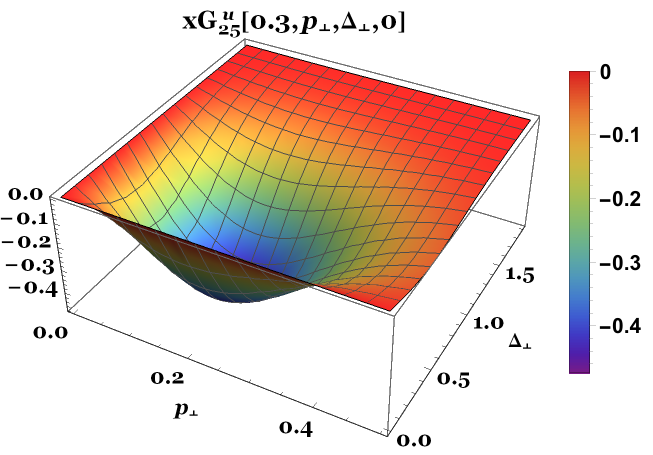}
			\hspace{0.05cm}
			(b)\includegraphics[width=7.3cm]{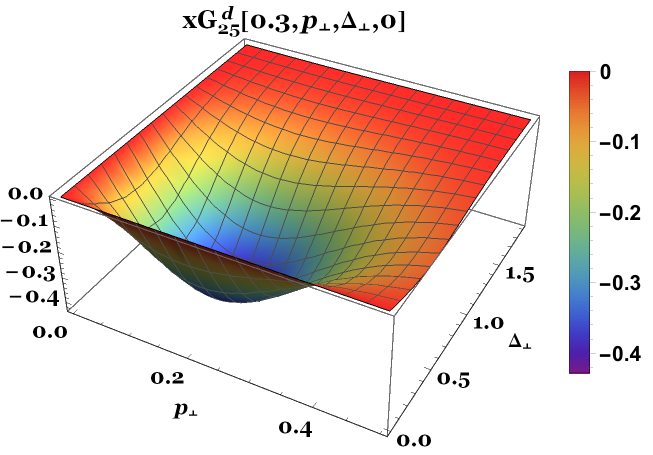}
			\hspace{0.05cm}
			(c)\includegraphics[width=7.3cm]{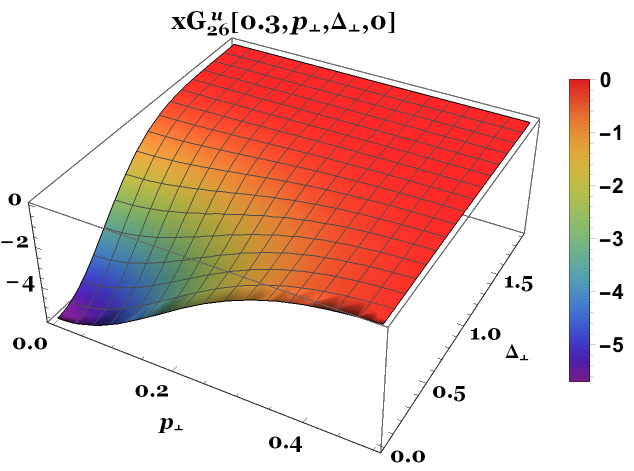}
			\hspace{0.05cm}
			(d)\includegraphics[width=7.3cm]{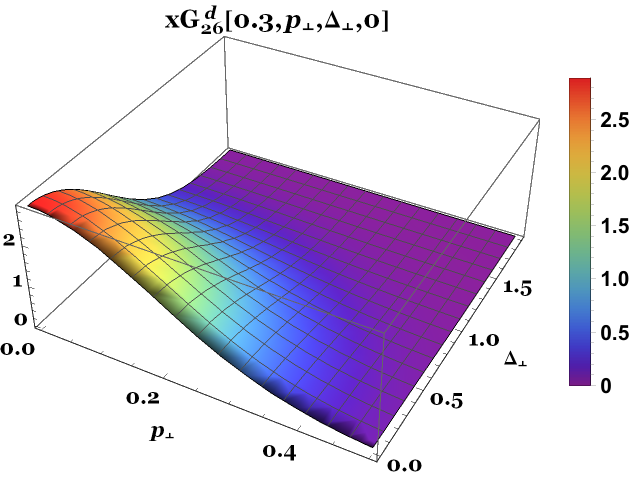}
			\hspace{0.05cm}
			(e)\includegraphics[width=7.3cm]{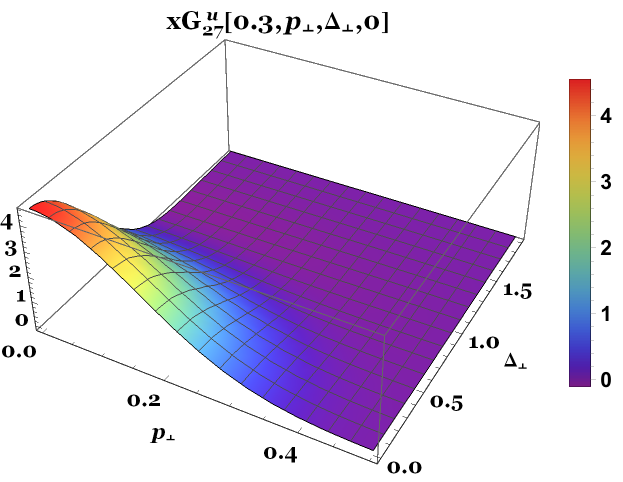}
			\hspace{0.05cm}
			(f)\includegraphics[width=7.3cm]{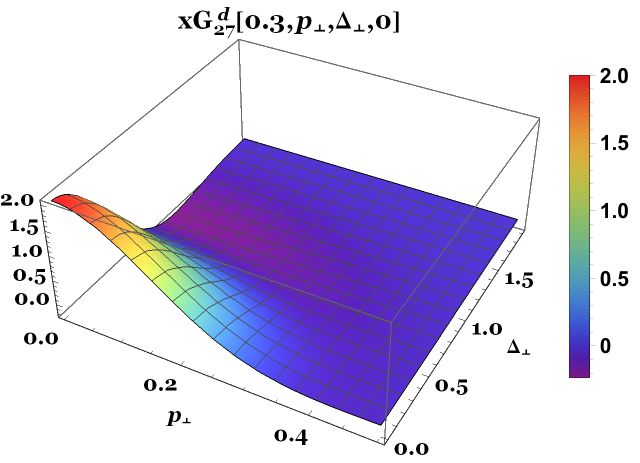}
			\hspace{0.05cm}
			(g)\includegraphics[width=7.3cm]{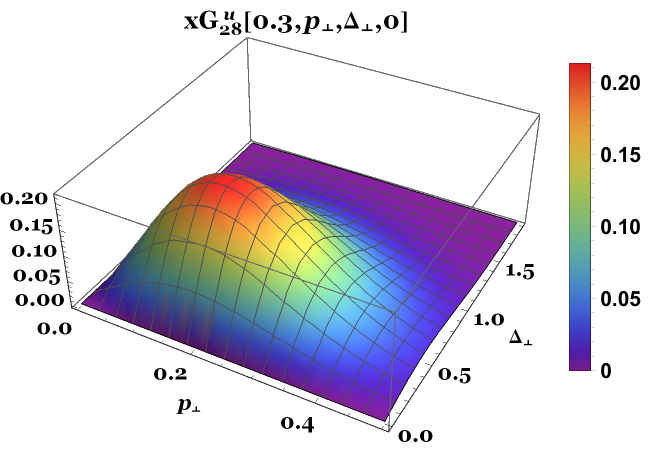}
			\hspace{0.05cm}
			(h)\includegraphics[width=7.3cm]{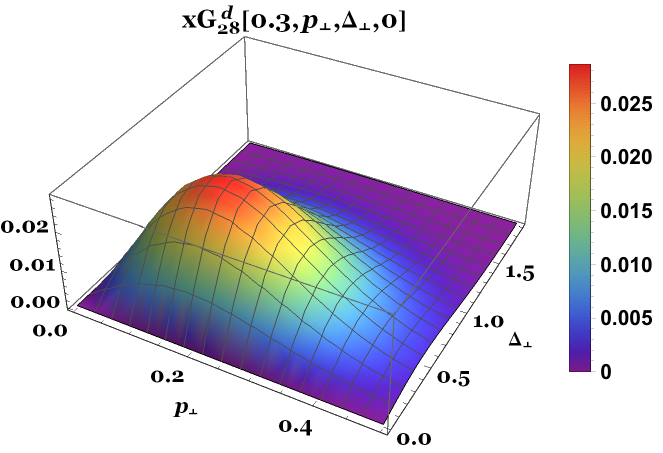}
			\hspace{0.05cm}\\
		\end{minipage}
		\caption{\label{fig3dPDG2} The sub-leading twist GTMDs 		
			$x G_{2,5}^{\nu}(x, p_{\perp},\Delta_{\perp},\theta)$,
			$x G_{2,6}^{\nu}(x, p_{\perp},\Delta_{\perp},\theta)$,
			$x G_{2,7}^{\nu}(x, p_{\perp},\Delta_{\perp},\theta)$, and
			$x G_{2,8}^{\nu}(x, p_{\perp},\Delta_{\perp},\theta)$
			are	plotted about ${ p_\perp}$ and ${{ \Delta_\perp}}$ for $x= 0.3$ keeping ${\bfp} \parallel {\Dp}$. In sequential order, $u$ and $d$ quarks are in the left and right columns.
		}
	\end{figure*}

\begin{figure*}
		\centering
		\begin{minipage}[c]{0.98\textwidth}
			(a)\includegraphics[width=7.3cm]{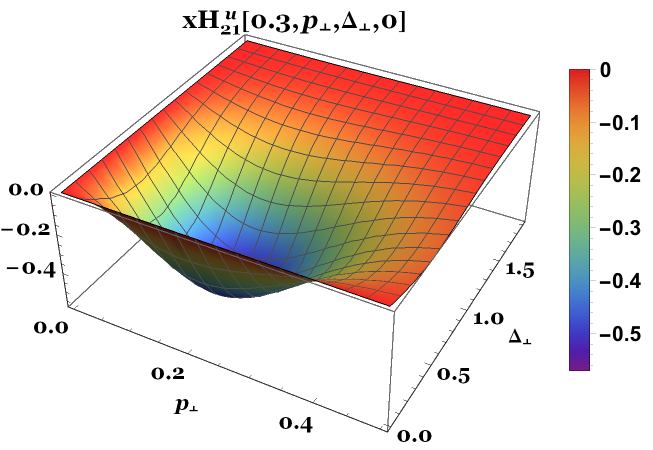}
			\hspace{0.05cm}
			(b)\includegraphics[width=7.3cm]{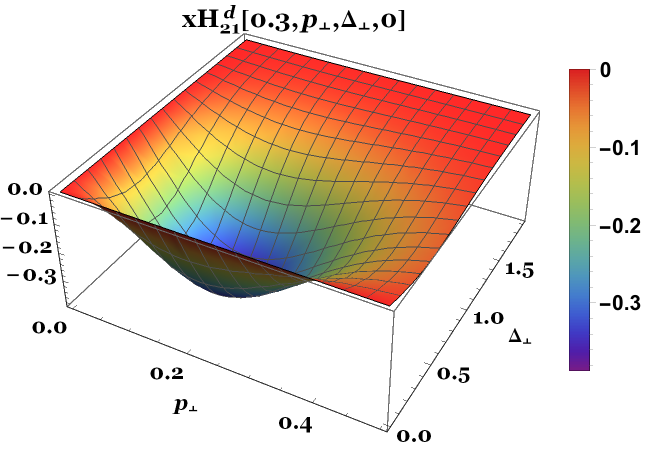}
			\hspace{0.05cm}
			(c)\includegraphics[width=7.3cm]{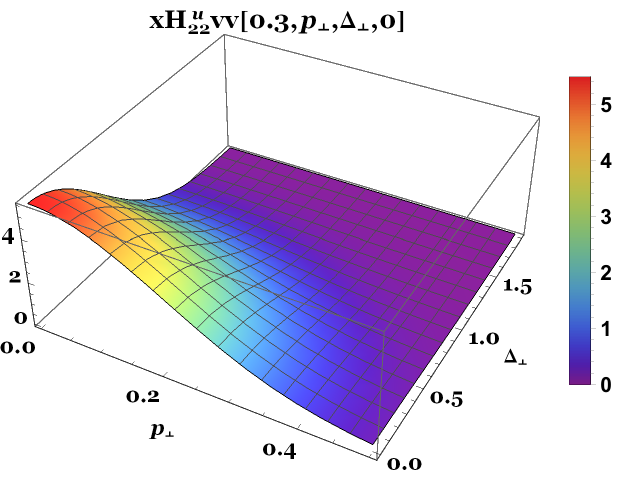}
			\hspace{0.05cm}
			(d)\includegraphics[width=7.3cm]{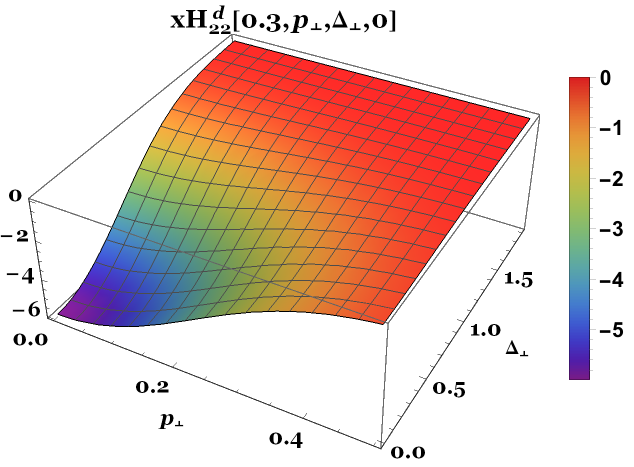}
			\hspace{0.05cm}
			(c)\includegraphics[width=7.3cm]{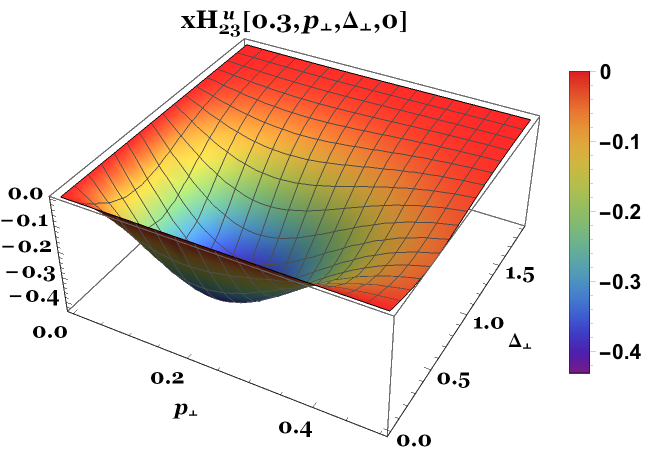}
			\hspace{0.05cm}
			(d)\includegraphics[width=7.3cm]{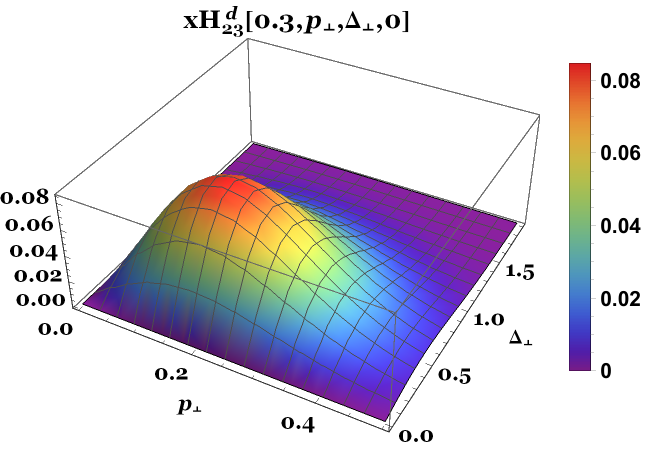}
			\hspace{0.05cm}
			\\
		\end{minipage}
		\caption{\label{fig3dPDH1} The sub-leading twist GTMDs 		
			$x H_{2,1}^{\nu}(x, p_{\perp},\Delta_{\perp},\theta)$,
			$x H_{2,2}^{\nu}(x, p_{\perp},\Delta_{\perp},\theta)$,
			and
			$x H_{2,3}^{\nu}(x, p_{\perp},\Delta_{\perp},\theta)$
				are	plotted about ${ p_\perp}$ and ${{ \Delta_\perp}}$ for $x= 0.3$ keeping ${\bfp} \parallel {\Dp}$. In sequential order, $u$ and $d$ quarks are in the left and right columns.
		}
	\end{figure*}

\begin{figure*}
		\centering
		\begin{minipage}[c]{0.98\textwidth}
			(a)\includegraphics[width=7.3cm]{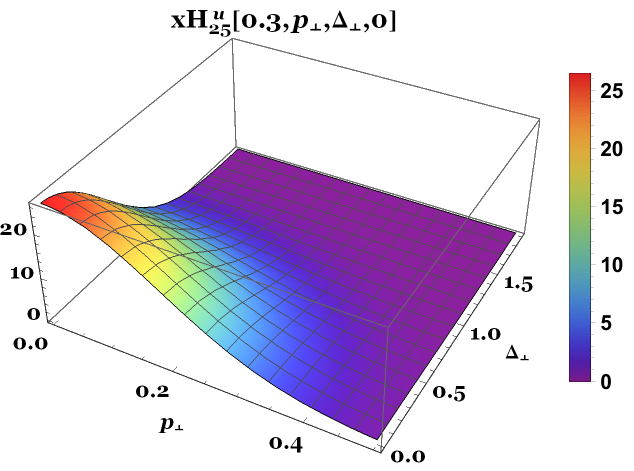}
			\hspace{0.05cm}
			(b)\includegraphics[width=7.3cm]{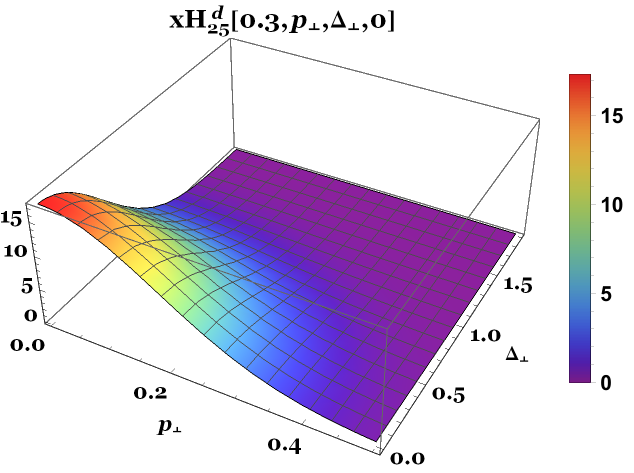}
			\hspace{0.05cm}
			(c)\includegraphics[width=7.3cm]{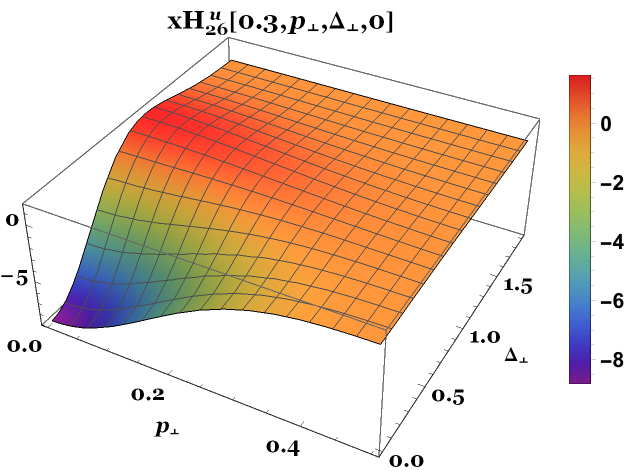}
			\hspace{0.05cm}
			(d)\includegraphics[width=7.3cm]{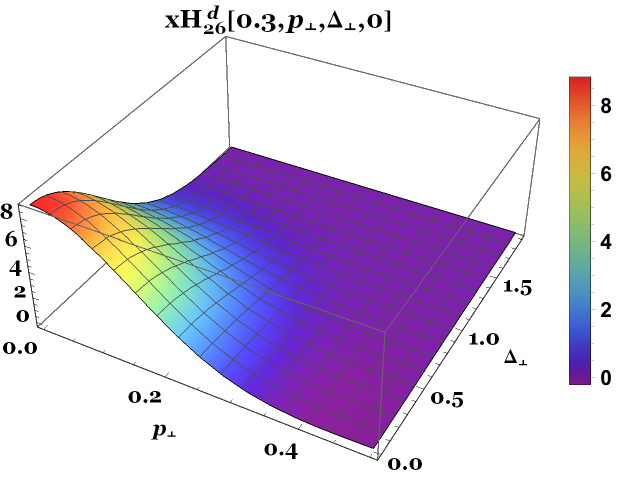}
			\hspace{0.05cm}
			(e)\includegraphics[width=7.3cm]{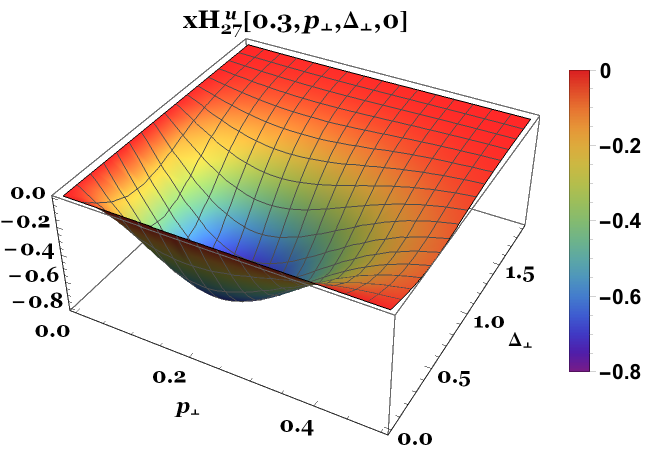}
			\hspace{0.05cm}
			(f)\includegraphics[width=7.3cm]{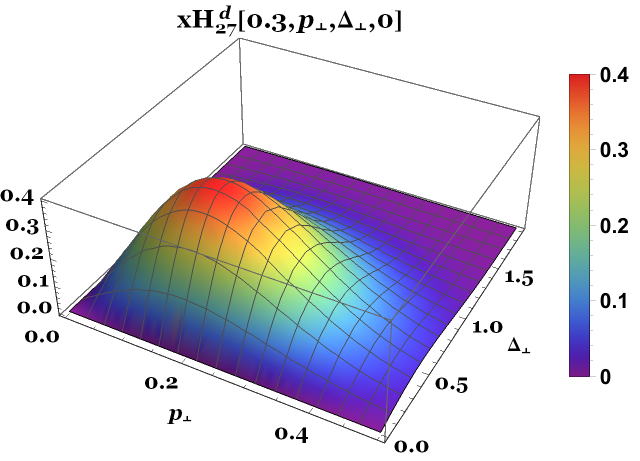}
			\hspace{0.05cm}
			(g)\includegraphics[width=7.3cm]{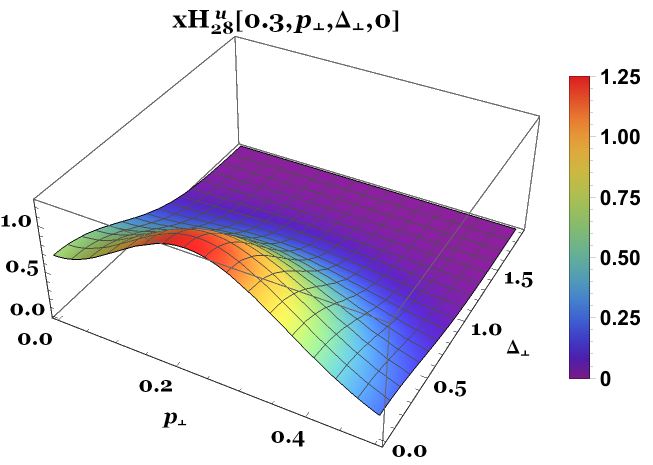}
			\hspace{0.05cm}
			(h)\includegraphics[width=7.3cm]{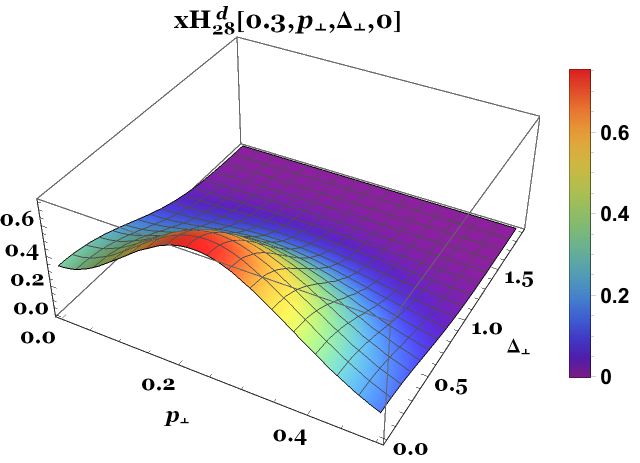}
			\hspace{0.05cm}\\
		\end{minipage}
		\caption{\label{fig3dPDH2} The sub-leading twist GTMDs 		
			$x H_{2,5}^{\nu}(x, p_{\perp},\Delta_{\perp},\theta)$,
			$x H_{2,6}^{\nu}(x, p_{\perp},\Delta_{\perp},\theta)$,
			$x H_{2,7}^{\nu}(x, p_{\perp},\Delta_{\perp},\theta)$, and
			$x H_{2,8}^{\nu}(x, p_{\perp},\Delta_{\perp},\theta)$
			are	plotted about ${ p_\perp}$ and ${{ \Delta_\perp}}$ for $x= 0.3$ keeping ${\bfp} \parallel {\Dp}$. In sequential order, $u$ and $d$ quarks are in the left and right columns.
		}
	\end{figure*}

To analyze sub-leading twist GTMDs with change in the transverse momentum of quark $\bfp$ and transverse momentum transfer between the beginning and the end state of quark $\Dp$ at the same time, we plot their $3$-dimensional deviations for $x= 0.3$ while keeping ${\bfp} \parallel {\Dp}$ in Figs. (\ref{fig3dPDE1}) to (\ref{fig3dPDH2}), following Ref. \cite{Sharma:2023tre}. In the next section, we will discuss these results in detail.
	%
%
	%
	%
		
	%
	\subsection{Transverse momentum dependent form factors (TMFFs)}\label{sstmff}
	
		
		\begin{figure*}
				\centering
				\begin{minipage}[c]{0.98\textwidth}
						(a)\includegraphics[width=7.3cm]{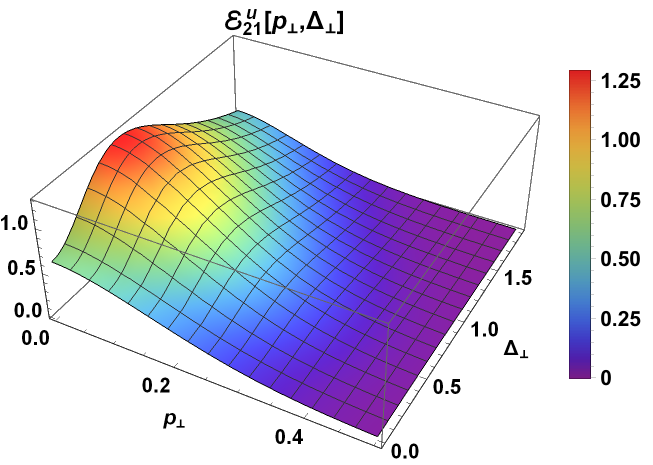}
						\hspace{0.05cm}
						(b)\includegraphics[width=7.3cm]{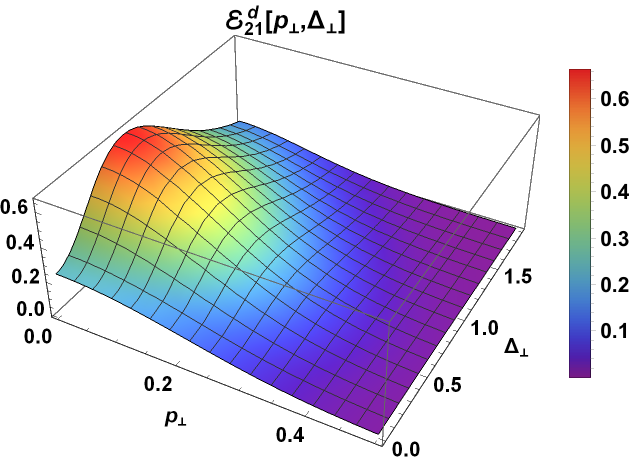}
						\hspace{0.05cm}
						(c)\includegraphics[width=7.3cm]{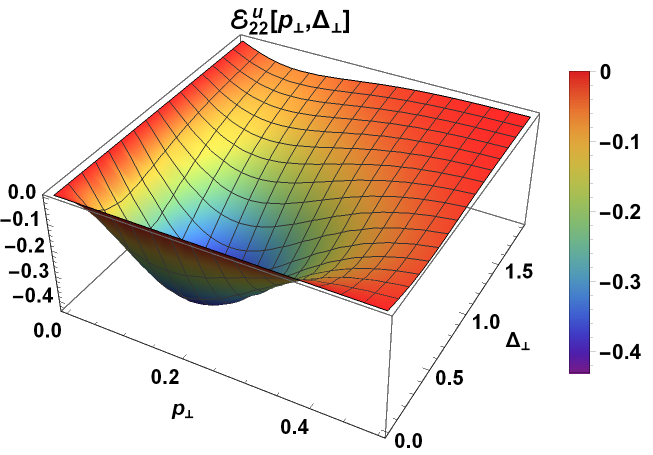}
						\hspace{0.05cm}
						(d)\includegraphics[width=7.3cm]{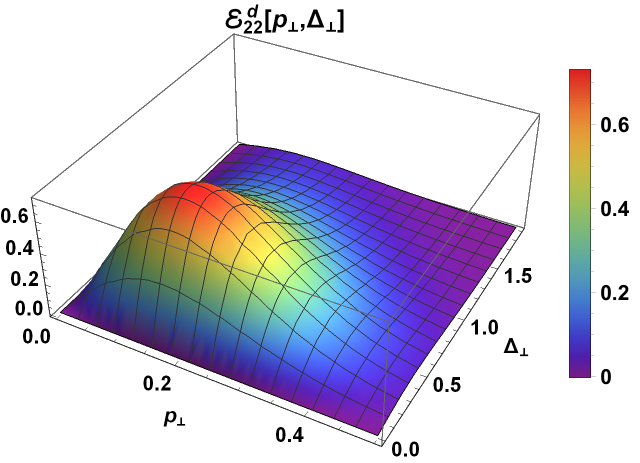}
						\hspace{0.05cm}
						(e)\includegraphics[width=7.3cm]{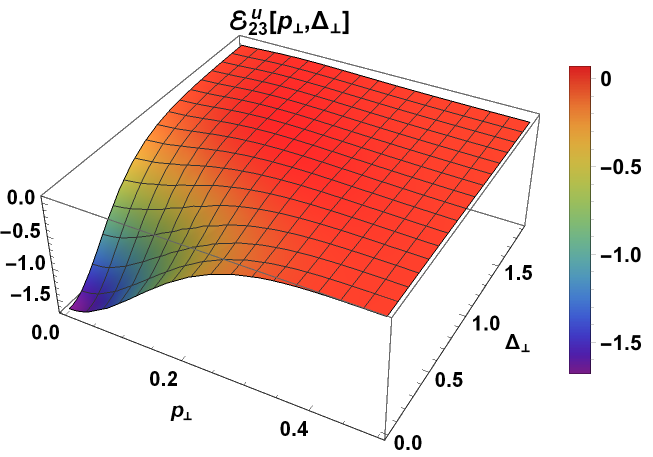}
						\hspace{0.05cm}
						(f)\includegraphics[width=7.3cm]{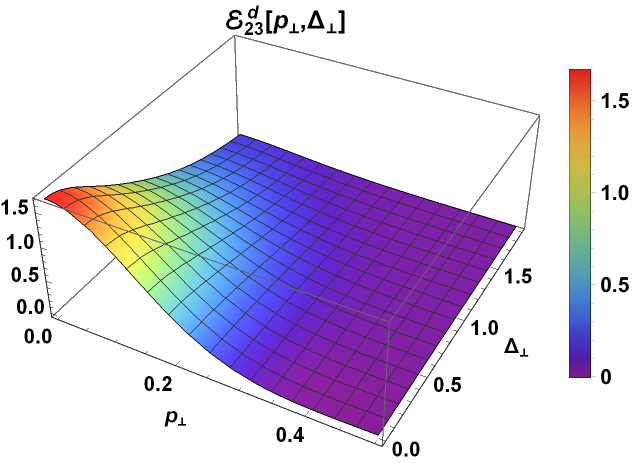}
						\hspace{0.05cm}
						(g)\includegraphics[width=7.3cm]{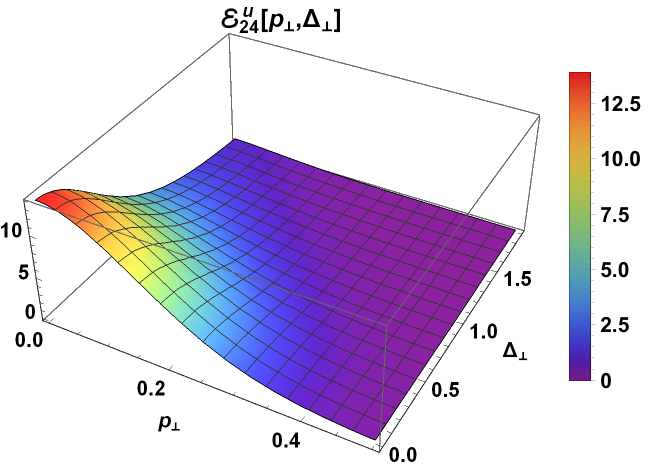}
						\hspace{0.05cm}
						(h)\includegraphics[width=7.3cm]{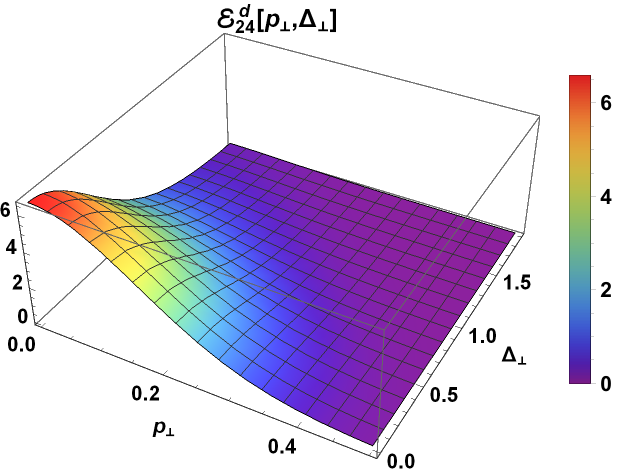}
						\hspace{0.05cm}\\
					\end{minipage}
				\caption{\label{fig3dTMFFE1} The sub-leading twist TMFFs 		
						$x E_{2,1}^{\nu}(x, p_{\perp})$,
						$x E_{2,2}^{\nu}(x, p_{\perp})$,
						$x E_{2,3}^{\nu}(x, p_{\perp})$, and
						$x E_{2,4}^{\nu}(x, p_{\perp})$
				are	plotted about ${ p_\perp}$ and ${{ \Delta_\perp}}$ keeping ${\bfp} \parallel {\Dp}$. In sequential order, $u$ and $d$ quarks are in the left and right columns.
					}
			\end{figure*}
		\begin{figure*}
				\centering
				\begin{minipage}[c]{0.98\textwidth}
						(a)\includegraphics[width=7.3cm]{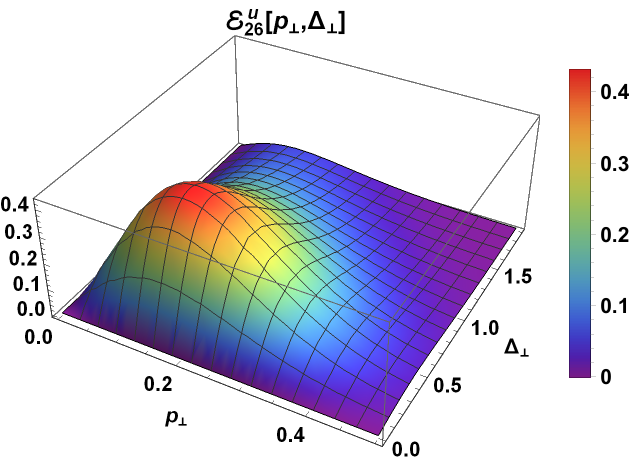}
						\hspace{0.05cm}
						(b)\includegraphics[width=7.3cm]{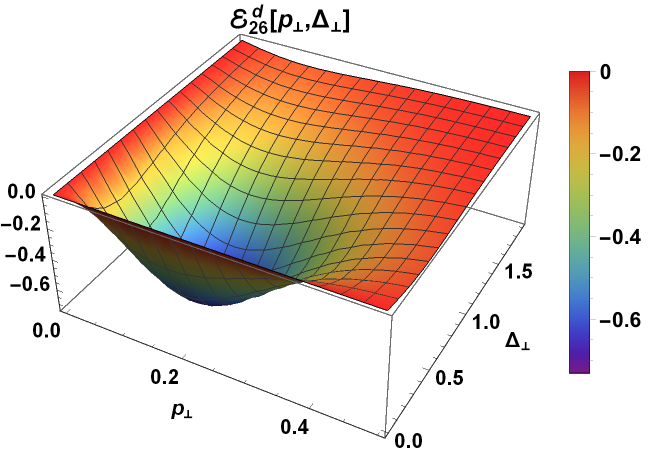}
						\hspace{0.05cm}
						(c)\includegraphics[width=7.3cm]{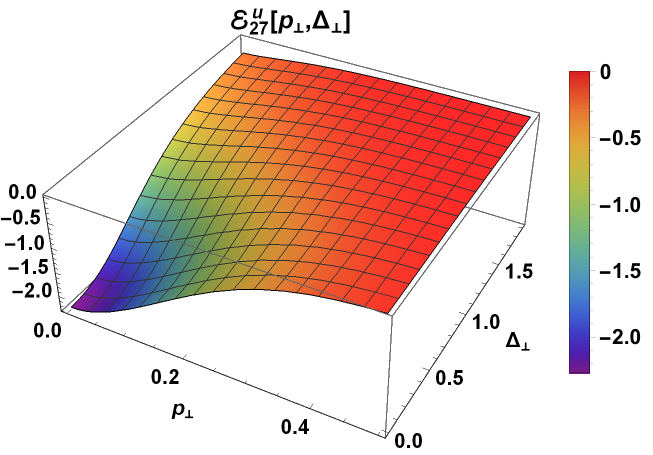}
						\hspace{0.05cm}
						(d)\includegraphics[width=7.3cm]{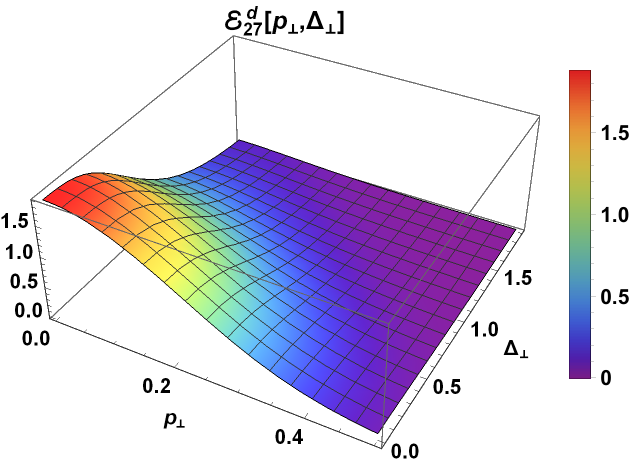}
						\hspace{0.05cm}
						(e)\includegraphics[width=7.3cm]{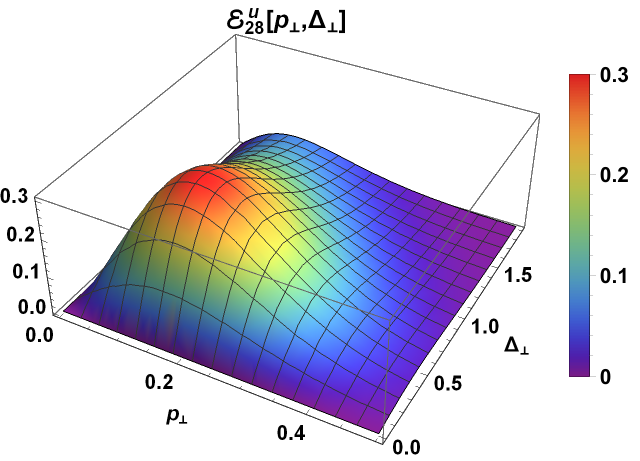}
						\hspace{0.05cm}
						(f)\includegraphics[width=7.3cm]{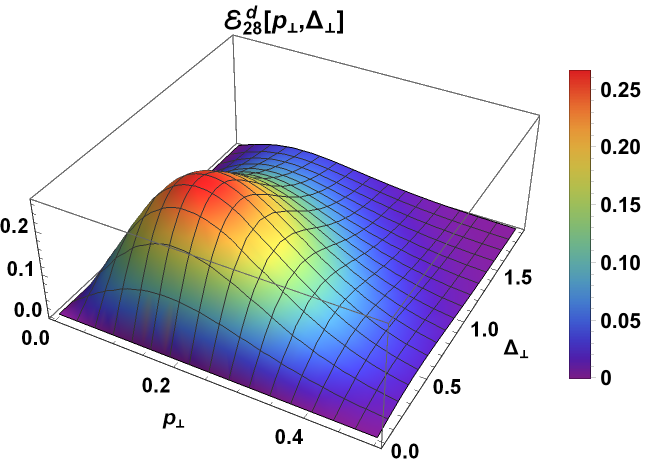}
						\hspace{0.05cm}
						\\
					\end{minipage}
				\caption{\label{fig3dTMFFE2} The sub-leading twist TMFFs 		
						$x E_{2,6}^{\nu}(x, p_{\perp})$,
						$x E_{2,7}^{\nu}(x, p_{\perp})$,
						and
						$x E_{2,8}^{\nu}(x, p_{\perp})$
						are	plotted about ${ p_\perp}$ and ${{ \Delta_\perp}}$ keeping ${\bfp} \parallel {\Dp}$. In sequential order, $u$ and $d$ quarks are in the left and right columns.
					}
			\end{figure*}
		\begin{figure*}
				\centering
				\begin{minipage}[c]{0.98\textwidth}
						(a)\includegraphics[width=7.3cm]{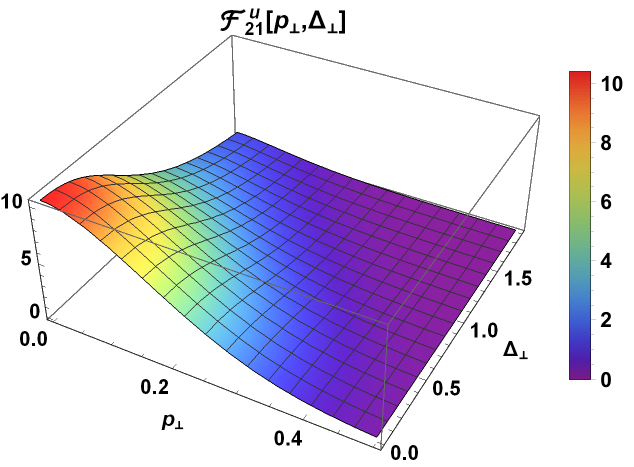}
						\hspace{0.05cm}
						(b)\includegraphics[width=7.3cm]{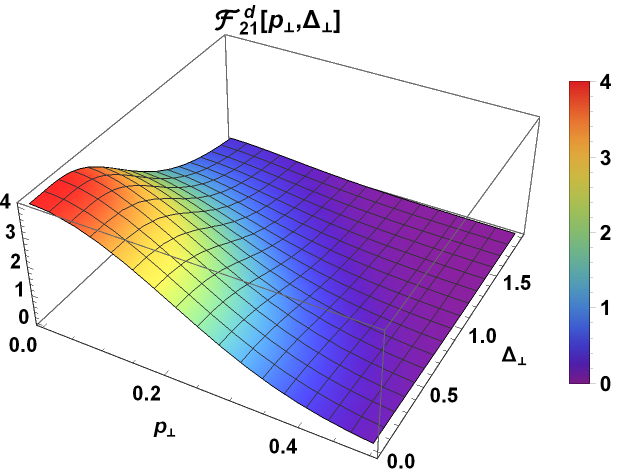}
						\hspace{0.05cm}
						(c)\includegraphics[width=7.3cm]{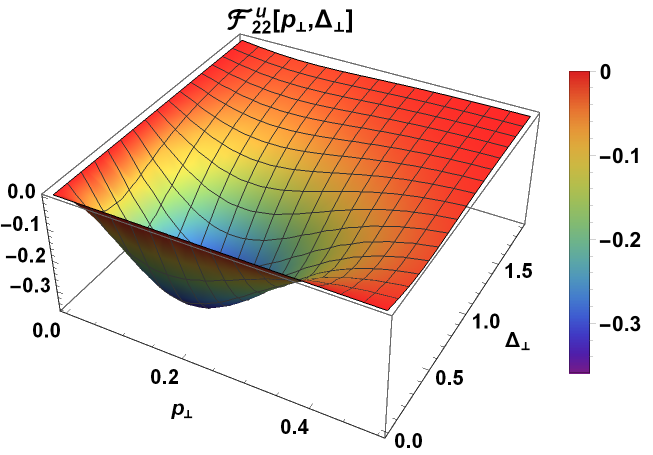}
						\hspace{0.05cm}
						(d)\includegraphics[width=7.3cm]{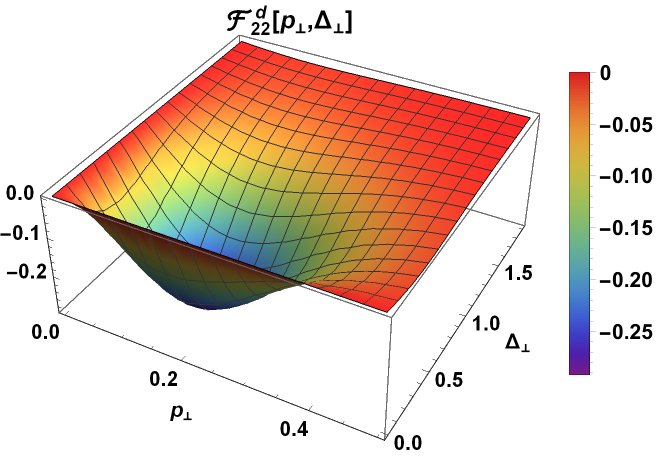}
						\hspace{0.05cm}
						(e)\includegraphics[width=7.3cm]{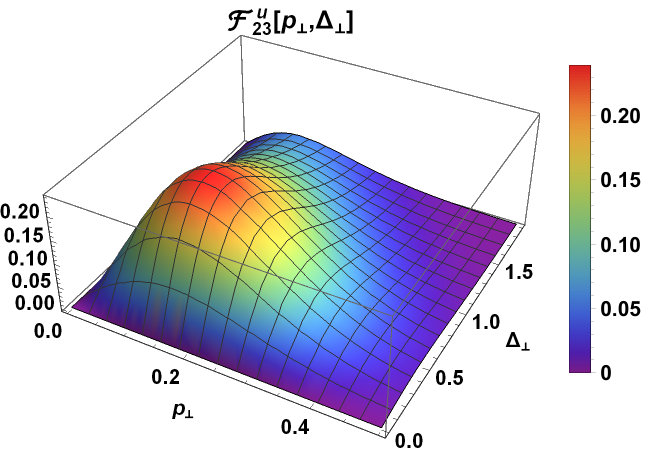}
						\hspace{0.05cm}
						(f)\includegraphics[width=7.3cm]{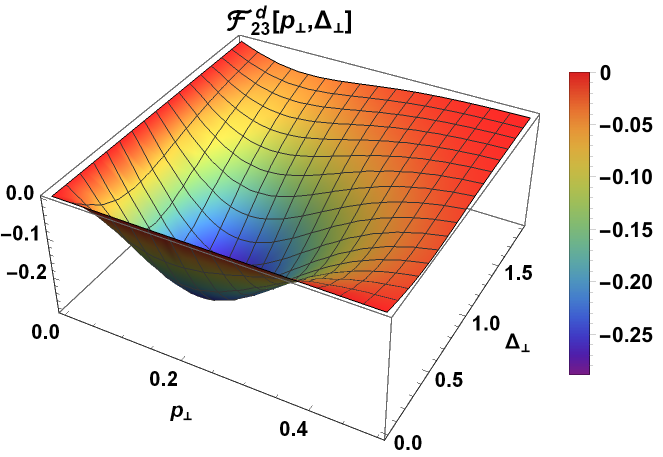}
						\hspace{0.05cm}
						(g)\includegraphics[width=7.3cm]{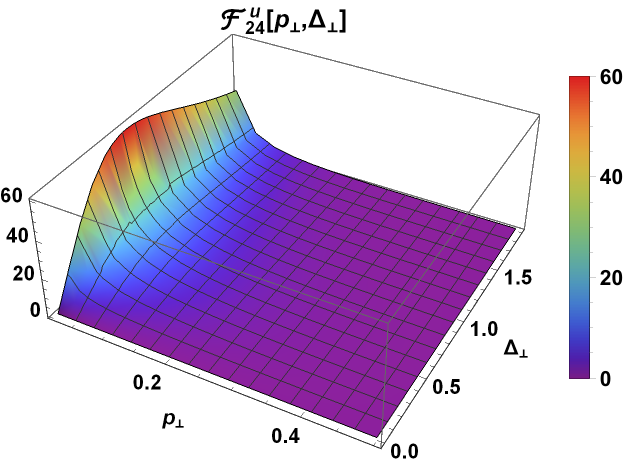}
						\hspace{0.05cm}
						(h)\includegraphics[width=7.3cm]{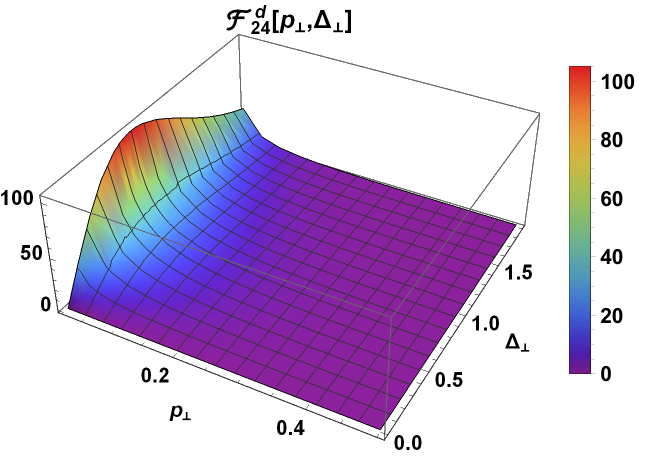}
						\hspace{0.05cm}\\
					\end{minipage}
				\caption{\label{fig3dTMFFF1} The sub-leading twist TMFFs 		
						$x F_{2,1}^{\nu}(x, p_{\perp})$,
						$x F_{2,2}^{\nu}(x, p_{\perp})$,
						$x F_{2,3}^{\nu}(x, p_{\perp})$, and
						$x F_{2,4}^{\nu}(x, p_{\perp})$
					are	about ${ p_\perp}$ and ${{ \Delta_\perp}}$ keeping ${\bfp} \parallel {\Dp}$. In sequential order, $u$ and $d$ quarks are in the left and right columns.
					}
			\end{figure*}
		\begin{figure*}
				\centering
				\begin{minipage}[c]{0.98\textwidth}
						(a)\includegraphics[width=7.3cm]{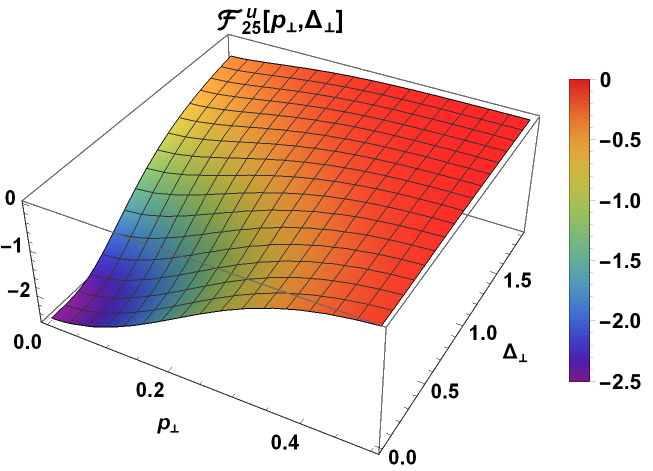}
						\hspace{0.05cm}
						(b)\includegraphics[width=7.3cm]{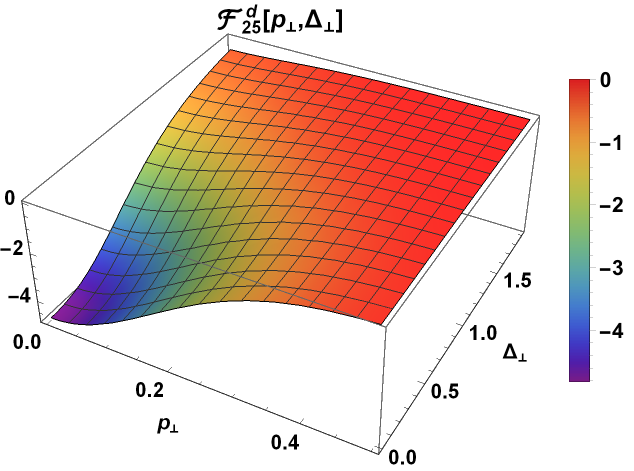}
						\hspace{0.05cm}
						(c)\includegraphics[width=7.3cm]{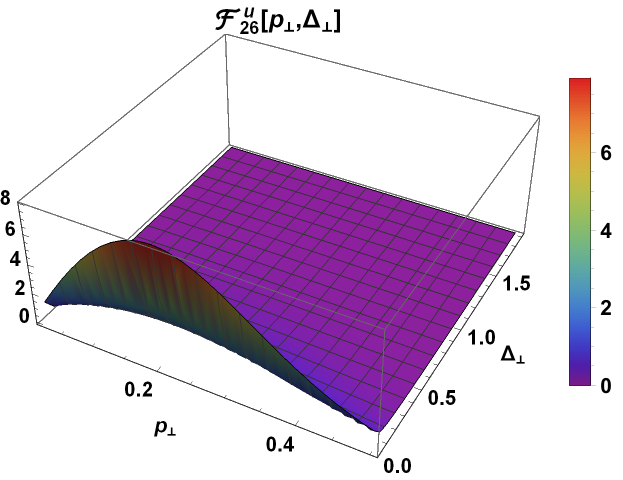}
						\hspace{0.05cm}
						(d)\includegraphics[width=7.3cm]{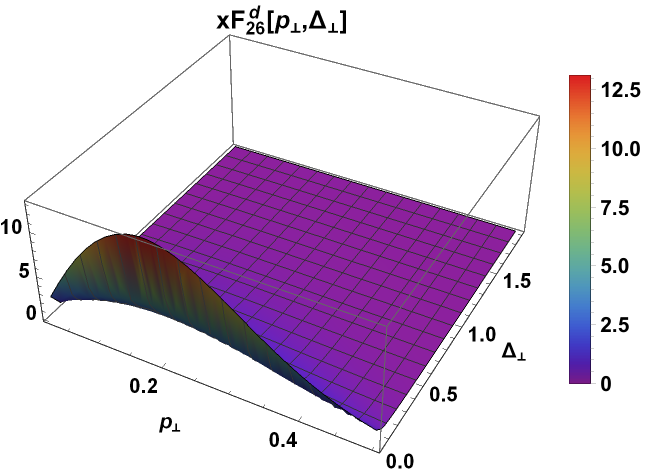}
						\hspace{0.05cm}
						(e)\includegraphics[width=7.3cm]{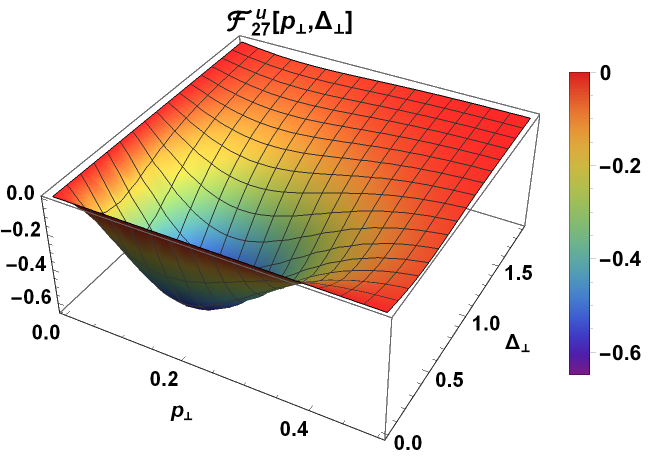}
						\hspace{0.05cm}
						(f)\includegraphics[width=7.3cm]{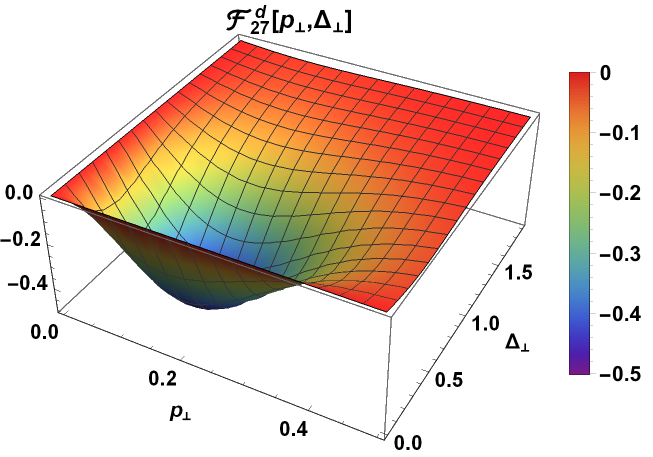}
						\hspace{0.05cm}
						(g)\includegraphics[width=7.3cm]{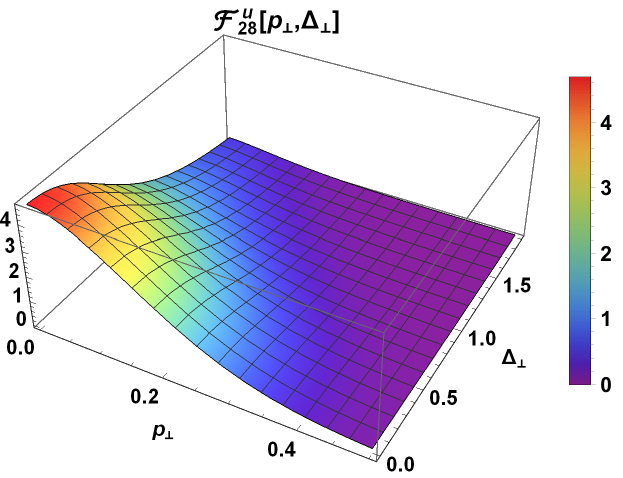}
						\hspace{0.05cm}
						(h)\includegraphics[width=7.3cm]{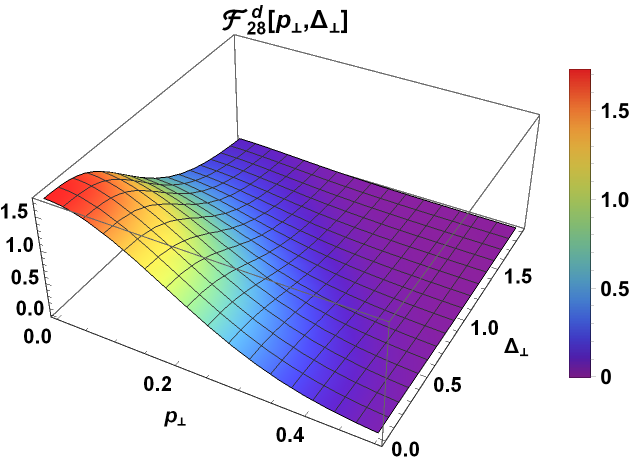}
						\hspace{0.05cm}\\
					\end{minipage}
				\caption{\label{fig3dTMFFF2} The sub-leading twist TMFFs 		
						$x F_{2,5}^{\nu}(x, p_{\perp})$,
						$x F_{2,6}^{\nu}(x, p_{\perp})$,
						$x F_{2,7}^{\nu}(x, p_{\perp})$, and
						$x F_{2,8}^{\nu}(x, p_{\perp})$
					are	plotted about ${ p_\perp}$ and ${{ \Delta_\perp}}$ keeping ${\bfp} \parallel {\Dp}$. In sequential order, $u$ and $d$ quarks are in the left and right columns.
					}
			\end{figure*}
		\begin{figure*}
				\centering
				\begin{minipage}[c]{0.98\textwidth}
						(a)\includegraphics[width=7.3cm]{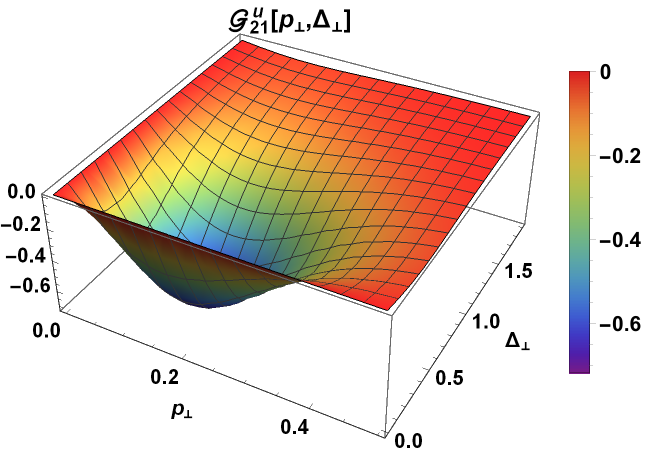}
						\hspace{0.05cm}
						(b)\includegraphics[width=7.3cm]{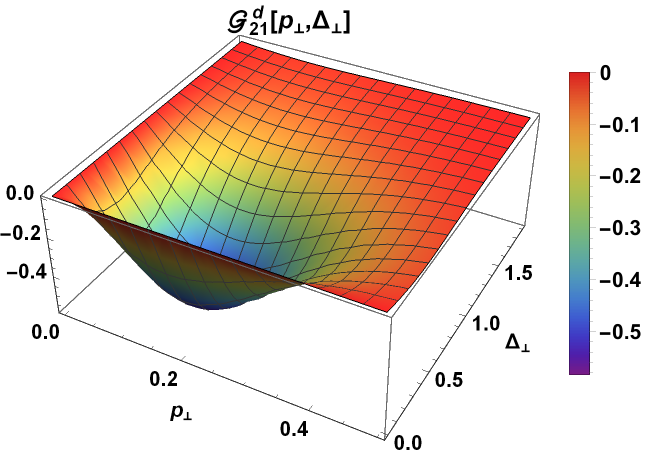}
						\hspace{0.05cm}
						(c)\includegraphics[width=7.3cm]{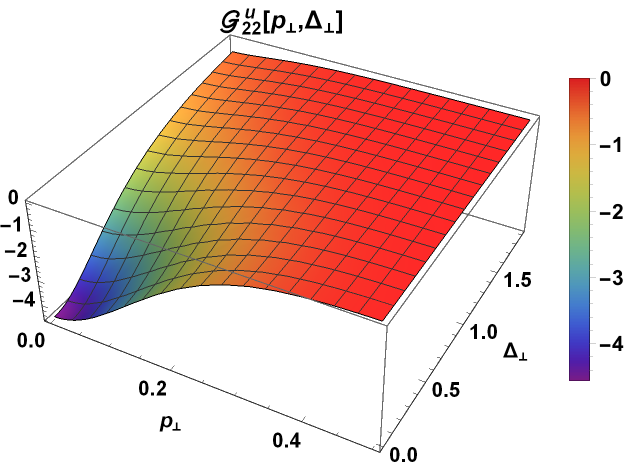}
						\hspace{0.05cm}
						(d)\includegraphics[width=7.3cm]{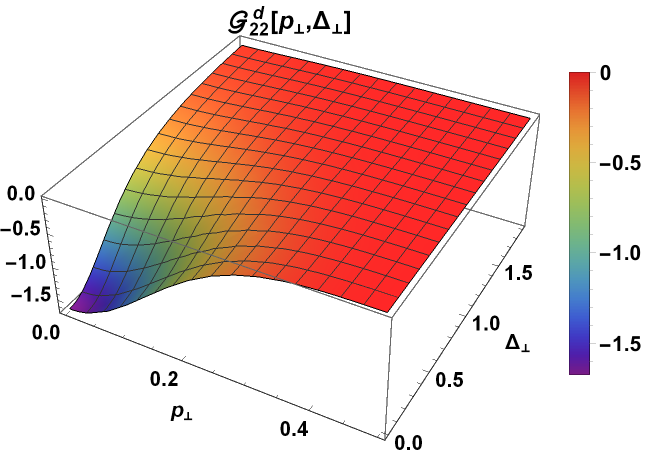}
						\hspace{0.05cm}
						(e)\includegraphics[width=7.3cm]{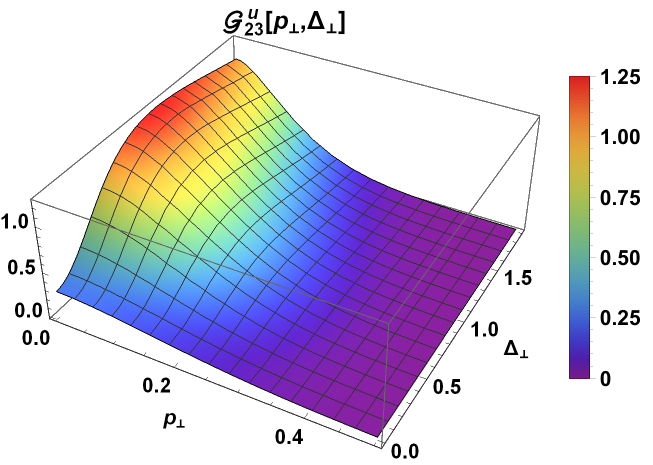}
						\hspace{0.05cm}
						(f)\includegraphics[width=7.3cm]{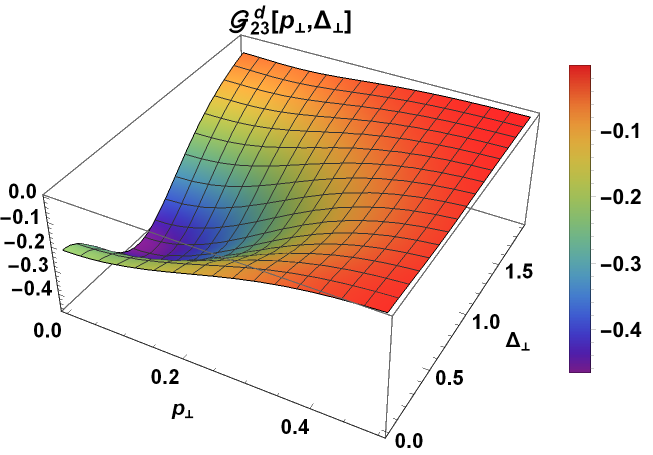}
						\hspace{0.05cm}
						(g)\includegraphics[width=7.3cm]{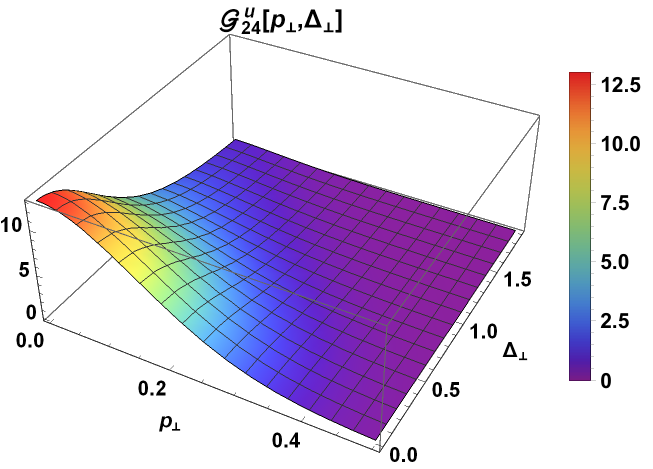}
						\hspace{0.05cm}
						(h)\includegraphics[width=7.3cm]{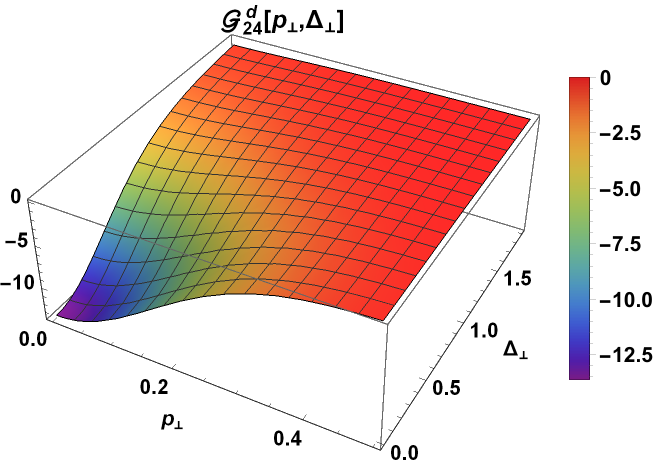}
						\hspace{0.05cm}\\
					\end{minipage}
				\caption{\label{fig3dTMFFG1} The sub-leading twist TMFFs 		
						$x G_{2,1}^{\nu}(x, p_{\perp})$,
						$x G_{2,2}^{\nu}(x, p_{\perp})$,
						$x G_{2,3}^{\nu}(x, p_{\perp})$, and
						$x G_{2,4}^{\nu}(x, p_{\perp})$
					are	about ${ p_\perp}$ and ${{ \Delta_\perp}}$ keeping ${\bfp} \parallel {\Dp}$. In sequential order, $u$ and $d$ quarks are in the left and right columns.
					}
			\end{figure*}
		\begin{figure*}
				\centering
				\begin{minipage}[c]{0.98\textwidth}
						(a)\includegraphics[width=7.3cm]{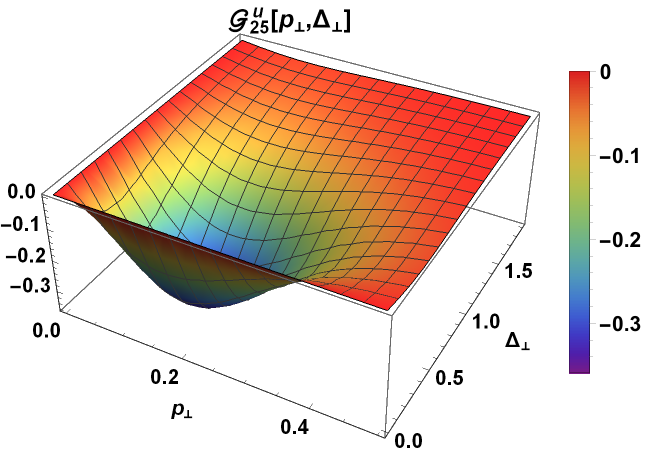}
						\hspace{0.05cm}
						(b)\includegraphics[width=7.3cm]{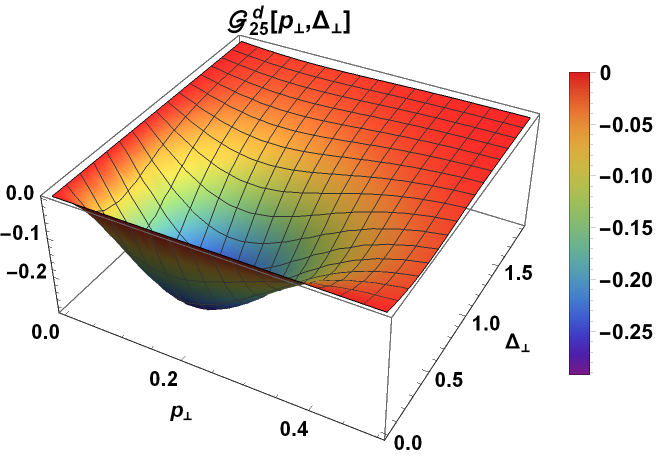}
						\hspace{0.05cm}
						(c)\includegraphics[width=7.3cm]{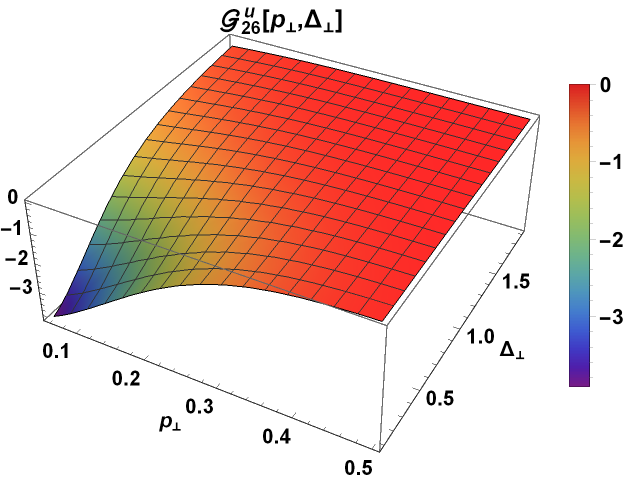}
						\hspace{0.05cm}
						(d)\includegraphics[width=7.3cm]{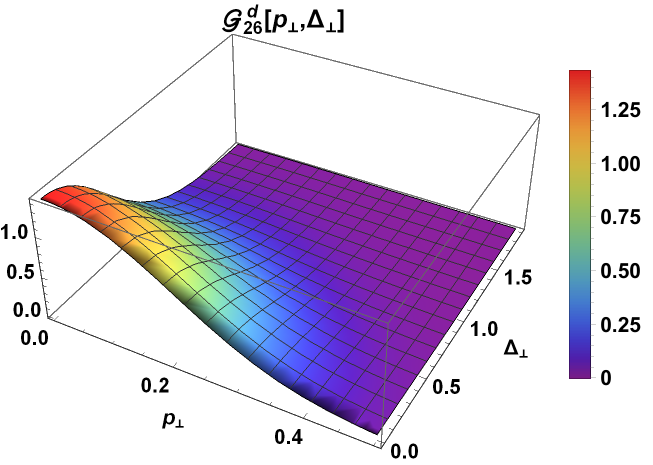}
						\hspace{0.05cm}
						(e)\includegraphics[width=7.3cm]{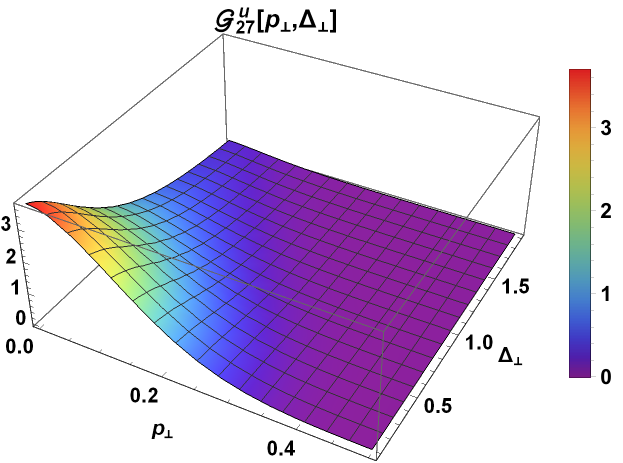}
						\hspace{0.05cm}
						(f)\includegraphics[width=7.3cm]{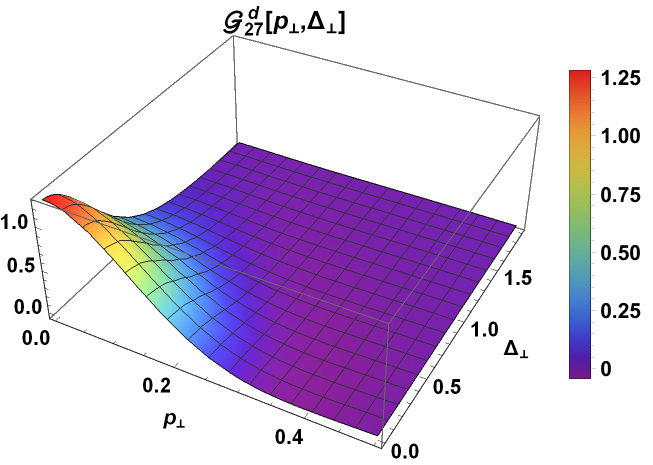}
						\hspace{0.05cm}
						(g)\includegraphics[width=7.3cm]{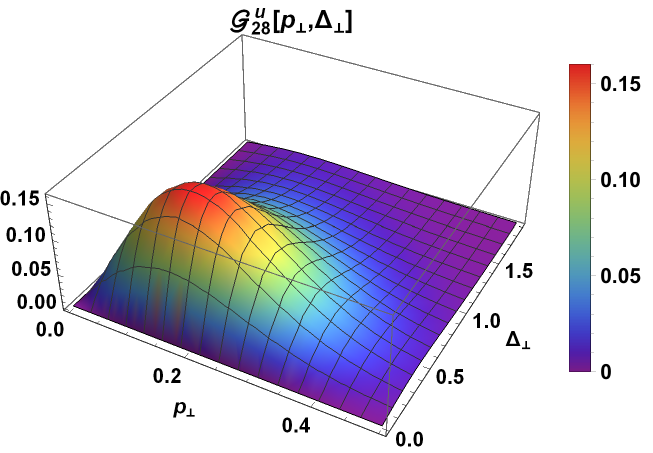}
						\hspace{0.05cm}
						(h)\includegraphics[width=7.3cm]{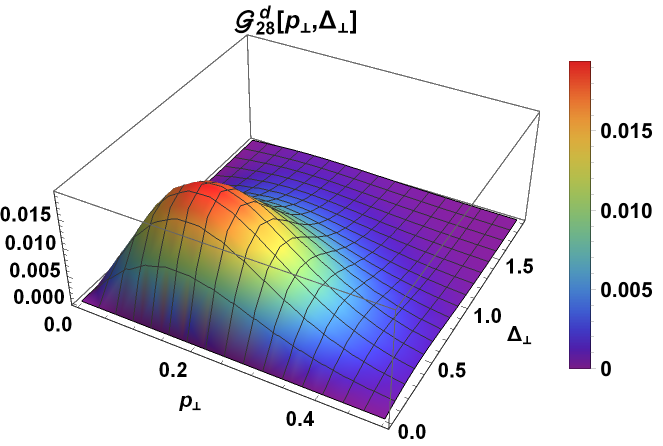}
						\hspace{0.05cm}\\
					\end{minipage}
				\caption{\label{fig3dTMFFG2} The sub-leading twist TMFFs 		
						$x G_{2,5}^{\nu}(x, p_{\perp})$,
						$x G_{2,6}^{\nu}(x, p_{\perp})$,
						$x G_{2,7}^{\nu}(x, p_{\perp})$, and
						$x G_{2,8}^{\nu}(x, p_{\perp})$
					are	plotted about ${ p_\perp}$ and ${{ \Delta_\perp}}$ keeping ${\bfp} \parallel {\Dp}$. In sequential order, $u$ and $d$ quarks are in the left and right columns.
					}
			\end{figure*}
		\begin{figure*}
				\centering
				\begin{minipage}[c]{0.98\textwidth}
						(a)\includegraphics[width=7.3cm]{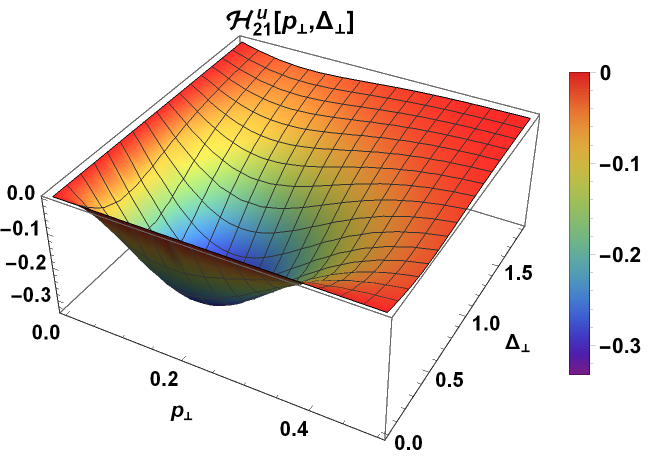}
						\hspace{0.05cm}
						(b)\includegraphics[width=7.3cm]{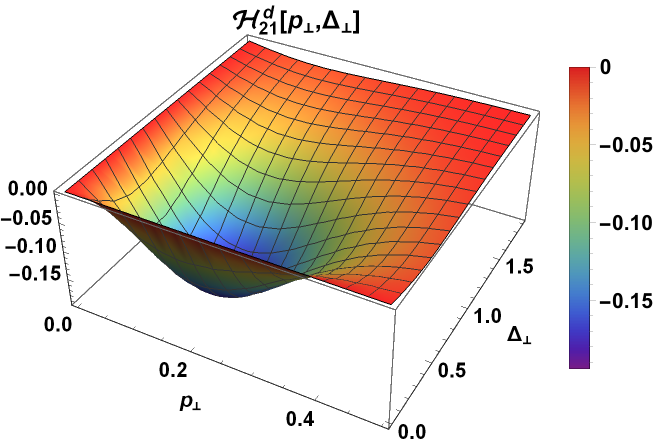}
						\hspace{0.05cm}
						(c)\includegraphics[width=7.3cm]{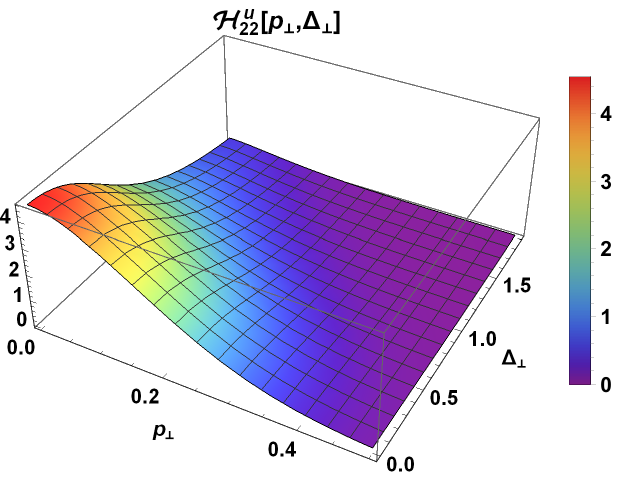}
						\hspace{0.05cm}
						(d)\includegraphics[width=7.3cm]{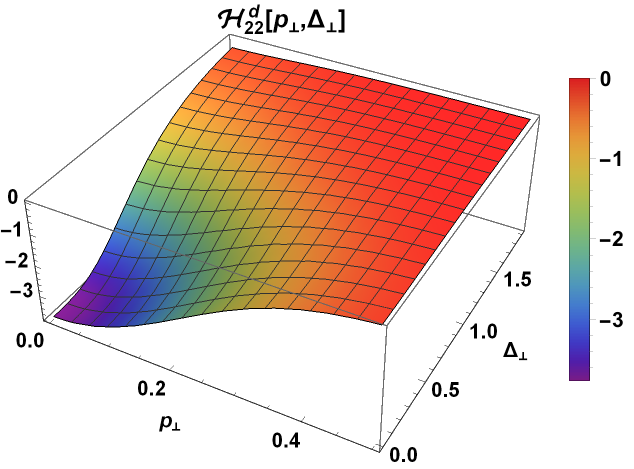}
						\hspace{0.05cm}
						(c)\includegraphics[width=7.3cm]{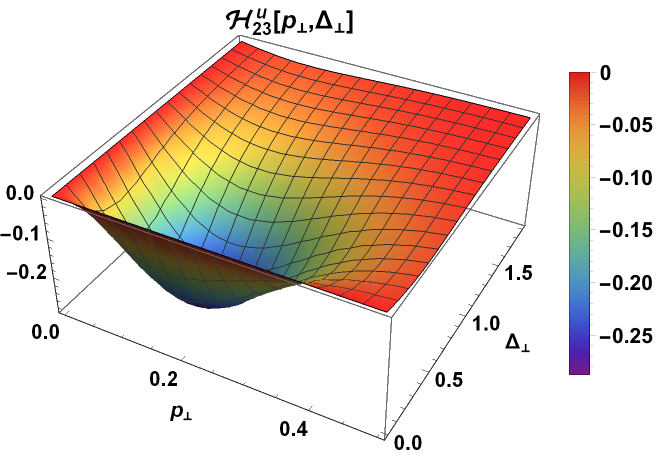}
						\hspace{0.05cm}
						(d)\includegraphics[width=7.3cm]{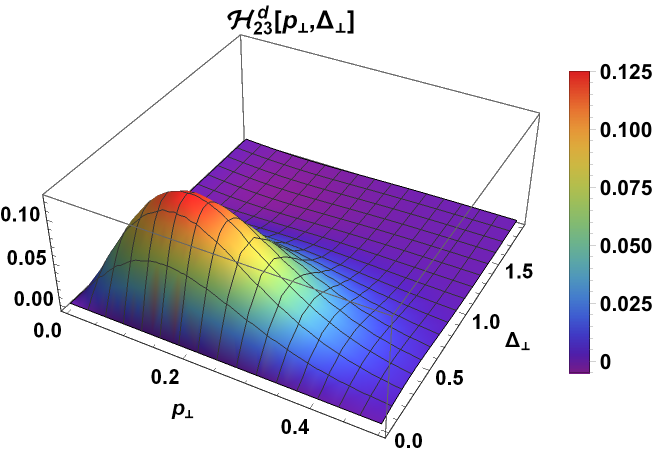}
						\hspace{0.05cm}
						\\
					\end{minipage}
				\caption{\label{fig3dTMFFH1} The sub-leading twist TMFFs 		
						$x H_{2,1}^{\nu}(x, p_{\perp})$,
						$x H_{2,2}^{\nu}(x, p_{\perp})$,
						and
						$x H_{2,3}^{\nu}(x, p_{\perp})$
					are	plotted about ${ p_\perp}$ and ${{ \Delta_\perp}}$ keeping ${\bfp} \parallel {\Dp}$. In sequential order, $u$ and $d$ quarks are in the left and right columns.
					}
			\end{figure*}
		\begin{figure*}
				\centering
				\begin{minipage}[c]{0.98\textwidth}
						(a)\includegraphics[width=7.3cm]{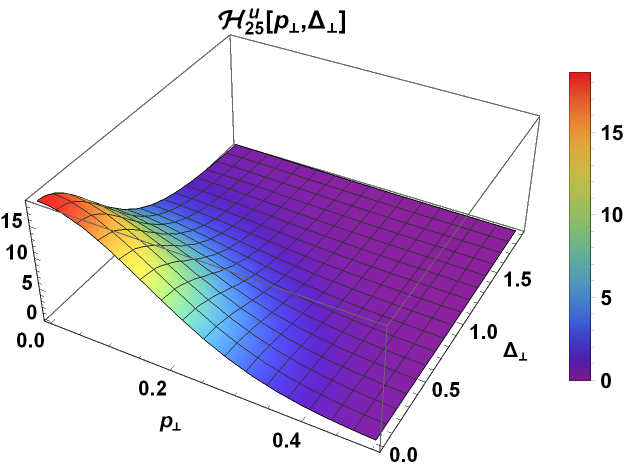}
						\hspace{0.05cm}
						(b)\includegraphics[width=7.3cm]{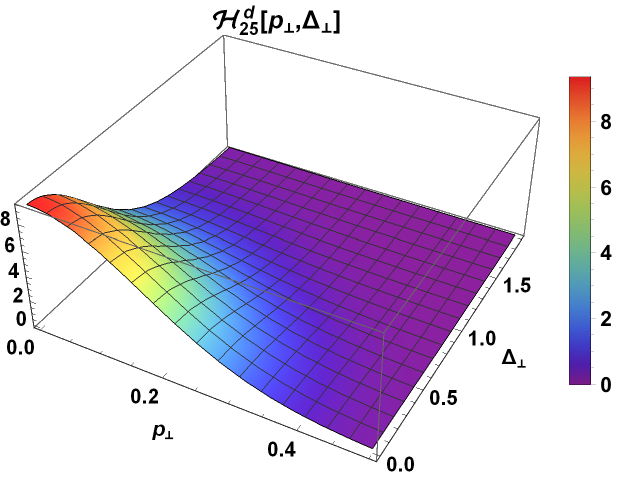}
						\hspace{0.05cm}
						(c)\includegraphics[width=7.3cm]{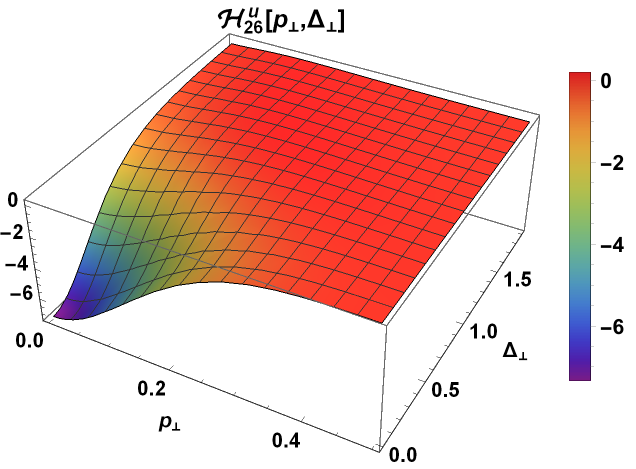}
						\hspace{0.05cm}
						(d)\includegraphics[width=7.3cm]{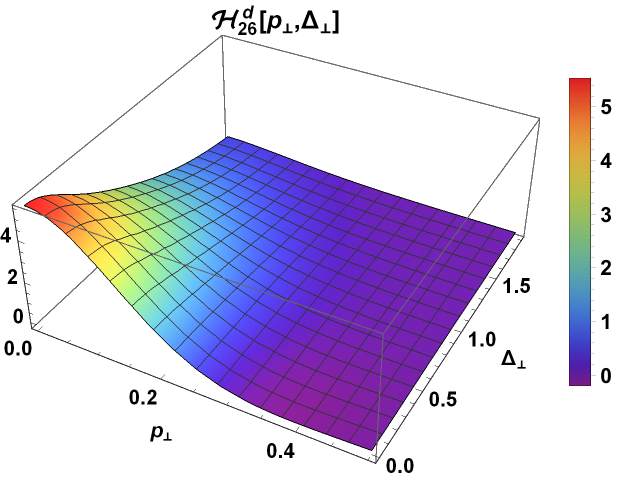}
						\hspace{0.05cm}
						(e)\includegraphics[width=7.3cm]{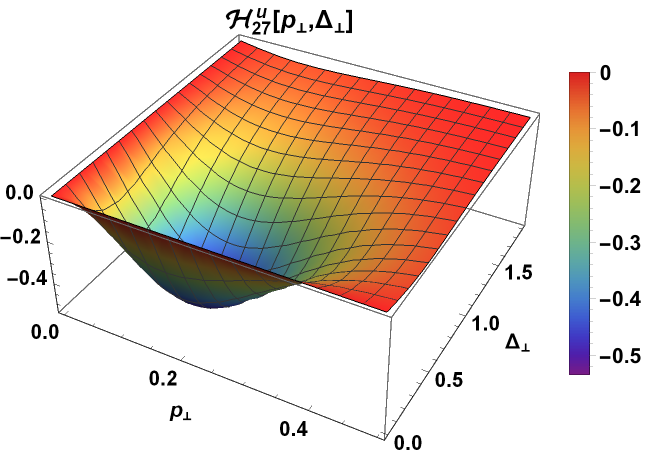}
						\hspace{0.05cm}
						(f)\includegraphics[width=7.3cm]{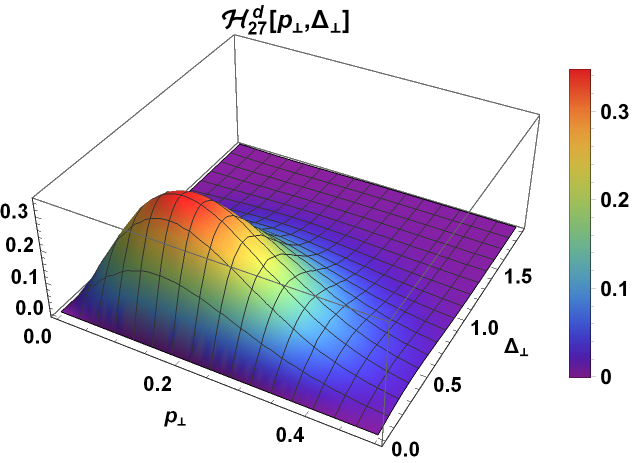}
						\hspace{0.05cm}
						(g)\includegraphics[width=7.3cm]{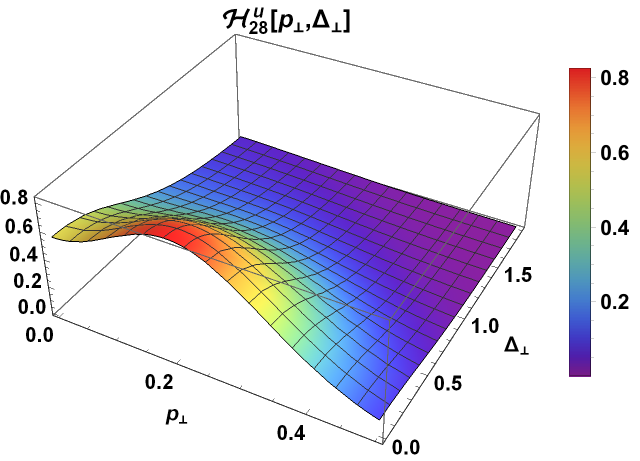}
						\hspace{0.05cm}
						(h)\includegraphics[width=7.3cm]{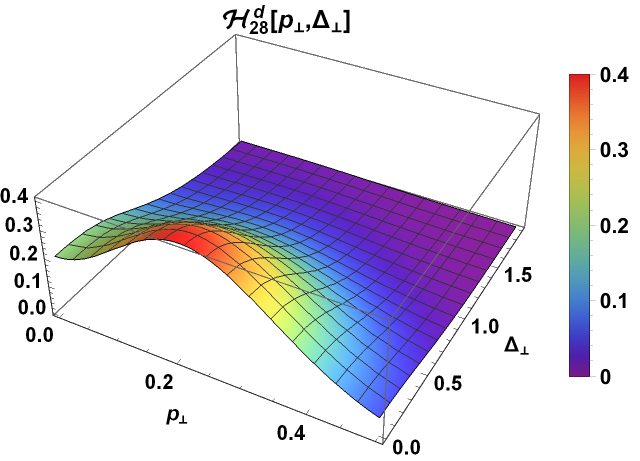}
						\hspace{0.05cm}\\
					\end{minipage}
				\caption{\label{fig3dTMFFH2} The sub-leading twist TMFFs 		
						$x H_{2,5}^{\nu}(x, p_{\perp})$,
						$x H_{2,6}^{\nu}(x, p_{\perp})$,
						$x H_{2,7}^{\nu}(x, p_{\perp})$, and
						$x H_{2,8}^{\nu}(x, p_{\perp})$
					are	plotted about ${ p_\perp}$ and ${{ \Delta_\perp}}$ keeping ${\bfp} \parallel {\Dp}$. In sequential order, $u$ and $d$ quarks are in the left and right columns.
					}
			\end{figure*}
  TMFFs have been  derived from GTMDs by integrating over the longitudinal momentum fraction $x$. In Figs. (\ref{fig3dTMFFE1}) to (\ref{fig3dTMFFH2}), TMFFs are plotted against the transverse momentum of the quark ${p_\perp}$ and the transverse momentum transfer to proton ${\Delta_\perp}$. Plots reveal that amplitude approaches zero as the transverse momentum transfer ${\Delta_\perp}$ is equal to or exceeds $1.5~\mathrm{GeV}$. This suggests that, at the model scale, such momentum transfer is not feasible for any polarization configuration. This observation is consistent with that for twist-$4$ GTMDs described in Ref. \cite{Sharma:2023tre}. In Fig. (\ref{fig3dTMFFE1}), TMFFs $E_{2,1},~E_{2,2},~E_{2,3}$, and $~E_{2,4}$ have been plotted, which correspond with the Dirac matrix structure $\Gamma=1$. TMFF $E_{2,1}$ has been plotted in Figs. \ref{fig3dTMFFE1} $(a)$ and \ref{fig3dTMFFE1} $(b)$. This TMFF shows quark flavor symmetry, as expected from the Eqs. \eqref{e21s} and \eqref{e21a}. When compared with the plots of GTMD $xE_{2,1}$ in Figs. \ref{fig3dPDE1} $(a)$ and \ref{fig3dPDE1} $(b)$, a small change in the amplitude of TMFF is observed for $u$ and $d$ quark, which arises as a result of the integration of GTMD $xE_{2,1}$ over the longitudinal momentum fraction $x$. TMFF $E_{2,2}$ is plotted in Figs. \ref{fig3dTMFFE1} $(c)$ and \ref{fig3dTMFFE1} $(d)$, which is found to change its polarity on the change of quark flavor from $u$ to $d$, in line with the previous observation of the corresponding GTMD. TMFF $E_{2,3}$ is found to show an anti-symmetric nature for the quark flavors $u$ and $d$ and is plotted in Figs. \ref{fig3dPDE1} $(e)$ and \ref{fig3dPDE1} $(f)$. In Figs. \ref{fig3dTMFFE1} $(g)$ and \ref{fig3dTMFFE1} $(h)$, TMFF $E_{2,4}$ has been plotted, and the plots suggest that this TMFF shows quark flavor symmetry. Now, we discuss the TMFFs $E_{2,6},~E_{2,7},$ and $~E_{2,8}$ corresponding to the Dirac matrix structure $\Gamma=\gamma_5$. TMFF $E_{2,6}$ is plotted in Figs. \ref{fig3dTMFFE2} $(a)$ and \ref{fig3dTMFFE2} $(b)$ for $u$ and $d$ quark, respectively. As expected, this TMFF shows behavior similar to TMFF $E_{2,2}$ but with opposite polarity. TMFF $E_{2,7}$ has been plotted in Figs. \ref{fig3dTMFFE2} $(c)$ and \ref{fig3dTMFFE2} $(d)$ and is found to be anti-symmetric for the quark flavors $u$ and $d$. TMFF $E_{2,8}$ plotted in Figs. \ref{fig3dTMFFE2} $(e)$ and \ref{fig3dTMFFE2} $(f)$ is found to be symmetric for the quark flavors $u$ and $d$. Next, we discuss the TMFFs $F_{2,1},~F_{2,3},~F_{2,4},$ and $~F_{2,7}$ pertaining to the Dirac matrix structure $\Gamma=\gamma^j$. These TMFFs are plotted in Figs. (\ref{fig3dTMFFF1}) and (\ref{fig3dTMFFF2}). TMFF $F_{2,1}$, which changes polarity of the distribution with the change in the quark flavor from $u$ to $d$, is plotted in Figs. \ref{fig3dTMFFF1} $(a)$ and \ref{fig3dTMFFF1} $(b)$ for the quarks $u$ and $d$, respectively. $3$-dimensional plots of TMFFs $F_{2,3}$ and $F_{2,4}$ are given in Figs. \ref{fig3dTMFFF1} $(e)$ to \ref{fig3dTMFFF1} $(h)$ for the quarks $u$ and $d$, accordingly. Observing the plots, TMFF $F_{2,3}$ shows quark flavor anti-symmetry, and TMFF $F_{2,4}$ exhibits divergence for zero momentum transfer. TMFF $F_{2,7}$ is plotted in Figs. \ref{fig3dTMFFF2} $(e)$ and \ref{fig3dTMFFF2} $(f)$ for $u$ and $d$ quark and is found to be quark flavor symmetric, indicating that this TMFF remains the same irrespective of the active quark flavor.
\par 
Considering the Dirac matrix structure $\Gamma=\gamma^j \gamma_5$, the TMFFs $G_{2,1},~G_{2,3},~G_{2,4}$, and $~G_{2,7}$ have been plotted in Figs. (\ref{fig3dTMFFG1}) and (\ref{fig3dTMFFG2}) for both quark flavors. In Figs. \ref{fig3dTMFFG1} $(a)$ and \ref{fig3dTMFFG1} $(b)$, TMFF $G_{2,1}$ has been plotted for $u$ and $d$ quark sequentially. Observing the plots, TMFF $G_{2,1}$ shows similar behavior for both active quark flavors. $3$-dimensional plots of TMFFs $G_{2,3}$ and $G_{2,4}$ are given in Figs. \ref{fig3dTMFFG1} $(e)$ to \ref{fig3dTMFFG1} $(h)$ for the quarks $u$ and $d$, accordingly. TMFF $G_{2,3}$ changes sign in amplitude when the quark flavor is changed, while TMFF $G_{2,4}$ exhibits anti-symmetric behavior. In Figs. \ref{fig3dTMFFG2} $(e)$ and \ref{fig3dTMFFG2} $(f)$, TMFF $G_{2,7}$ is plotted sequentially for $u$ and $d$ quark. TMFF $G_{2,7}$ is symmetric for both quark flavors. The highest amplitude of the active $d$ quark distribution was found to be approximately one-third of the highest amplitude of the active $u$ quark distribution for this TMFF.
\par
Finally, we conclude our discussion by examining TMFFs $H_{2,1},~H_{2,2},~H_{2,6}$, and $~H_{2,8}$ corresponding to the Dirac matrix structures $\Gamma=i\sigma^{ij} \gamma_5$ and $\Gamma=i\sigma^{+-} \gamma_5$. These TMFFs have been plotted in Figs. (\ref{fig3dTMFFH1}) and (\ref{fig3dTMFFH2}) for $u$ and $d$ quark, sequentially. TMFF $H_{2,1}$ and $H_{2,2}$ are shown in Figs. \ref{fig3dTMFFH1} $(a)$ to \ref{fig3dTMFFH1} $(f)$ for $u$ and $d$ quark, respectively. TMFF $H_{2,1}$ maintains the same polarity when changing the active quark flavor, while TMFF $H_{2,2}$ exhibits exact anti-symmetric behavior. TMFF $H_{2,6}$ is plotted in Figs. \ref{fig3dTMFFH2} $(a)$ and \ref{fig3dTMFFH2} $(b)$ for $u$ and $d$ quark, respectively. It changes polarity with a change in quark flavor from $u$ to $d$."It reverses polarity with the transition in quark flavor from $u$ to $d$ . In Figs. \ref{fig3dTMFFH2} $(e)$ and \ref{fig3dTMFFH2} $(f)$, TMFF $H_{2,8}$ is plotted, showing quark flavor symmetry as observed from the plots.
%
%
%
	\section{Conclusion}\label{seccon}
	\noindent
In this work, we investigate the sub-leading twist GTMDs of the proton by using the LFQDM. We obtain the sub-leading twist GTMDs by integrating the off-diagonal, un-integrated quark-quark GPCF correlator over the quark's light-front energy. We present the explicit equations of these sub-leading twist GTMDs by employing the sub-leading twist Dirac matrix structure in a quark-quark GTMD correlator and solving the associated parametrization equations. Our analysis includes all potential active quark-diquark configurations corresponding to isoscalar-scalar, isoscalar-vector, and isovector-vector proton states, considering both $u$ and $d$ flavors for the active quark. Notably, our results reveal that the GTMDs $E_{2,5}$ and $H_{2,4}$, associated with the Dirac matrix structures $1$ and $\Gamma = i\sigma^{ij} \gamma_5$, respectively, vanish within the LFQDM framework. We explore the multi-dimensional nature of these sub-leading twist GTMDs through $2$-dimensional and $3$-dimensional plots, varying one or two variables while holding others fixed or integrated. Specifically, we analyze GTMDs corresponding to Dirac matrix structures $\Gamma = 1, \gamma_5$, and their interplay with various TMDs as outlined in Eqs. \eqref{tmd5} to \eqref{tmd16}, emphasizing their theoretical, experimental, and phenomenological relevance. The sub-leading twist GTMDs derived in our study correspond to the sub-leading twist TMDs in the limit of zero momentum transfer ($\Delta = 0$), as referenced in Ref. \cite{sstwist3}. Furthermore, we observe that the amplitude of nearly all sub-leading twist TMFFs diminishes as the transverse momentum transfer $\Dp$ exceeds $1.5 \mathrm{GeV}$. This finding imposes constraints on the possible momentum transfer $\Dp$ at the model scale for all polarization configurations of the sub-leading twist TMFFs.
\par
In summary, this study enhances our understanding of proton structure by elucidating the complex dynamics captured by sub-leading twist GTMDs within the LFQDM framework. These insights are crucial for future research in hadronic physics, particularly in exploring higher-order twist effects on parton dynamics. While the experimental observation of GTMDs has traditionally involved indirect extraction from processes like DVCS and SIDIS, recent studies \cite{Echevarria:2022ztg, Bhattacharya:2017cal} suggest direct access to GTMDs through the DDY process. This should motivate the design of sophisticated experiments to probe the proton's multi-dimensional structure in future research endeavors.
%
%
	\section{Acknowledgement}
H.D. would like to thank  the Science and Engineering Research Board, Anusandhan-National Research Foundation, Government of India under the scheme SERB-POWER Fellowship (Ref No. SPF/2023/000116) for financial support.

	%

	

	%
\end{document}